\RequirePackage{fix-cm}
\documentclass[onecolumn,natbib]{svjour3}
\smartqed  

\pdfoutput=1

\usepackage{rotating}
\usepackage{graphicx}

\usepackage{empheq}

\usepackage{times}
\usepackage{amstext, amsmath, amssymb}
\usepackage{subfig}

\usepackage{paralist}
\usepackage{epstopdf}

\usepackage[utf8]{inputenc}
\usepackage[T1]{fontenc}
\usepackage{textcomp}

\usepackage[%
    pdftex,
    bookmarks=true,         
    unicode=false,          
    pdftoolbar=true,        
    pdfmenubar=true,        
    pdffitwindow=false,     
    pdfstartview={FitH},    
    pdftitle={An Analytic Expression for the Distribution of the Generalized Shiryaev-Roberts Diffusion},    
    pdfauthor={Aleksey S. Polunchenko and Grigory Sokolov},     
    pdfsubject={An Analytic Expression for the Distribution of the Generalized Shiryaev-Roberts Diffusion},   
    pdfcreator={Aleksey S. Polunchenko},   
    pdfproducer={MikTeX}, 
    pdfkeywords={Generalized Shiryaev-Roberts procedure;Kolmogorov forward equation;Markov diffusion processes;Quickest change-point detection;Sequential analysis}, 
    pdfnewwindow=true,      
    colorlinks=false,       
    linkcolor=red,          
    citecolor=green,        
    filecolor=magenta,      
    urlcolor=cyan           
]{hyperref}

\newcommand{\ignore}[1]{}

\renewcommand{\Pr}{\mathbb{P}} 
\DeclareMathOperator{\EV}{\mathbb{E}} 

\DeclareMathOperator{\LR}{\Lambda}

\DeclareMathOperator{\Res}{\mathrm{Res}}

\newcommand{\abs}[1]{\left\vert#1\right\vert}

\newcommand{\T}{T}

\renewcommand{\le}{\leqslant} 
\renewcommand{\ge}{\geqslant}

\DeclareMathOperator{\ONE}{\mathchoice{\rm 1\mskip-4.2mu l}{\rm 1\mskip-4.2mu l}{\rm 1\mskip-4.6mu l}{\rm 1\mskip-5.2mu l}}
\newcommand{\indicator}[1]{\ONE_{\left\{#1\right\}}}

\spnewtheorem*{remark*}{Remark}{\bf}{\rm}

\graphicspath{{./gfx/}}

\journalname{Methodol Comput Appl Probab}
\begin{document}

\title{An Analytic Expression for the Distribution of the Generalized Shiryaev--Roberts Diffusion}
\subtitle{The Fourier Spectral Expansion Approach}

\titlerunning{Exact Distribution of the Generalized Shiryaev--Roberts Diffusion}        

\author{Aleksey~S.~Polunchenko\and~Grigory~Sokolov}
\authorrunning{Polunchenko and Sokolov} 

\institute{A.S. Polunchenko \at
    Department of Mathematical Sciences\\
    State University of New York at Binghamton\\
    Binghamton, NY 13902--6000, USA\\
              Tel.: +1-607-777-6906\\
              Fax: +1-607-777-2450\\
              \email{aleksey@binghamton.edu}           
    \and
    G. Sokolov \at
    Department of Mathematical Sciences\\
    State University of New York at Binghamton\\
    Binghamton, NY 13902--6000, USA\\
              Tel.: +1-607-777-4239\\
              Fax: +1-607-777-2450\\
              \email{gsokolov@binghamton.edu}           
}

\date{Received: date / Accepted: date}

\maketitle

\begin{abstract}
%
We consider the quickest change-point detection problem where the aim is to detect the onset of a pre-specified drift in ``live''-monitored standard Brownian motion; the change-point is assumed unknown (nonrandom). The topic of interest is the distribution of the Generalized Shryaev--Roberts (GSR) detection statistic set up to ``sense'' the presence of the drift. Specifically, we derive a closed-form formula for the transition probability density function (pdf) of the time-homogeneous Markov diffusion process generated by the GSR statistic when the Brownian motion under surveillance is ``drift-free'', i.e., in the pre-change regime; the GSR statistic's (deterministic) nonnegative headstart is assumed arbitrarily given. The transition pdf formula is found analytically, through direct solution of the respective Kolmogorov forward equation via the Fourier spectral method to achieve separation of the spacial and temporal variables. The obtained result generalizes the well-known formula for the (pre-change) stationary distribution of the GSR statistic: the latter's stationary distribution is the temporal limit of the distribution sought in this work. To conclude, we exploit the obtained formula numerically and briefly study the pre-change behavior of the GSR statistic versus three factors:\begin{inparaenum}[\itshape(a)]\item drift-shift magnitude, \item time, and \item the GSR statistic's headstart\end{inparaenum}.

\keywords{Generalized Shiryaev--Roberts procedure\and
Kolmogorov forward equation\and
Markov diffusion processes\and
Quickest change-point detection\and
Sequential analysis
} 
%

%
%
%
%
%
%
\subclass{MSC 62L10\and MSC 60G10\and MSC 62M15\and MSC 60J60}
\end{abstract}

\section{Introduction}
\label{sec:intro}
Sequential (quickest) change-point detection is concerned with the development and evaluation of reliable statistical procedures for early detection of unanticipated changes that may (or may not) occur online in the characteristics of a ``live''-monitored (random) process. Specifically, the latter is observed continuously with the intent to ``flag an alarm'' in the event (and as soon as) the behavior of the process starts to suggest the process may have (been) statistically changed. The alarm is to be flagged as quickly as is possible within a set tolerable level of the ``false positive'' risk. See, e.g.,~\cite{Shiryaev:Book78}, \cite{Basseville+Nikiforov:Book93}, \cite{Poor+Hadjiliadis:Book09}, \cite{Veeravalli+Banerjee:AP2013}, \cite{Tartakovsky+etal:Book2014} and the references therein.

A change-point detection procedure is identified with a stopping time, $\T$, that is adapted to the filtration, $(\mathcal{F}_t)_{t\ge0}$, generated by the observed process, $(X_t)_{t\ge0}$; the semantics of $\T$ is that it constitutes a rule to stop and declare that the statistical profile of the observed process may have (been) changed. A ``good'' (i.e., optimal or nearly optimal) detection procedure is one that minimizes (or nearly minimizes) the desired detection delay penalty-function, subject to a constraint on the false alarm risk. For an overview of the major optimality criteria, see, e.g.,~\cite{Tartakovsky+Moustakides:SA10}, \cite{Polunchenko+Tartakovsky:MCAP2012}, \cite{Polunchenko+etal:JSM2013},~\cite{Veeravalli+Banerjee:AP2013} or~\cite[Part~II]{Tartakovsky+etal:Book2014}.

\begin{figure}[h]
    \centering
    \includegraphics[width=0.9\textwidth]{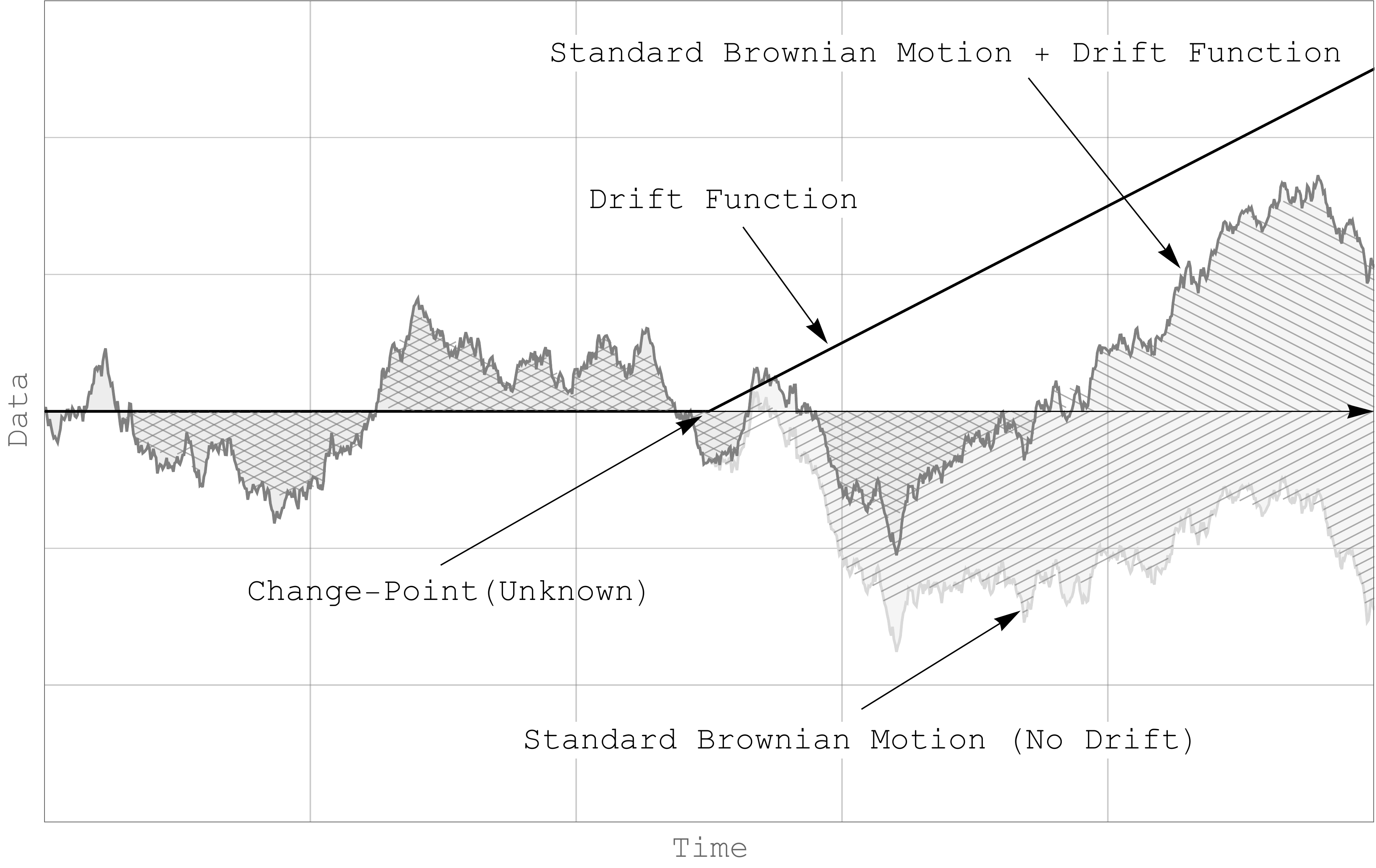}
    \caption{Standard Brownian motion gaining a persistent drift.}
    \label{fig:BM-change-point-scenario}
\end{figure}

This work focuses on the popular minimax setup of the basic change-point detection problem where the observed process, $(X_t)_{t\ge0}$, is standard Brownian motion that at an unknown (nonrandom) time moment $\nu$---referred to as the change-point---may (or may not) experience an abrupt and permanent change in the drift, from a value of zero initially, i.e., $\EV[dX_t]=0$ for $t\in[0,\nu]$, to a known value $\mu\neq0$ following the change-point, i.e., $\EV[dX_t]=\mu$ for $t\in(\nu,\infty)$. This is illustrated in Figure~\ref{fig:BM-change-point-scenario}. The goal is to find out---as quickly as is possible within an {\it a~priori} set level of the ``false positive'' risk---whether the drift of the process is no longer zero. See, e.g.,~\cite{Pollak+Siegmund:B85},~\cite{Shiryaev:RMS1996,Shiryaev:Bachelier2002},~\cite{Moustakides:AS2004},~\cite{Feinberg+Shiryaev:SD2006}, and~\cite{Burnaev+etal:TPA2009}.

More formally, under the above Brownian motion change-point scenario, the observed process, $(X_t)_{t\ge0}$, is governed by the stochastic differential equation (SDE):
\begin{align}\label{eq:BM-change-point-model}
dX_t
&=
\mu\indicator{t>\nu}dt+dB_t,\;t\ge0,\;\text{with}\;X_0=0,
\end{align}
where $(B_t)_{t\ge0}$ is standard Brownian motion (i.e., $\EV[dB_t]=0$, $\EV[(dB_t)^2]=dt$, and $B_0=0$), $\mu\neq0$ is the known post-change drift value, and $\nu\in[0,\infty]$ is the unknown (nonrandom) change-point; here and onward, the notation $\nu=0$ ($\nu=\infty$) is to be understood as the case when the drift is in effect {\it ab initio} (or never, respectively).

To perform change-point detection under model~\eqref{eq:BM-change-point-model}, the standard approach has been to employ Page's~\citeyearpar{Page:B54} Cumulative Sum (CUSUM) ``inspection scheme''. This choice may be justified by the fact (established by~\citealt{Beibel:AS1996}, by~\citealt{Shiryaev:RMS1996} and by~\citealt{Moustakides:AS2004}) that the CUSUM scheme is strictly minimax-optimal in the sense of~\cite{Lorden:AMS71}; the discrete-time equivalent of this result was first established by~\cite{Moustakides:AS86}, although an alternative proof was also later offered by~\cite{Ritov:AS90} who used a game-theoretic argument.

However, when one is interested in minimax optimality as defined by~\cite{Pollak:AS85}, a sensible alternative to using the CUSUM scheme would be to devise the Generalized Shiryaev--Roberts (GSR) procedure. The latter is due to~\cite{Moustakides+etal:SS11} and is a headstarted version of the classical quasi-Bayesian Shiryaev--Roberts (SR) procedure that emerged from the independent work of~\cite{Shiryaev:SMD61,Shiryaev:TPA63} and that of~\cite{Roberts:T66}. With Pollak's~\citeyearpar{Pollak:AS85} definition of minimax optimality in mind, the motivation to prefer the GSR procedure over the CUSUM chart stems from the results obtained (for the discrete-time analogue of the problem) by~\cite{Tartakovsky+Polunchenko:IWAP10,Polunchenko+Tartakovsky:AS10} and then also by~\cite{Tartakovsky+etal:TPA2012} who showed that the GSR procedure with a carefully designed headstart may be faster (in Pollak's~\citeyear{Pollak:AS85} sense) than the CUSUM scheme; as a matter of fact,~\cite{Tartakovsky+Polunchenko:IWAP10,Polunchenko+Tartakovsky:AS10} proved the GSR procedure (with a ``finetuned'' headstart) to be the fastest (i.e., the best one can do) in two specific (discrete-time) scenarios. For an attempt to extend these results to the Brownian motion scenario, see, e.g.,~\cite{Burnaev:ARSAIM2009}.

More specifically, the GSR procedure calls for stopping as soon as the GSR detection statistic, $(R_t^r)_{t\ge0}$, hits a critical level known as the detection threshold. The latter is set so as have the ``false positive'' risk at a desired ``height''. Let $\Pr_\infty$ ($\Pr_0$) denote the probability measure (distribution law) generated by the observed process, $(X_t)_{t\ge0}$, under the assumption that $\nu=\infty$ ($\nu=0$); note that $\Pr_\infty$ is the Wiener measure. Let $\left.\Pr_{\infty}\right|_{\mathcal{F}_t}$ ($\left.\Pr_{0}\right|_{\mathcal{F}_t}$) be the restriction of probability measure $\Pr_\infty$ ($\Pr_{0}$) to the filtration $\mathcal{F}_t$. Further, define
\begin{align*}
\LR_t
&\triangleq
\dfrac{d\left.\Pr_{0}\right|_{\mathcal{F}_t}}{d\left.\Pr_{\infty}\right|_{\mathcal{F}_t}},\;t\ge0,
\end{align*}
i.e., the Radon--Nikod\'{y}m derivative of $\left.\Pr_{0}\right|_{\mathcal{F}_t}$ with respect to $\left.\Pr_{\infty}\right|_{\mathcal{F}_t}$. It is well-known that for the Brownian motion scenario under consideration
\begin{align*}
\LR_t
&=
\exp\left\{\mu X_t-\dfrac{\mu^2}{2} t\right\},\;\text{so that}\; d\LR_t=\mu\LR_tdX_t,\;\LR_0=1;
\end{align*}
cf.~\cite{Shiryaev:Book1999} and~\cite{Liptser+Shiryaev:Book2001}. The process $\{\LR_t\}_{t\ge0}$ manifests the likelihood ratio to test the hypothesis $\mathcal{H}_0\colon\nu=0$ against the alternative $\mathcal{H}_\infty\colon\nu=\infty$, and is the key ingredient of the CUSUM statistic as well as of the GSR statistic, $(R_t^r)_{t\ge0}$. Specifically, tailored to the Brownian motion scenario at hand, the GSR statistic, $(R_t^r)_{t\ge0}$, is of the form
\begin{align}\label{eq:Rt_r-def}
\begin{split}
R_t^r
&\triangleq
r\LR_t+\int_0^t\dfrac{\LR_t}{\LR_s}\,ds\\
&=
r\exp\left\{\mu X_t-\dfrac{\mu^2t}{2}\right\}+\int_0^t\exp\left\{\mu(X_t-X_s)-\dfrac{\mu^2(t-s)}{2}\right\}ds,\;t\ge0,
\end{split}
\end{align}
where $R_0^r=r\ge0$ is the headstart (a deterministic point selected so as to optimize the GSR procedure's performance; see, e.g.,~\citealt{Tartakovsky+Polunchenko:IWAP10,Polunchenko+Tartakovsky:AS10,Moustakides+etal:SS11,Tartakovsky+etal:TPA2012,Polunchenko+Sokolov:EnT2014}). When $R_0^r=r=0$, it is said that there is no headstart. The GSR statistic, $(R_t^r)_{t\ge0}$, with no headstart is equivalent to the classical SR statistic. Consequently, the GSR procedure whose statistic has no headstart is equivalent to the classical SR procedure. Hence, the labels ``{\em Generalized} SR statistic'' and ``{\em Generalized} SR procedure'', which appear to have both been coined by~\cite{Tartakovsky+etal:TPA2012}.

We are now in a position to formulate the specific problem addressed in this paper: to obtain an explicit closed-form formula for $p_\infty(y,t|x,s)\triangleq d\Pr_\infty(R_t^r\le y|R_s^r=x)/dy$, $0\le s<t<\infty$, $x,y\ge0$, i.e., for the $\Pr_{\infty}$-transition probability density function (pdf) of the GSR statistic $(R_t^r)_{t\ge0}$ with the headstart, $R_0^r=r\ge0$, assumed given. That is, in this paper the GSR statistic, $(R_t^r)_{t\ge0}$, is effectively let ``run loose'' over the entire space $(R_t^r,t)\in[0,\infty)\times[0,\infty)$ with no detection threshold imposed, and the goal is to find the exact transition pdf of the one-dimensional Markov diffusion process $(R_t^r)_{t\ge0}$ under probability measure $\Pr_\infty$. More specifically, observe that, by It\^{o} formula, the $\Pr_\infty$-differential of the GSR diffusion $(R_t^r)_{t\ge0}$ is $dR_t^r=dt+\mu R_t^r dB_t$, where $R_0^r=r\ge0$, whence $(R_t^r)_{t\ge0}$ is seen to be time-homogeneous. Therefore, $p_\infty(y,t|x,s)$ depends on $s$ and $t$ only through the difference $t-s\ge0$, and it suffices to find $p_\infty(x,t|r)\triangleq p_\infty(x,t|r,0)$, $x,r,t\ge0$. Hence, we shall concentrate on finding $p_\infty(x,t|r)$, $x,r,t\ge0$. Also, to lighten the notation, from now on we shall drop the ``$\infty$'' in the subscript of $p_{\infty}(x,t|r)$ and simply write $p(x,t|r)$.


The principal approach we intend to undertake to find $p(x,t|r)$ consists in solving directly the respective Kolmogorov forward equation
\begin{align}\label{eq:SR-Rt-density-Kolmogorov-fwd-PDE}
\dfrac{\partial}{\partial t}p(x,t|r)
&=
-\dfrac{\partial}{\partial x}p(x,t|r)
+\dfrac{\mu^2}{2}\dfrac{\partial^2}{\partial x^2}\big[x^2p(x,t|r)\big],\; x,t,r\ge0,
\end{align}
subject to\begin{inparaenum}[\itshape(a)]\item the (natural) normalization constraint
\begin{align}\label{eq:normalization-constraint}
\int_{0}^{\infty} p(x,t|r)\,dx
&=
1
\end{align}
valid for all $r,t\ge0$, and \item the initial condition $\lim_{t\to0+}p(x,t|r)=\delta(x-r)$ valid for all $x$\end{inparaenum}; here and onward $\delta(z)$ denotes the Dirac delta function so that ``$\lim_{t\to0+}p(x,t|r)=\delta(x-r)$'' is to be understood as equality of distributions. We note also that naturally $p(x,t|r)\ge0$ for all $r,x,t\ge0$. The initial condition and the normalization constraint~\eqref{eq:normalization-constraint} are also to be complemented by two boundary conditions of spacial (i.e., in the $x$ variable) type. These conditions are provided in Section~\ref{sec:main-result} which is also the section where we solve~\eqref{eq:SR-Rt-density-Kolmogorov-fwd-PDE} explicitly and thus obtain the main result of this paper.

More specifically, to recover $p(x,t|r)$ from~\eqref{eq:SR-Rt-density-Kolmogorov-fwd-PDE} we devise the so-called Fourier spectral method, a separation-of-variables-type technique the general theory for which was developed by~\cite{Weyl:MA1910} and by~\cite{Titchmarsh:Book1962}. See also, e.g.,~\cite{Levitan:Book1950}, \cite{McKean:TAMS1956}, \cite{Dunford+Schwartz:Book1963}, \cite{Levitan+Sargsjan:Book1975}. The Weyl--Titchmarsh theory was then applied and sharpened further in the contexts of probability and stochastic processes by~\cite{Kac:BSMSP1951,Kac:Book1959},~\cite{Feller:AM1952},~\cite{Ito+McKean:Book1974}. With the specifics deferred to Section~\ref{sec:main-result}, the Fourier method allows to separate the spacial variable, $x$, and the temporal variable, $t$, and thus reduce the original equation~\eqref{eq:SR-Rt-density-Kolmogorov-fwd-PDE} to a (singular) Sturm--Liouville problem. The fundamental solutions of the latter, in turn, allow to explicitly construct the Green's function associated with the corresponding Sturm--Liouville (linear, differential) operator. As the resolvent of the Sturm--Liouville operator, the Green's function provides an exhaustive characterization of the operator's properties, including its spectrum. In particular, the Green's function is directly connected to the transition pdf $p(x,t|r)$: the former is the Laplace transform (taken with respect to time) of the latter. Therefore, getting $p(x,t|r)$ is a matter of inverting its (temporal) Laplace transform given by the Green's function. The inversion can accomplished by virtue of Cauchy's Residue Theorem. The poles (or branch cuts) of the Green's function yield the eigenvalues of the Sturm--Liouville operator and the corresponding residues determine the contributions of the eigenfunctions to the pdf $p(x,t|r)$.

The formula for $p(x,t|r)$ found using the Fourier method in Section~\ref{sec:main-result} is a generalization of the result of~\cite{Shiryaev:SMD61,Shiryaev:TPA63} who obtained the limit $\rho(x)\triangleq\lim_{t\to+\infty}p(x,t|r)$, i.e., the stationary solution of equation~\eqref{eq:SR-Rt-density-Kolmogorov-fwd-PDE}. This solution is effectively the stationary distribution of the GSR statistic, $(R_t^r)_{t\ge0}$, under the ``no-drift'' hypothesis; see also, e.g.,~\cite[Remark~4.7,~pp.~466--467]{Feinberg+Shiryaev:SD2006} and~\cite[Remark~2,~p.~529]{Burnaev+etal:TPA2009}. For a proof of existence of this limiting invariant distribution for any initial point $R_0^r=r\ge0$, see, e.g.,~\cite[Proposition~3,~p.~271]{Pollak+Siegmund:B85}. All these results are reviewed at greater length in Section~\ref{sec:literature-overview} which is intended to survey the relevant prior literature. Remarkably, the Kolmogorov forward equation~\eqref{eq:SR-Rt-density-Kolmogorov-fwd-PDE} (possibly subject to different conditions and constraints) has arisen in other disciplines as well, notably in physics (in particular, in quantum mechanics), chemistry, and in mathematical finance. It should therefore come as no surprise that in these fields the respective solution has been obtained using completely different techniques (e.g., Feynman path integrals; see~\citealt{Feynman:RMPh1948}) and with no reference to the GSR statistic. In Section~\ref{sec:literature-overview} we briefly review these results as well, as they may cast light on a possible interpretation of the GSR statistic.

The plan for the remainder of the paper is as follows. First, in Section~\ref{sec:literature-overview} we give a brief account of the relevant prior literature. The main section of the paper is Section~\ref{sec:main-result}, which is where we first set up the corresponding Kolmogorov forward equation~\eqref{eq:SR-Rt-density-Kolmogorov-fwd-PDE} more formally, and then solve it explicitly, i.e., obtain analytically a closed-form formula for the $\Pr_\infty$-transition pdf of the GSR statistic. Next, in Section~\ref{sec:discussion} the obtained formula is reconciled with several previously published parallel results. In Section~\ref{sec:numerical-study} we exploit the found pdf $p(x,t|r)$ numerically and offer a brief numerical study of the statistical behavior exhibited by the GSR statistic in the ``drift-free'' regime. To carry out the study, we implemented the obtained pdf formula in {\em Mathematica}\footnote{{\em Mathematica} is a popular software package developed by Wolfram Research, Inc. as a programming environment for scientific computing. See on the Web at~\url{http://www.wolfram.com/mathematica/}.}. Lastly, Section~\ref{sec:conclusion} is meant to draw a line under the entire paper.

\section{Overview of the Relevant Prior Literature}
\label{sec:literature-overview}

As can be gathered from the introduction, the centerpiece of this work is the solution of the Kolmogorov forward equation~\eqref{eq:SR-Rt-density-Kolmogorov-fwd-PDE}. This problem---in one or another form, shape, and context---has been addressed in the literature before, and this section's objective is to provide a brief overview of the relevant prior results.

One of the first to arrive at a closely related problem was~\cite{Wong:SPMPE1964}. The actual objective was to find the transition pdf of a stochastic process with a given {\em stationary} (as $t\to+\infty$) distribution. Specifically, the point of departure for~\cite{Wong:SPMPE1964} was the celebrated work of~\cite{Kolmogorov:MA1931}, where it was noticed for the first time that it is possible to construct a diffusion process whose stationary distribution is a member of the Pearson~\citeyearpar{Pearson:PhTRSL1895} distribution family\footnote{To be more precise,~\cite{Pearson:PhTRSL1895} only obtained and dealt with a special case of what later became known as the Pearson (first-order separable differential) equation, the basis for the Pearson distribution family.}. However, since~\cite{Kolmogorov:MA1931} didn't carry out the actual construction,~\cite{Wong:SPMPE1964} effectively picked up where~\cite{Kolmogorov:MA1931} left off, and built on to the work of~\cite{Karlin+McGregor:JMAA1960} to derive a number of processes with a Pearson-type stationary distribution; for an overview of analytically treatable Kolmogorov--Pearson diffusions see, e.g.,~\cite{Avram+etal:TechRep2012}. One of the Pearson distributions considered by~\cite{Wong:SPMPE1964} was the inverted Gamma distribution, which is a generalized extreme value Fr\'{e}chet--Gumbel-type distribution. Incidentally, this latter distribution was demonstrated by~\cite{Shiryaev:SMD61,Shiryaev:TPA63} to be the stationary distribution of the classical SR statistic (special case of the GSR statistic with no headstart) set up for the Brownian motion change-point scenario~\eqref{eq:BM-change-point-model} when the observed process, $(X_t)_{t\ge0}$, is still drift-free; see also, e.g.,~\cite[Remark~4.7,~pp.~466--467]{Feinberg+Shiryaev:SD2006} and~\cite[Remark~2,~p.~529]{Burnaev+etal:TPA2009}. For a proof of existence of this distribution see, e.g.,~\cite[Proposition~3,~p.~271]{Pollak+Siegmund:B85} and~\cite[Section~2.2,~p.~3]{Peskir:Shiryaev2006}. Hence, effectively~\cite[Section~2.F,~p.~271]{Wong:SPMPE1964} inadvertently discovered the SDE whose solution (in a special case) is the GSR diffusion\footnote{We also warn of a typo on page~271 of~\cite[Section~2.F,~p.~271]{Wong:SPMPE1964} in the second formula from the bottom of the page (the formula is unnumbered). Specifically, as given by~\cite[Section~2.F,~p.~271]{Wong:SPMPE1964}, the formula (in the original notation) is
\begin{align*}
{}_2F_{0}(-\alpha-i\mu,-\alpha+i\mu,-x)
&=
x^{\alpha+i\mu}\,\Psi\left(-\frac{\alpha}{2}-i\mu,1-2i\mu,\frac{1}{x}\right),
\end{align*}
where $i\triangleq\sqrt{-1}$, ${}_2F_0$ is the generalized hypergeometric series, and $\Psi$ is the Tricomi confluent hypergeometric function. The typo is that the first argument of the $\Psi$-function in the right-hand side should be $-\alpha-i\mu$, not $-\alpha/2-i\mu$. That is, the correct identity is
\begin{align*}
{}_2F_{0}(-\alpha-i\mu,-\alpha+i\mu,-x)
&=
x^{\alpha+i\mu}\,\Psi\left(-\alpha-i\mu,1-2i\mu,\frac{1}{x}\right),
\end{align*}
as can be confirmed, e.g., with~\cite[Section~13.1,~p.~504]{Abramowitz+Stegun:Handbook1964}.}.
One ``problem'' with Wong's~\citeyearpar{Wong:SPMPE1964} work, however, is that it is rather concise; in fact, it is so brief that it is, in essence, a ``cook book'' of ``ready-made'' formulae with practically no derivation thereof offered. That said,~\cite{Wong:SPMPE1964} does provide the general recipe to find the transition pdf formulae, and the recipe is to solve the respective Kolmogorov forward equation directly through the Fourier spectral method which we mentioned emerged from the Weyl--Titchmarsh theory; see~\cite{Weyl:MA1910} and~\cite{Titchmarsh:Book1962}. It is this approach and method that we intend to employ in Section~\ref{sec:main-result} below, but with no key details omitted. Another ``issue'' with Wong's~\citeyearpar{Wong:SPMPE1964} work is that little attention is paid to the amenability of the obtained transition pdf formulae to numerical evaluation, an aspect important for applications. We provide a short numerical study of the found distribution carried out with the aid of a {\em Mathematica} script we prepared.

A more recent and more relevant example of prior work on our problem would be the work of~\cite{Peskir:Shiryaev2006}. Just like us in the present paper,~\cite{Peskir:Shiryaev2006} also focused on (effectively) the GSR statistic, and considered a problem even more general than the one treated in this work. Specifically,~\cite{Peskir:Shiryaev2006} was after the fundamental solution of the following Kolmogorov forward equation:
\begin{align}\label{eq:Peskir-equation}
\dfrac{\partial}{\partial t}\tilde{p}(x,t|y)
&=
-\dfrac{\partial}{\partial x}\big[(1+a x)\tilde{p}(x,t|y)\big]
+b\dfrac{\partial^2}{\partial x^2}\big[x^2 \tilde{p}(x,t|y)\big],\;t\ge0,
\end{align}
where $a\in\mathbb{R}$ and $b\ge0$ are given constants. This equation is more general than equation~\eqref{eq:SR-Rt-density-Kolmogorov-fwd-PDE} that we are after: the two coincide when $a=0$ and $b=\mu^2/2$. However, even when $a\neq0$, the foregoing equation is still closely connected to the GSR diffusion: if $a=\theta\mu$, then the equation governs the transition pdf of the GSR statistic, $(R_t^r)_{t\ge0}$, when the drift of the observed process $(X_t)_{t\ge0}$ is $\theta$ for all $t\ge0$, but the GSR statistic is set up to ``anticipate'' the drift to be $\mu$. Moreover,~\cite{Peskir:Shiryaev2006} did not restrict attention to the case when $x,y\ge0$ (as we do in this work), but allowed $x,y\in\mathbb{R}$. One can therefore argue that~\cite{Peskir:Shiryaev2006} was, in effect, concerned with the transition pdf of the GSR statistic (with the headstart given and arbitrary from the entire real line) in a regime general enough to include as special cases\begin{inparaenum}[\itshape(a)]\item the pre-change regime, \item the post-change regime, and \item the post-change regime with drift-shift-misspecification, i.e., when the actual drift-shift magnitude is different from the one ``anticipated'' by the GSR statistic\end{inparaenum}; this latter interpretation can be given based on, e.g., the work of~\cite{Pollak+Siegmund:B85}, where the performance of the original SR procedure was studied in the minimax Brownian motion context with a possible drift-shift-misspecification. Peskir's~\citeyearpar{Peskir:Shiryaev2006} approach to solving equation~\eqref{eq:Peskir-equation} involved obtaining the Laplace transform (the Green's function) of the equation's fundamental solution, and then inverting it to get the sought transition pdf itself. However, although~\cite{Peskir:Shiryaev2006} did succeed in obtaining the Laplace transform (which in the pre-change regime tuned out to be the Laplace transform of the Hartman--Watson distribution; see~\citealt{Hartman+Watson:AP1974}), he was able to invert it explicitly in the pre-change regime only and only for the zero headstart (i.e., for the classical SR statistic). As per the hope expressed by~\cite{Peskir:Shiryaev2006} for someone to ``run the last leg'' and complete his work, we generalize the result of~\cite{Peskir:Shiryaev2006} for the pre-change regime to an arbitrary (nonnegative) headstart.

We now recall the remark we made in the introduction that the very same Kolmogorov forward equation~\eqref{eq:SR-Rt-density-Kolmogorov-fwd-PDE} that we are after in this paper has also appeared in such disciplines as physics and mathematical finance. Specifically, in physics, the equation is known as the Fokker--Planck equation (after~\citealt{Fokker:APh1941} and~\citealt{Planck:PAW1917}), and there is voluminous literature devoted to the solution of this equation (using completely different methods). To be more specific, equation~\eqref{eq:SR-Rt-density-Kolmogorov-fwd-PDE} arises in physics (as well as in chemistry), e.g., in connection with the so-called Morse~\citeyearpar{Morse:PhysRev1929} potential. The latter is a harmonic oscillator model widely used to describe the variation of energy with respect to the internuclear distance in a diatomic molecule. See, e.g.,~\cite{Grosche:AP1988} and~\cite{Morse+Feshbach:Book1953}. More specifically, the respective equation that governs these energy variations is equivalent to our Kolmogorov forward equation~\eqref{eq:SR-Rt-density-Kolmogorov-fwd-PDE} up to the initial and boundary conditions. It is known as one of the very few equations in quantum mechanics that permit analytical solution. The standard solution strategy used in physics to handle this equation (as well as Fokker--Plank equations in general) is to exploit the Feynman~\citeyearpar{Feynman:RMPh1948} path integral framework. On a higher level, the physical equivalent of our Kolmogorov forward equation~\eqref{eq:SR-Rt-density-Kolmogorov-fwd-PDE} describes the position of a particle moving around in an inhomogeneous environment, driven by a combination of random forces (e.g., thermal noise). In this context, equation~\eqref{eq:SR-Rt-density-Kolmogorov-fwd-PDE} has been studied, e.g., by~\cite{Monthus+Comtet:JPhIF1994} and by~\cite{Comtet+Monthus:JPhA1996}.

Another area to which the Kolmogorov forward equation~\eqref{eq:SR-Rt-density-Kolmogorov-fwd-PDE} is no stranger is mathematical finance. The financial interpretation of the equation is that it describes the price of a certain financial derivative as a function of time and the derivative parameters. Moreover, Markov processes (especially in dimension one) are often used in finance to model the time-evolution of the ``state of the economy''; the latter is essentially the financial equivalent of the random environment inside which the physical particle is moving, if one is to relate to the physical analogy. The Weyl--Titchmarsh spectral theory as well as the Feynman path integral approach have both been used in mathematical finance to price financial instruments and to describe the evolution of the underlying economic state. Excellent references assessing the state-of-the-art of financial derivative pricing methods would be~\cite{Borodin+Salminen:Book2002},~\cite{Linetsky:MF2006,Linetsky:HandbookChapter2007}, and the references therein. By way of example, the equation considered by~\cite{Peskir:Shiryaev2006}, i.e., equation~\eqref{eq:Peskir-equation}, arises in finance in the context of pricing the so-called Asian options and Merton's cash dividends; see, e.g.,~\cite{Linetsky:IJTAF2004,Linetsky:OR2004}. As an economic model, equations akin to~\eqref{eq:SR-Rt-density-Kolmogorov-fwd-PDE} or to~\eqref{eq:Peskir-equation} arise in connection to the so-called Hull--White volatility model (see~\citealt{Hull+White:JF1987}) which pillars on the same type of SDEs that governs the GSR statistic. For a thorough mathematical treatment of the Hull--White volatility model, see, e.g.,~\cite{Fatone+etal:IJMNTA2013}. Moreover, diffusion processes similar to the GSR statistic have also been encountered as stochastic interest rate models, and have been used to find the present value of annuities. See, e.g.,~\cite{Vanneste+etal:IME1994,DeSchepper+etal:IME1994,DeSchepper+Goovaerts:IME1999}. In the latter papers, the authors devised the Feynman path integral approach to solve the equivalent of equation~\eqref{eq:SR-Rt-density-Kolmogorov-fwd-PDE}. We note that in Feynman path integral framework the solution need not integrate to one, as is required in our case by the normalization constraint~\eqref{eq:normalization-constraint}. As a result, the obtained solution simply dies off to zero as time goes on, and no nontrivial stationary distribution is exhibited, which is in stark contrast with the aforementioned result of~\cite{Shiryaev:SMD61,Shiryaev:TPA63} concerning the (nontrival) stationary distribution of the GSR statistic.

One more area where the Kolmogorov forward equation~\eqref{eq:SR-Rt-density-Kolmogorov-fwd-PDE} has appeared is stochastic processes. In the latter field, one of the key methods to deal with the equation is the Feynman--Kac formula that emerged from the pioneering work of~\cite{Feynman:RMPh1948} and that of~\cite{Kac:TAMS1949,Kac:BSMSP1951,Kac:Book1959}. The Feynman--Kac formula establishes a bridge between parabolic partial differential equations (PDEs) and stochastic processes; we note that equation~\eqref{eq:SR-Rt-density-Kolmogorov-fwd-PDE} is a parabolic PDE. Another possible approach to finding the transition pdf of the GSR diffusion, $(R_t^r)_{t\ge0}$, is to derive it directly from the definition of the process, viz. from formula~\eqref{eq:Rt_r-def}. Specifically, the idea is to exploit the {\em linear} connection between $(R_t^r)_{t\ge0}$ and the likelihood ratio process $\{\LR_t\}_{t\ge0}$, and then capitalize on the fact that $\LR_t$ is a geometric (exponential) Brownian motion, a well-studied type of diffusions. The aforementioned work of~\cite{Peskir:Shiryaev2006} contains a number of references where precisely this approach was devised to derive the distribution of effectively the GSR diffusion (although for obvious reasons no connection to the GSR statistic was made). See, e.g.,~\cite{Yor:AAP1992},~\cite{Donati-Martin+etal:RMI2001},~\cite{Dufresne:AAP2001} and~\cite{Schroder:AAP2003}. However, the obtained distribution is only for the case of no headstart, i.e., assuming the starting point of the diffusion is zero.

We conclude this section with a remark that the above is just a small sample of examples of contexts where the Kolmogorov forward equation arises: the ``footprint'' of the equation is much larger and spans many more disciplines. It would be a daunting task to review all of the relevant applications of the equation in a holistic manner in a single paper.

\section{The Main Result}
\label{sec:main-result}

This section is the centerpiece of this work. It is intended to provide a solution to the main problem of this paper: to obtain a closed-form formula for the $\Pr_\infty$-transition pdf of the GSR diffusion, $(R_t^r)_{t\ge0}$, defined by~\eqref{eq:Rt_r-def} above.

To get started, recall that under probability measure $\Pr_\infty$, the GSR diffusion, $(R_t^r)_{t\ge0}$, solves the SDE: $dR_t^r=dt+\mu R_t^r dB_t$, $t\ge0$, where $R_r^t=r\ge0$. Since the drift function $a(x,t)\equiv1$ and the diffusion coefficient $b(x,t)=\mu x$ are both independent of time $t\ge0$, the GSR diffusion, $(R_t^r)_{t\ge0}$, is time-homogeneous. Consequently, according to the fundamental work of~\cite{Kolmogorov:MA1931}, the sought transition pdf, $p(x,t|r)$, $x,r,t\ge0$, satisfies the Kolmogorov forward equation:
\begin{align}\label{eq:SR-Rt-pdf-Kolmogorov-fwd-PDE-2}
\dfrac{\partial}{\partial t}p(x,t|r)
&=
-\dfrac{\partial}{\partial x}p(x,t|r)
+\dfrac{\mu^2}{2}\dfrac{\partial^2}{\partial x^2}\big[x^2p(x,t|r)\big],
\;\; x,r,t\ge0,
\end{align}
subject to\begin{inparaenum}[\itshape(a)]\item the (natural) normalization constraint
\begin{align}\label{eq:normalization-constraint1}
\int_{0}^{\infty}p(x,t|r)\,dx
&=
1,
\;\; r,t\ge0,
\end{align}
and \item the initial condition $\lim_{t\to0+}p(x,t|r)=\delta(x-r)$ valid for all $x$\end{inparaenum}. Since equation~\eqref{eq:SR-Rt-pdf-Kolmogorov-fwd-PDE-2} is a PDE of order one in time $t$ and order two in space $x$, and is to hold over the entire space $(x,t)\in[0,\infty)\times[0,\infty)$, the {\em temporal} initial condition and the normalization constraint are to be complemented by two {\em spacial} boundary conditions: one at $x=0$ (or as $x\to0+$) and one as $x\to+\infty$. To obtain these conditions let
\begin{align*}
J(x,t|r)
&\triangleq
p(x,t|r)-\dfrac{\mu^2}{2}\dfrac{\partial}{\partial x}\big[x^2 p(x,t|r)\big]
\end{align*}
be the probability current, so that equation~\eqref{eq:SR-Rt-pdf-Kolmogorov-fwd-PDE-2} can be rewritten more compactly as
\begin{align}\label{eq:SR-Rt-density-cont-eqn}
\dfrac{\partial}{\partial t}p(x,t|r)
+
\dfrac{\partial}{\partial x}J(x,t|r)
&=
0,
\end{align}
which is effectively a (one-dimensional) continuity equation: it constitutes the law of conservation of probability, an analogue of the well-known law of conservation of energy from physics. For further notational brevity, we shall omit the headstart $r$ in $p(x,t|r)$ and simply write $p(x,t)$ throughout this section, unless it is necessary to emphasize the dependence on $r$.

To obtain the first boundary condition, observe that since $R_0^r=r\ge0$ by assumption, it can be deduced from definition~\eqref{eq:Rt_r-def} that $R_t^r\ge0$ almost surely under any probability measure and for any $t\ge0$. If $R_t^r$ is interpreted as the position of a hypothetical ``particle'' at time instance $t$, then the nonnegativity of $R_t^r$ for all $t\ge0$ is to say that the particle is to never leave the nonnegative half-plane. Put another way, no particle flow through the $x$-axis is permitted. As a result, the probability current $J(x,t)$ through the $x$-axis must be zero at all times, i.e., $J(0,t)=0$ for all $t\ge0$, or explicitly
\begin{align}\label{eq:bnd-cond1}
p(x,t)-\dfrac{\mu^2}{2}\dfrac{\partial}{\partial x}\big[x^2 p(x,t)\big]
&=
0,
\;\;x\to0+,
\end{align}
which is the first boundary condition.

The second boundary condition can be obtained by integrating both sides of~\eqref{eq:SR-Rt-density-cont-eqn} with respect to $x$ over $[0,+\infty)$ and then using the normalization constraint~\eqref{eq:normalization-constraint1}. Specifically, this yields that $J(+\infty,t)-J(0,t)=0$ for all $t\ge0$,. Therefore, the probability current through $x=+\infty$ must match the probability current through $x=0$ at all times. Since in the latter case the probability current is zero, the second boundary condition is:
\begin{align}\label{eq:bnd-cond2}
p(x,t)-\dfrac{\mu^2}{2}\dfrac{\partial}{\partial x}\big[x^2 p(x,t)\big]
&=
0,
\;\;x\to+\infty.
\end{align}

To solve equation~\eqref{eq:SR-Rt-pdf-Kolmogorov-fwd-PDE-2}, let us first consider the equation's stationary solution. Specifically, the stationarity here is in the temporal sense, i.e., $p(x,t)$ is independent of $t$, which is the case in the limit as $t\to+\infty$. Let $\rho(x)\triangleq\lim_{t\to+\infty}p(x,t)$ denote the corresponding limit. For a proof that $\rho(x)$ exists, see, e.g.,~\cite[Proposition~3,~p.~271]{Pollak+Siegmund:B85}. Since in the stationary regime $\partial p(x,t)/\partial t\equiv\partial\rho(x)/\partial t\equiv0$ for all $x$, equation~\eqref{eq:SR-Rt-density-cont-eqn} simplifies to $\partial J(x,t)/\partial x=0$. The latter equation, in turn, is to say that the probability current, $J(x,t)$, as a function of time, is constant, and in view of the two boundary conditions established above, the value of that constant is zero. Hence, we obtain
\begin{align*}
\rho(x)-\dfrac{\mu^2}{2}\dfrac{d}{d x}\big[x^2\rho(x)\big]
&=
0.
\end{align*}

The solution to the foregoing equation is well-known and is given by the Fr\'{e}chet-type distribution density
\begin{align}\label{eq:rho-answer}
\rho(x)
&=
e^{-\tfrac{2}{\mu^2 x}}\dfrac{2}{\mu^2 x^2}\indicator{x\ge0}.
\end{align}

This result was first obtained by~\cite{Shiryaev:SMD61,Shiryaev:TPA63} as the stationary distribution of the original SR statistic (with zero headstart). See also, e.g.,~\cite{Feinberg+Shiryaev:SD2006} and~\cite[Remark~2,~p.~529]{Burnaev+etal:TPA2009}. As a {\em stationary} distribution, not only is $\rho(x)$ independent of $t$, but it is also independent of the headstart $r$. Therefore, $\rho(x)$ given by~\eqref{eq:rho-answer} is the stationary distribution of the GSR statistic, $(R_t^r)_{t\ge0}$, as well. It is also noteworthy that the distribution~\eqref{eq:rho-answer} is a special case of the extreme value inverse Gamma distribution, and in a completely different context (viz. finance) and using different techniques has been discovered, e.g., by~\cite[Theorem~1,~p.~224]{Milevsky:IME1997}. It is also a special case of~\cite[Formula~(74),~p.~264]{Comtet+etal:JAP1998}.

Let us now attack equation~\eqref{eq:SR-Rt-pdf-Kolmogorov-fwd-PDE-2} for arbitrary (finite) time $0\le t<\infty$. To that end, let us first heuristically outline our strategy. The main idea is to suppose that $p(x,t)$ is of the form $p(x,t)=\rho(x)\,\psi(x)\,\tau(t)$ where $\rho(x)$ is the stationary density given by~\eqref{eq:rho-answer}, and $\psi(x)$ and $\tau(t)$ are to be found. In other words, the idea is to assume that the spacial variable, $x$, and the temporal variable, $t$, can be separated. If that were the case, then the substitution $p(x,t)=\rho(x)\,\psi(x)\,\tau(t)$ would bring equation~\eqref{eq:SR-Rt-pdf-Kolmogorov-fwd-PDE-2} into the following form:
\begin{align*}
\dfrac{\tau'(t)}{\tau(t)}
&=
\dfrac{1}{\rho(x)\,\psi(x)}\left(-\dfrac{d}{dx}\big[\rho(x)\,\psi(x)\big]+\dfrac{\mu^2}{2}\dfrac{d^2}{d x^2}\big[x^2\rho(x)\,\psi(x)\big]\right),
\;\; x,t\ge0.
\end{align*}

Since the $x$ and $t$ variables are now on different sides of the equation, in order for the two sides to be equal to one another {\em irrespective} of $x$ and $t$, the two sides are {\em both} to be equal to the same {\em constant}, say $\lambda$. Therefore, the substitution $p(x,t)=\rho(x)\,\psi(x)\,\tau(t)$ effectively splits the original PDE~\eqref{eq:SR-Rt-pdf-Kolmogorov-fwd-PDE-2} into two ordinary differential equations (ODEs):
\begin{align}\label{eq:two-ODEs}
\dfrac{\tau'(t)}{\tau(t)}
&=
\lambda
\;\;\text{and}\;\;
\dfrac{1}{\rho(x)\,\psi(x)}\left(-\dfrac{d}{dx}\big[\rho(x)\,\psi(x)\big]+\dfrac{\mu^2}{2}\dfrac{d^2}{d x^2}\big[x^2\rho(x)\,\psi(x)\big]\right)=\lambda,
\end{align}
for some $\lambda$. The first of these two equations is straightforward to solve: the corresponding general nontrivial solution is a multiple of the exponential function $e^{\lambda t}$, $t\ge0$; note that this function is never zero. For the second equation, observe first that from~\eqref{eq:rho-answer} it can be readily concluded that $\mu^2 x^2 \rho'(x)=2\rho(x)\,(1-\mu^2 x)$. As a result, we obtain
\begin{align*}
\begin{split}
-\dfrac{d}{dx}&\big[\rho(x)\,\psi(x)\big]+\dfrac{\mu^2}{2}\dfrac{d^2}{d x^2}\big[x^2\rho(x)\,\psi(x)\big]=\\
&=
-\dfrac{d}{dx}\big[\rho(x)\,\psi(x)\big]+\\
&\qquad\qquad
+
\dfrac{d}{d x}\Bigg[\mu^2 x\,\rho(x)\,\psi(x)+\dfrac{\mu^2 x^2}{2}\rho'(x)\,\psi(x)+\\
&\qquad\qquad\qquad\qquad
+\dfrac{\mu^2 x^2}{2}\rho(x)\,\psi'(x)\Bigg]\\
&\stackrel{\text{(a)}}{=}
-\dfrac{d}{dx}\big[\rho(x)\,\psi(x)\big]+\\
&\qquad\qquad
+
\dfrac{d}{d x}\Bigg[\mu^2 x\,\rho(x)\,\psi(x)+\rho(x)\,\psi(x)-\mu^2 x\,\rho(x)\,\psi(x)+\\
&\qquad\qquad\qquad\qquad
+\dfrac{\mu^2 x^2}{2}\rho(x)\,\psi'(x)\Bigg]\\
&=
-\dfrac{d}{dx}\big[\rho(x)\,\psi(x)\big]+
\dfrac{d}{d x}\Bigg[\rho(x)\,\psi(x)+\dfrac{\mu^2 x^2}{2}\rho(x)\,\psi'(x)\Bigg]\\
&=
\dfrac{\mu^2}{2}\dfrac{d}{d x}\big[x^2\rho(x)\,\psi'(x)\big],
\end{split}
\end{align*}
where $\text{(a)}$ is due to the aforementioned identity $\mu^2 x^2 \rho'(x)=2\rho(x)\,(1-\mu^2 x)$. With this in mind, the second of the two ODEs~\eqref{eq:two-ODEs} becomes
\begin{align}\label{eq:eigen-eqn-two}
\dfrac{\mu^2}{2}\dfrac{d}{d x}\big[x^2\rho(x)\,\psi'(x)\big]
&=
\lambda\,\rho(x)\,\psi(x),
\end{align}
which can be recognized as the {\em characteristic} equation of the linear differential operator:
\begin{align}\label{eq:operator-D-def}
\mathcal{D}
&=
\dfrac{\mu^2}{2\,\rho(x)}\dfrac{d}{d x}\,x^2\rho(x)\,\dfrac{d}{dx}
\end{align}
i.e., equation~\eqref{eq:eigen-eqn-two} determines the eigenvalues $\lambda$ and the corresponding eigenfunctions $\psi(x)$ of the operator $\mathcal{D}$. By exactly the same argument it can be shown that the two boundary conditions~\eqref{eq:bnd-cond1}--\eqref{eq:bnd-cond2} under the substitution $p(x,t)=\rho(x)\,\psi(x)\,\tau(t)$ convert to
\begin{align}\label{eq:bdd-cond}
\dfrac{\mu^2}{2}\,x^2\,\rho(x)\,\psi'(x)
&=
0
\;\;
\text{to be satisfied as $x\to0+$ and as $x\to+\infty$},
\end{align}
where to get rid of $\tau(t)$ we used the fact that $\tau(t)\neq 0$ for any $t\ge0$.

Complemented with the two boundary conditions~\eqref{eq:bdd-cond}, equation~\eqref{eq:eigen-eqn-two} is a Sturm--Loiuville problem. Therefore, by attempting to separate the $x$ and $t$ variables we reduced the original equation~\eqref{eq:SR-Rt-pdf-Kolmogorov-fwd-PDE-2} to the Sturm--Loiuville problem~\eqref{eq:eigen-eqn-two} subject to two boundary conditions~\eqref{eq:bdd-cond}. To emphasize the dependence of $\psi(x)$ on $\lambda$ let from now on $\psi(x,\lambda)$ denote the solution (eigenfunction) corresponding to the eigenvalue $\lambda$. As soon as all eigenvalue-eigenfunction pairs $\{\lambda,\psi(x,\lambda)\}$ of the operator $\mathcal{D}$ are found, the solution $p(x,t)$ to the original equation~\eqref{eq:SR-Rt-pdf-Kolmogorov-fwd-PDE-2} is obtained as follows:
\begin{align}\label{eq:pdf-spectral-expansion}
p(x,t|r=y)
&=
\rho(x)\sum_{k} C_{k}(y)\,e^{\lambda_{k} t}\,\psi(x,\lambda_{k}),
\end{align}
where the sum is over those of the eigenvalues that make the solution satisfy the boundary conditions; here $C_{k}(y)$ are selected from the initial conditions. We note that the sum in the right-hand side of~\eqref{eq:pdf-spectral-expansion} is not to be interpreted literally, for the spectrum of the operator $\mathcal{D}$ may include both discrete as well as continuous components. As a matter of fact, as we shall see below, for the equation~\eqref{eq:SR-Rt-pdf-Kolmogorov-fwd-PDE-2} that we are after the spectrum does have components of either type, with a gap in between. As a result, expansion~\eqref{eq:pdf-spectral-expansion} generalizes to
\begin{align}\label{eq:pdf-spectral-expansion2}
p(x,t|r=y)
&=
\rho(x)\left\{\sum_{k} C_{k}(y)\,e^{\lambda_{k} t}\,\psi(x,\lambda_{k})+\int C_{\lambda}(y)\,e^{\lambda t}\,\psi(x,\lambda)\,d\lambda\right\},
\end{align}
where the sum reflects the contribution of the discrete part of the spectrum and the integral accounts for the contribution of the continuous component of the spectrum; cf.~\cite[Formula~(14), p.~266]{Wong:SPMPE1964}.

The above discussion suggests that determining the eigenvalues $\lambda$, i.e., the spectrum of the operator $\mathcal{D}$, is the essence of the entire separation of variables approach to solve equation~\eqref{eq:SR-Rt-pdf-Kolmogorov-fwd-PDE-2}. To determine the eigenvalues, note that by integrating~\eqref{eq:eigen-eqn-two} through with respect to $x$ over the interval $[0,+\infty)$, and making use of the boundary conditions~\eqref{eq:bdd-cond}, we obtain
\begin{align*}
\lambda\int_0^{\infty}\rho(x)\,\psi(x,\lambda)\,dx
&=
0,
\end{align*}
whence it follows that
\begin{align}\label{eq:eig-fcn-int-to-zero}
\text{either $\lambda=0$ or }&\int_0^{\infty}\rho(x)\,\psi(x,\lambda)\,dx
=
0,
\end{align}
which provides a rule whereby one is to ``filter out'' the eigenvalues (and then also the corresponding eigenfunctions). In fact, it is easy to see that $\lambda=0$ always ``works'', and the corresponding eigenfunction is $\psi(x,0)=1$. The corresponding contribution to the density $p(x,t)$ expanded as in~\eqref{eq:pdf-spectral-expansion} can be seen to be $\rho(x)$ given by~\eqref{eq:rho-answer}, which is the stationary solution of equation~\eqref{eq:SR-Rt-pdf-Kolmogorov-fwd-PDE-2}.

We also remark that the eigenfunctions corresponding to different eigenvalues are orthogonal in the sense that
\begin{align}\label{eq:eigenfun-orthonorm}
\int_{0}^{\infty}\rho(x)\,\psi(x,\lambda_i)\,\psi(x,\lambda_j)\,dx
&=
\delta(\lambda_i-\lambda_j),
\end{align}
if $\lambda_i$ and $\lambda_j$ are both from the continuous part of the spectrum. For the discrete part of the spectrum the Delta function in right-hand side above is to be replaced with Kronecker's Delta $\delta_{ij}\triangleq\indicator{i=j}$. We note that this orthogonality property assumes the eigenfunctions are of unit length, i.e., $\|\psi(\cdot,\lambda)\|^2=1$, where the length is defined relative to the ``weight function'' $\rho(x)$, i.e.,
\begin{align}\label{eq:norm-def}
\|\psi(\cdot,\lambda)\|^2
&\triangleq
\int_{0}^{\infty}\rho(x)\,\psi^2(x,\lambda)\,dx.
\end{align}

Next observe that from multiplying~\eqref{eq:eigen-eqn-two} through by $\psi(x,\lambda)$ and integrating both sides of the resulting equation with respect to $x$ over $[0,\infty)$, we obtain
\begin{align*}
\dfrac{\mu^2}{2}\int_{0}^{\infty}\psi(x,\lambda)\,\dfrac{d}{dx}\left[x^2\,\rho(x)\,\dfrac{d}{dx}\psi(x,\lambda)\right]dx
-
\lambda\int_{0}^{\infty}\rho(x)\,\psi^2(x,\lambda)\,dx
&=
0,
\end{align*}
which, after recognizing the second term in the left-hand side as $\|\psi(\cdot,\lambda)\|^2$, i.e., the squared norm~\eqref{eq:norm-def} of $\psi(x,\lambda)$, reduces further to
\begin{align*}
\lambda\,\|\psi(\cdot,\lambda)\|^2
&=
\dfrac{\mu^2}{2}\int_{0}^{\infty}\psi(x,\lambda)\,\dfrac{d}{dx}\left[x^2\,\rho(x)\,\dfrac{d}{dx}\psi(x,\lambda)\right]dx,
\end{align*}
or
\begin{align}\label{eq:lambda-step2}
\lambda
&=
\dfrac{\mu^2}{2}\int_{0}^{\infty}\psi(x,\lambda)\,\dfrac{d}{dx}\left[x^2\,\rho(x)\,\dfrac{d}{dx}\psi(x,\lambda)\right]dx,
\end{align}
because without loss of generality $\psi(x,\lambda)$ may be assumed to be of unit length in the sense of~\eqref{eq:norm-def}, i.e., $\|\psi(\cdot,\lambda)\|^2=1$.

Next, integration by parts applied to the integral in the right-hand side of~\eqref{eq:lambda-step2} reduces the latter to
\begin{align*}
\lambda
&=
\dfrac{\mu^2}{2}\int_{0}^{\infty}\psi(x,\lambda)\,\dfrac{d}{dx}\left[x^2\,\rho(x)\,\dfrac{d}{dx}\psi(x,\lambda)\right]dx\\
&=
\dfrac{\mu^2}{2}\left\{\left.\psi(x,\lambda)\left[x^2\,\rho(x)\,\dfrac{d}{dx}\psi(x,\lambda)\right]\right|_{0}^{\infty}-\int_{0}^{\infty}x^2\,\rho(x)\left[\dfrac{d}{dx}\psi(x,\lambda)\right]^2dx\right\}\\
&=
-\dfrac{\mu^2}{2}\int_0^{\infty}x^2\,\rho(x)\left[\dfrac{d}{dx}\psi(x,\lambda)\right]^2dx,
\end{align*}
where we also used the boundary conditions~\eqref{eq:bdd-cond}. The obtained result implies that $\lambda\le 0$, i.e., the spectrum must be concentrated in the nonpositive half of the real line.

Now, multiplying~\eqref{eq:pdf-spectral-expansion} through by $\psi(x,\lambda_j)$ and integrating with respect to $x$ over the interval $[0,\infty)$, we obtain
\begin{align*}
\int_{0}^{\infty}p(x,t|r=y)\,\psi(x,\lambda_j)\,dx
&=
\sum_{i} C_{i}(y)\, e^{\lambda_{i} t}\left[\int_{0}^{\infty}\rho(x)\,\psi(x,\lambda_{i})\,\psi(x,\lambda_j)\,dx\right],
\end{align*}
whence, in view of the orthogonality property~\eqref{eq:eigenfun-orthonorm}, one can conclude that
\begin{align*}
C_{k}(y)\,e^{\lambda_{k} t}
&=
\int_{0}^{\infty}p(x,t|r=y)\,\psi(x,\lambda_k)\,dx,
\end{align*}
and because $C_{k}(y)$ is to be independent of $t$, evaluating the integral above at $t\to 0+$ and making use of the initial condition $\lim_{t\to0+}p(x,t|r=y)=\delta(x-y)$ we obtain
\begin{align*}
C_{k}(y)
&=
\int_{0}^{\infty}\delta(x-y)\,\psi(x,\lambda_{k})\,dx=\psi(y,\lambda_{k}).
\end{align*}

Therefore, the expansion~\eqref{eq:pdf-spectral-expansion} becomes
\begin{align}\label{eq:pdf-spectral-expansion1}
p(x,t|r=y)
&=
\rho(x)\sum_{k} e^{\lambda_{k} t}\,\psi(x,\lambda_{k})\,\psi(y,\lambda_{k}),
\end{align}
and we observe the symmetry $p(x,t|r=y)/\rho(x)=p(y,t|r=x)/\rho(y)$. Correspondingly, the more general expansion~\eqref{eq:pdf-spectral-expansion2} becomes
\begin{align}\label{eq:pdf-spectral-expansion3}
p(x,t|r=y)
&=
\rho(x)\left\{\sum_{k} e^{\lambda_{k} t}\,\psi(x,\lambda_{k})\,\psi(y,\lambda_{k})+\int e^{\lambda t}\,\psi(x,\lambda)\,\psi(y,\lambda)\,d\lambda\right\},
\end{align}
which is exactly~\cite[Formula~(14),~p.~266]{Wong:SPMPE1964}. It is important to note that the expansions~\eqref{eq:pdf-spectral-expansion1} and~\eqref{eq:pdf-spectral-expansion2} require the eigenfunctions $\psi(x,\lambda)$ to be of unit length in the sense of definition~\eqref{eq:norm-def}, i.e., $\|\psi(\cdot,\lambda)\|^2=1$.

At this point observe that by multiplying, e.g., expansion~\eqref{eq:pdf-spectral-expansion1} through by $e^{-\lambda t}$ and then integrating both sides of the result with respect to $t$ over the interval $[0,\infty)$, we obtain
\begin{align}\label{eq:pdf-Laplace-transform}
\begin{split}
\mathfrak{L}\big\{p(x,t|r=y);t\to\lambda\big\}
&\triangleq
\int_{0}^{\infty} e^{-\lambda t}\,p(x,t|r=y)\,dt\\
&=
\rho(x)\sum_{k}\dfrac{\psi(x,\lambda_{k})\,\psi(y,\lambda_{k})}{\lambda-\lambda_{k}},
\end{split}
\end{align}
which provides an explicit expression for the (temporal) Laplace transform of the sought pdf $p(x,t|r)$

The series appearing in the right-hand side of~\eqref{eq:pdf-Laplace-transform} is the cornerstone of the general method introduced by~\cite{Weyl:MA1910} and by~\cite{Titchmarsh:Book1962} and then developed further, e.g., by~\cite{McKean:TAMS1956} and by~\cite[Section~4.11,~pp.~149--150]{Ito+McKean:Book1974} to solve a large class of PDEs, including equation~\eqref{eq:SR-Rt-pdf-Kolmogorov-fwd-PDE-2}. Specifically, the series appearing in the right-hand side of~\eqref{eq:pdf-Laplace-transform} is the Green's function associated with the differential operator $\mathcal{D}$ given by~\eqref{eq:operator-D-def}. Consider a one-dimensional time-homogeneous Markov diffusion $(Y_t)_{t\ge0}$ that satisfies the SDE: $dY_t=a(Y_t)\,dt+b(Y_t)\,dB_t$, $t\ge0$, with $Y_0=y$. Note that the SDE that governs the GSR statistic $(R_t^r)_{t\ge0}$ is a special case of the SDE for $(Y_t)_{t\ge0}$. Let $\tilde{p}(x,t|y)$ denote the transition pdf associated with the process $(Y_t)_{t\ge0}$. Introduce the differential operator
\begin{align*}
\mathcal{G}
&\triangleq
\dfrac{1}{2} b^2(x)\dfrac{\partial^2}{\partial x^2}+a(x)\dfrac{\partial}{\partial x}
\end{align*}
with
\begin{align*}
\mathcal{G}^*
&\triangleq
\dfrac{1}{2}\dfrac{\partial^2}{\partial x^2}b^2(x)-\dfrac{\partial}{\partial x}a(x)
\end{align*}
being the corresponding adjoint operator. According to the work of~\cite{Kolmogorov:MA1931}, the pdf $\tilde{p}(x,t|y)$ satisfies the Kolmogorov forward equation as well as the Kolmogorov backward equation. In terms of the operators $\mathcal{G}$ and $\mathcal{G}^*$, the forward equation can be compactly written as $\mathcal{G}^*\circ\tilde{p}=\tilde{p}_{t}$, and the operator form of the backward equation is $-\mathcal{G}\circ \tilde{p}=p_{t}$. One of the fundamental properties of the operators $\mathcal{G}$ and $\mathcal{G}^*$ is that they can be parameterized as follows
\begin{align}\label{eq:operator-G-param}
\mathcal{G}
&=
\dfrac{1}{\mathfrak{m}(x)}\dfrac{d}{dx}\dfrac{1}{\mathfrak{s}(x)}\dfrac{d}{dx}
\;\;\text{and}\;\;
\mathcal{G}^*
=
\dfrac{d}{dx}\dfrac{1}{\mathfrak{s}(x)}\dfrac{d}{dx}\dfrac{1}{\mathfrak{m}(x)},
\end{align}
where the function $\mathfrak{s}(x)$ is the solution of the ODE $\mathcal{G}\circ\mathfrak{s}(x)=0$ and the function $\mathfrak{m}(x)$ is found from the ODE $\mathcal{G}^*\circ\mathfrak{m}(x)=0$. The function $\mathfrak{s}(x)$ is known as the scale measure, and the function $\mathfrak{m}(x)$ is referred to as the speed measure. The two ODEs that define the scale and the speed measures make the operators $\mathcal{G}$ and $\mathcal{G}^*$ self-adjoint with respect to, respectively, the scale measure and the speed measure. It is easy to see that
\begin{align}\label{eq:speed+scale-def}
\mathfrak{s}(x)
&\triangleq
\exp\left\{-\int\dfrac{2 a(x)}{b^2 (x)}\,dx\right\}
\;\;\text{and}\;\;
\mathfrak{m}(x)\triangleq\dfrac{2}{b^2(x)\,\mathfrak{s}(x)},
\end{align}
and, moreover, observe that the equation $\mathcal{G}^*\circ\mathfrak{m}=0$ is precisely the equation on the stationary density (as $t\to+\infty$) of the process $(Y_t)_{t\ge0}$. That is, the speed measure $\mathfrak{m}(x)$ determines the stationary distribution (if one exists) of the process $(Y_t)_{t\ge0}$. In particular, since for our equation~\eqref{eq:SR-Rt-pdf-Kolmogorov-fwd-PDE-2} we have $a(x)\equiv 1$ and $b(x)=\mu x$, then  from~\eqref{eq:speed+scale-def} we obtain
\begin{align}\label{eq:speed+scale-answer}
\mathfrak{s}(x)
&=
e^{\tfrac{2}{\mu^2 x}}
\;\;\text{and}\;\;
\mathfrak{m}(x)
=
e^{-\tfrac{2}{\mu^2 x}}\dfrac{2}{\mu^2 x^2},
\end{align}
i.e., the speed measure $\mathfrak{m}(x)$, as expected, coincides with the stationary density $\rho(x)$ given by~\eqref{eq:rho-answer}. Therefore, by comparing the parametrization~\eqref{eq:operator-G-param} of the operator $\mathcal{G}$ and the definition~\eqref{eq:operator-D-def} of the operator $\mathcal{D}$, we see that the characteristic equation $\mathcal{G}\circ\varphi(x)=\lambda\,\varphi(x)$ is equivalent to the equation $\mathcal{D}\circ\varphi(x)=\lambda\,\varphi(x)$ given explicitly by~\eqref{eq:eigen-eqn-two}.

The right-hand side of~\eqref{eq:pdf-Laplace-transform} can be recognized as the Green's function $G_{\lambda}(x,y)$ associated with the differential operator $\mathcal{D}$ given by~\eqref{eq:operator-D-def}. More specifically, the Green's function is defined as the solution of the equation $\mathcal{D}\circ G_{\lambda}(x,y)=\delta(x-y)$. The Green's function has a simple physical interpretation: if the operator $\mathcal{D}$ describes the evolution in time of the state of a system, then the Green's function $G_{\lambda}(x,y)$ captures the response of the system to a unit ``shock''.

More importantly, the Green's function $G_{\lambda}(x,y)$ is obtainable (see, e.g.,~\citealt[p.~19]{Borodin+Salminen:Book2002}) explicitly via the formula
\begin{align}\label{eq:green-fn-def}
\begin{split}
G_{\lambda}(x,y)
&=
\dfrac{\mathfrak{m}(x)}{w_{\lambda}}\,
\varphi_{+}(\min\{x,y\},\lambda)\,\varphi_{-}(\max\{x,y\},\lambda)\\
&=
\dfrac{\mathfrak{m}(x)}{w_{\lambda}}%
\begin{cases}
\varphi_{+}(y,\lambda)\,\varphi_{-}(x,\lambda),\;\;\text{if $y\le x$;}\\
\varphi_{+}(x,\lambda)\,\varphi_{-}(y,\lambda),\;\;\text{if $y\ge x$,}
\end{cases}
\end{split}
\end{align}
where $\mathfrak{m}(x)$ is the speed measure~\eqref{eq:speed+scale-def}, $\varphi_{+}(x,\lambda)$ and $\varphi_{-}(x,\lambda)$ are two independent solutions of the homogeneous (second order) Sturm--Loiuville equation $\mathcal{G}\circ\varphi(x)=\lambda\,\varphi(x)$, and
\begin{align}\label{eq:w_lambda_def}
w_{\lambda}
&=
-\frac{1}{\mathfrak{s}(x)}\mathcal{W}\left\{\varphi_{+}(x,\lambda),\varphi_{-}(x,\lambda)\right\},
\end{align}
with $\mathcal{s}(x)$ being the scale measure~\eqref{eq:speed+scale-def} and $\mathcal{W}\left\{f(x),g(x)\right\}$ denoting the Wronskian
\begin{align}\label{eq:Wronskian-def}
\mathcal{W}\left\{f(x),g(x)\right\}
&\triangleq
\det\begin{bmatrix}
f(x)&g(x)\\
f'(x)&g'(x)
\end{bmatrix}
=
f(x)\,g'(x)-g(x)\,f'(x).
\end{align}

We would like to note that the $\pm$ notation in $\varphi_{+}(x,\lambda)$ and $\varphi_{-}(x,\lambda)$ is to indicate that the $+$ function is increasing in $x$ and the $-$ function is decreasing in $x$. This distinction makes the two functions $\varphi_{+}(x,\lambda)$ and $\varphi_{-}(x,\lambda)$ unique, although up to a constant factor, which is irrelevant because of the division by the Wronskian $w_{\lambda}$. Also, note that the scale measure $\mathfrak{s}(x)$-adapted Wronskian $w_{\lambda}$ is independent of $x$.

Once the Green's function is found from formula~\eqref{eq:green-fn-def}, it is a matter of taking the inverse Laplace transform of it with respect to $\lambda$ to recover the density $\tilde{p}(x,t|y)$. Specifically, the inversion formula is
\begin{align}\label{eq:inv-Laplace-transform}
\tilde{p}(x,t|y)
&=
\mathfrak{L}^{-1}\big\{G_{\lambda}(x,y);\lambda\to t\big\}
\triangleq
\dfrac{1}{2\pi\imath}\int_{\gamma-\imath\infty}^{\gamma+\imath\infty} e^{\lambda t}\,G_{\lambda}(x,y)\,d\lambda,
\end{align}
where $G_{\lambda}(x,y)$ is to be understood as the analytic continuation of the Green's function~\eqref{eq:green-fn-def} and the integration is performed along the contour parallel to the imaginary axis and located in the right half-plane. The standard practice to carry out the integration is to invoke the Residue Theorem. As a matter of fact, by comparing~\eqref{eq:pdf-Laplace-transform} and~\eqref{eq:inv-Laplace-transform} one can see that the poles of the Green's function will yield the eigenvalues while the respective residues will contribute to the value of the integral, i.e., to the pdf $\tilde{p}(x,t|y)$. Obtaining the eigenvalues as the poles of the Green's function is generally easier than using~\eqref{eq:eig-fcn-int-to-zero} above. This is the cornerstone of the Weyl--Titchmarsh framework to treat the Sturm--Liouville problem; see~\cite{Weyl:MA1910} and~\cite{Titchmarsh:Book1962}.

A few remarks are now in order. First, note that our normalization constraint~\eqref{eq:normalization-constraint1} assumes that the pdf $p(x,t|r)$---or the more general pdf $\tilde{p}(x,t|y)$---is defined with respect to the Lebesgue measure $dx$ and not with respect to the speed measure $\mathfrak{m}(x)\,dx$ given by~\eqref{eq:speed+scale-answer}, which we obtained from the definition~\eqref{eq:speed+scale-def}. However, the speed measure $\mathfrak{m}(x)\,dx$ is included in the formula~\eqref{eq:green-fn-def} for the Green's function $G_{\lambda}(x,y)$. This is slightly different from, e.g.,~\cite{Borodin+Salminen:Book2002}, who define the pdf with respect to the speed measure $\mathfrak{m}(x)\,dx$, so that the formula~\eqref{eq:green-fn-def} for the Green's function does not have the factor $\mathfrak{m}(x)$.

Second, note that by construction~\eqref{eq:green-fn-def} the Green's function $G_{\lambda}(x,y)$ exhibits the following property: $G_{\lambda}(x,y)/\mathfrak{m}(x)$ is symmetric with respect to interchanging $x$ and $y$, i.e., $G_{\lambda}(x,y)/\mathfrak{m}(x)\equiv G_{\lambda}(y,x)/\mathfrak{m}(y)$ for all $x,y\ge0$. A direct consequence of this is that pdf $p(x,t|r)$---or the more general pdf $\tilde{p}(x,t|y)$---also exhibits the same type of symmetry, i.e., ${p}(x,t|y)/\mathfrak{m}(x)\equiv {p}(y,t|x)/\mathfrak{m}(y)$. In fact, this is consistent with the expansions~\eqref{eq:pdf-spectral-expansion2}--\eqref{eq:pdf-spectral-expansion3} which also yield this result.

In addition, it is also noteworthy that since in the formula~\eqref{eq:green-fn-def} for the Green's function $G_{\lambda}(x,y)$ the eigenfunctions are standardized by their Wronskian $w_{\lambda}$, it follows that the eigenfunctions $\psi_{\pm}(x,\lambda)$ need not be ``one unit long'' with the length defined by the norm~\eqref{eq:norm-def}. This is in stark contrast with~\cite[Formula~(14),~p.~266]{Wong:SPMPE1964} which is the expansion~\eqref{eq:pdf-spectral-expansion3} we derived above.

We shall now apply the Green's function framework to our equation~\eqref{eq:SR-Rt-pdf-Kolmogorov-fwd-PDE-2}, which we already reduced to equation~\eqref{eq:eigen-eqn-two}. To that end, let us first define $u\triangleq u(x)=2/(\mu^2 x)$ so that $x=2/(\mu^2 u)$ and rewrite equation~\eqref{eq:eigen-eqn-two} with $x$ changed to $u$. Specifically, since
\begin{align*}
u_{x}
&\triangleq
\dfrac{d u(x)}{d x}
=
-\dfrac{\mu^2 u^2}{2}
\;\;\text{and}\;\;
\dfrac{2}{\mu^2 x^2}=\dfrac{\mu^2 u^2}{2},
\end{align*}
so that
\begin{align*}
\psi_{x}(x)
&=
-\dfrac{\mu^2 u^2}{2}\psi_{u}(u)
\;\;\text{and}\;\;
\psi_{xx}(x)
=
u^2\left(\dfrac{\mu^2 u}{2}\right)^2\psi_{uu}(u)+2u\left(\dfrac{\mu^2 u}{2}\right)^2\psi_{u}(u),
\end{align*}
we can see that under the change of variables $x\mapsto u\triangleq u(x)$ equation~\eqref{eq:eigen-eqn-two} takes the form
\begin{align}\label{eq:eigenfcn-eqn-u}
u^2\,\psi_{uu}(u)+u\,(2-u)\,\psi_{u}(u)+\lambda\,\dfrac{2}{\mu^2}\,\psi(u)
&=
0,
\end{align}
which under the substitution
\begin{align}\label{eq:eigfun-eqn-sub}
\psi(u)
&=
\dfrac{e^{\tfrac{u}{2}}}{u}\,\xi(u),
\;\;
\text{with $\xi(u)$ at least twice differentiable},
\end{align}
converts further to
\begin{align}\label{eq:eigenfcn-eqn-u2}
\xi_{uu}(u)+\left\{-\dfrac{1}{4}+\dfrac{1}{u}+\dfrac{1/4-\alpha^2(\lambda)}{u^2}\right\}\xi(u)
&=
0,
\end{align}
where
\begin{align}\label{eq:alpha-def}
\alpha
&\triangleq
\alpha(\lambda)
=
\sqrt{\dfrac{1}{4}+\dfrac{2\lambda}{\mu^2}}
\;\;\text{and}\;\;
\lambda
\triangleq
\lambda(\alpha)
=
\frac{\mu^2}{8}(4\alpha^2-1).
\end{align}

Equation~\eqref{eq:eigenfcn-eqn-u2} is a particular case of the so-called Whittaker equation
\begin{align}\label{eq:Whittaker-eqn}
\xi_{zz}(z)+\left\{-\dfrac{1}{4}+\dfrac{a}{z}+\dfrac{1/4-b^2}{z^2}\right\}\xi(z)
&=
0,
\;\; a,b,z\in\mathbb{C},
\end{align}
which is a self-adjoint equation that emerged from the work of Whittaker~\citeyearpar{Whittaker:BAMS1904}; see, e.g.,~\cite[Chapter~I]{Buchholz:Book1969}. Specifically, Whittaker~\citeyearpar{Whittaker:BAMS1904} introduced his equation as a form of the Kummer equation, which is satisfied by the confluent hypergeometric functions. See, e.g.,~\cite{Slater:Book1960}. Whittaker's~\citeyearpar{Whittaker:BAMS1904} equation~\eqref{eq:Whittaker-eqn} is used to define the now-well-known two Whittaker functions $W_{a,b}(z)$ and ${M}_{a,b}(z)$ as its two independent (fundamental) solutions; note that Whittaker's equation~\eqref{eq:Whittaker-eqn} is a second-order ODE. Since our equation~\eqref{eq:eigenfcn-eqn-u2} is a special case of Whittaker's equation~\eqref{eq:Whittaker-eqn}, the eigenfunctions $\varphi(x)$ are expressible through the Whittaker $W$ and $M$ functions with appropriately chosen indices and argument. Hence, let us briefly pause our derivation of the density $p(x,t|r)$ and summarize certain essential properties of the two Whittaker functions.

Let us first consider the Whittaker $M_{a,b}(z)$ function. It is defined provided $2b\neq -2n-1$, $n\in\mathbb{N}\cup\{0\}$. Under the latter assumption $M_{a,b}(z)$ is an analytic function, whatever be $a,z\in\mathbb{C}$. Otherwise, if the condition on the second index $b$ is violated, the Whittaker $M_{a,b}(z)$ function experiences a simple pole. To avoid this problem, the {\em regularized} Whittaker function $\mathcal{M}_{a,b}(z)\triangleq M_{a,b}(z)/\Gamma(1+2b)$ was introduced, which is a well-behaved function for any $b\in\mathbb{C}$. Here and onward $\Gamma(z)$ denotes the well-known Gamma function. Since $\mathcal{M}_{a,b}(z)$ is analytic for all $a,b,z\in\mathbb{C}$, and as a multiple of the $M_{a,b}(z)$ function also solves Whittaker's equation~\eqref{eq:Whittaker-eqn}, we shall from now on deal with the $\mathcal{M}_{a,b}(z)$ function.

The Whittaker $W_{a,b}(z)$ function is defined from the $M_{a,b}(z)$ function as follows:
\begin{align}\label{eq:Whittaker-fncs-identity}
W_{a,b}(z)
&=
\dfrac{\Gamma(-2b)}{\Gamma(1/2-b-a)}M_{a,b}(z)+\dfrac{\Gamma(2b)}{\Gamma(1/2+b-a)}M_{a,-b}(z);
\end{align}
cf., e.g.,~\cite[Identity~13.1.34,~p.~505]{Abramowitz+Stegun:Handbook1964}. The $W_{a,b}(z)$ function is analytic for all $a,b,z\in\mathbb{C}$. It can be readily observed from the definition that $W_{a,b}(z)=W_{a,-b}(z)$, and this symmetry will play an important role in the sequel.

Another important property of the two Whittaker functions is their Wronskian. Specifically, according to, e.g.,~\cite[Formula~(33),~p.~25]{Buchholz:Book1969}, we have
\begin{align}\label{eq:Whittaker-funcs-Wronskian}
\mathcal{W}\big\{W_{a,b}(z),\mathcal{M}_{a,b}(z)\big\}
&=
\dfrac{1}{\Gamma(b-a+1/2)},
\end{align}
whence one can see that when $b-a+1/2$ is either zero or a negative integer, the Gamma function in the denominator has a simple pole, and the Whittaker functions become dependent. In that case, the Whittaker functions degenerate to a type of polynomial known as the Laguerre polynomial; the latter polynomials are constructed from the standard monomial basis $\{1,x,x^2,\ldots,x^n,\ldots\}$ by the Gram--Schmidt procedure and form an orthonormal basis on $x\in\mathbb{R}^{+}$ with respect to the measure $e^{-x}dx$.

Let us now return to our equation~\eqref{eq:eigenfcn-eqn-u}. From~\eqref{eq:Whittaker-eqn},~\eqref{eq:eigenfcn-eqn-u2}, and~\eqref{eq:eigfun-eqn-sub}, one can see that the two solutions to our  equation~\eqref{eq:eigenfcn-eqn-u} are as follows:
\begin{align}\label{eq:eig-fn-answer-no-class-u}
\psi_1(u,\lambda)
&=
\dfrac{e^{\tfrac{u}{2}}}{u}\,W_{1,\alpha}(u)
\;\;\text{and}\;\;
\psi_2(u,\lambda)
=
\dfrac{e^{\tfrac{u}{2}}}{u}\,\mathcal{M}_{1,\alpha}(u)
\;\;\text{with}\;\; u=\dfrac{2}{\mu^2 x},
\end{align}
whence, by performing the back substitution $x=2/(\mu^2 u)$, we finally arrive at the {\em nonnormalized} eigenfunctions
\begin{align}\label{eq:eig-fn-answer-no-class}
\psi_1(x,\lambda)
&=
e^{\tfrac{1}{\mu^2 x}}\,\dfrac{\mu^2 x}{2}\,W_{1,\alpha}\left(\dfrac{2}{\mu^2 x}\right)
\,\text{and }
\psi_2(x,\lambda)
=
e^{\tfrac{1}{\mu^2 x}}\,\dfrac{\mu^2 x}{2}\,\mathcal{M}_{1,\alpha}\left(\dfrac{2}{\mu^2 x}\right),
\end{align}
where $x\ge0$ and $\alpha\triangleq\alpha(\lambda)$ is given by~\eqref{eq:alpha-def}.

Let us now classify the two eigenfunctions, i.e., identify the increasing one and the decreasing one. This requires looking at the behavior of each of the two eigenfunctions at the boundaries, viz. as $x\to0+$ and as $x\to+\infty$. To that end, observe that, on the one hand, since
\begin{align*}
\begin{split}
W_{a,b}(x)
&\sim
\dfrac{\Gamma(2b)}{\Gamma(b-a+1/2)}\,x^{-b+\tfrac{1}{2}}\,e^{-\tfrac{x}{2}}
\;\;\text{as}\;\; x\to0+,
\;\;\text{and}\\
\mathcal{M}_{a,b}(x)
&\sim
\dfrac{1}{\Gamma(1+2b)}\,x^{b+\tfrac{1}{2}}\,e^{-\tfrac{x}{2}}
\;\;\text{as}\;\; x\to0+,
\end{split}
\end{align*}
it follows at once that $\lim_{x\to+\infty}\psi_1(x,\lambda)=\infty$ and $\lim_{x\to+\infty}\psi_2(x,\lambda)=0$. On the other hand, from the asymptotic properties
\begin{align*}
\begin{split}
W_{a,b}(x)
&\sim
x^{a}\,e^{-\tfrac{x}{2}}
\;\;\text{as}\;\; x\to+\infty,
\;\;\text{and}\\
\mathcal{M}_{a,b}(x)
&\sim
\dfrac{1}{\Gamma(b-a+1/2)}\,x^{-a}\,e^{\tfrac{x}{2}}
\;\;\text{as}\;\;
x\to+\infty,
\end{split}
\end{align*}
we obtain that $\lim_{x\to0+}\psi_1(x,\lambda)=1$ and $\lim_{x\to0+}\psi_2(x,\lambda)=\infty$.

Therefore, we find that $\psi_{+}(x,\lambda)\equiv\psi_1(x,\lambda)$ and $\psi_{-}(x,\lambda)\equiv\psi_2(x,\lambda)$, or more explicitly
\begin{align}\label{eq:eig-fn-answer}
\psi_{+}(x,\lambda)
&=
e^{\tfrac{1}{\mu^2 x}}\,\dfrac{\mu^2 x}{2}\,W_{1,\alpha}\left(\dfrac{2}{\mu^2 x}\right)
\,\text{and }
\psi_{-}(x,\lambda)
=
e^{\tfrac{1}{\mu^2 x}}\,\dfrac{\mu^2 x}{2}\,\mathcal{M}_{1,\alpha}\left(\dfrac{2}{\mu^2 x}\right),
\end{align}
where $x\ge0$ and $\alpha\triangleq\alpha(\lambda)$ is given by~\eqref{eq:alpha-def}; cf., e.g.,~\cite[Subsection~4.6.1, p.~276--276]{Linetsky:HandbookChapter2007}.

The next step is to find the Wronskian $\mathcal{W}\big\{\psi_{+}(x,\lambda),\psi_{-}(x,\lambda)\big\}$ and then obtain the speed-measure $\mathfrak{s}(x)$-adapted Wronskian $w_{\lambda}$ given by~\eqref{eq:w_lambda_def}; recall that for $f(x)$ and $g(x)$ sufficiently smooth the Wronskian $\mathcal{W}\big\{f(x),g(x)\big\}$ is defined by~\eqref{eq:Wronskian-def}. To that end, it will prove convenient to temporarily go back to the variable $u\triangleq u(x)=2/(\mu^2 x)$. Specifically, since $\psi_{+}(x,\lambda)\equiv\psi_1(x,\lambda)$ and $\psi_{-}(x,\lambda)\equiv\psi_2(x,\lambda)$, then from~\eqref{eq:eig-fn-answer-no-class-u} we have
\begin{align*}
\begin{split}
\mathcal{W}\big\{\psi_{+}(u,\lambda),\psi_{-}(u,\lambda)\big\}
&=
\mathcal{W}\left\{\dfrac{e^{\tfrac{u}{2}}}{u}\,W_{1,\alpha}(u),\dfrac{e^{\tfrac{u}{2}}}{u}\,\mathcal{M}_{1,\alpha}(u)\right\}\\
&
\stackrel{\text{(a)}}{=}
-\mathcal{W}\left\{\dfrac{e^{\tfrac{u}{2}}}{u}\,\mathcal{M}_{1,\alpha}(u),\dfrac{e^{\tfrac{u}{2}}}{u}\,W_{1,\alpha}(u)\right\}\\
&
\stackrel{\text{(b)}}{=}
-\dfrac{e^{u}}{u^2}\,\mathcal{W}\big\{\mathcal{M}_{1,\alpha}(u),W_{1,\alpha}(u)\big\}\\
&
\stackrel{\text{(c)}}{=}
\dfrac{e^{u}}{u^2\Gamma(\alpha-1/2)},
\end{split}
\end{align*}
where $\text{(a)}$ is because $\mathcal{W}\big\{f(x),g(x)\big\}=-\mathcal{W}\big\{g(x),f(x)\big\}$, which is easy to see directly from the definition~\eqref{eq:Wronskian-def} of the Wronskian, for $\text{(b)}$ we used \cite[Formula~(35), p.~26]{Buchholz:Book1969}, as per which $\mathcal{W}\big\{f(z)\,\mathcal{M}_{a,b}(z),f(z)\,W_{a,b}(z)\big\}=
f^2(z)\,\mathcal{W}\big\{\mathcal{M}_{a,b}(z),W_{a,b}(z)\big\}$, and $\text{(c)}$ is due to~\eqref{eq:Whittaker-funcs-Wronskian}.

Next, from the foregoing and the obvious identity $\mathcal{W}\big\{f(x),g(x)\big\}=u_{x}\mathcal{W}\big\{f(u),g(u)\big\}$, for the Wronskian in terms of $x$ we obtain
\begin{align*}
\mathcal{W}\big\{\psi_{+}(x,\lambda),\psi_{-}(x,\lambda)\big\}
&=
-e^{\tfrac{2}{\mu^2 x}}\,\dfrac{\mu^2}{2\,\Gamma(\alpha-1/2)},
\end{align*}
and consequently using~\eqref{eq:speed+scale-answer} and~\eqref{eq:w_lambda_def} it follows that
\begin{align}\label{eq:w_lambda-answer}
w_{\lambda}
&\triangleq
\dfrac{1}{\mathfrak{s}(x)}\,\mathcal{W}\big\{\psi_{+}(x,\lambda),\psi_{-}(x,\lambda)\big\}
=
\dfrac{\mu^2}{2\,\Gamma(\alpha-1/2)},
\end{align}
where $\alpha\triangleq\alpha(\lambda)$ is given by~\eqref{eq:alpha-def}.

We are now ready to write down the Green's function in a closed form. Specifically, together~\eqref{eq:w_lambda-answer},~\eqref{eq:eig-fn-answer} and~\eqref{eq:green-fn-def} yield
\begin{align}\label{eq:Greens-fcn-answer}
\begin{split}
G_{\lambda}(x,y)
&=
e^{-\tfrac{1}{\mu^2 x}}\,\dfrac{2}{\mu^2 x}\,e^{\tfrac{1}{\mu^2 y}}\,\dfrac{\mu^2 y}{2}\,\Gamma\left(\alpha-\frac{1}{2}\right)\times\\
&\quad\times\,W_{1,\alpha}\left(\dfrac{2}{\mu^2\min\{x,y\}}\right)\mathcal{M}_{1,\alpha}\left(\dfrac{2}{\mu^2\max\{x,y\}}\right),\;x,y\ge0,
\end{split}
\end{align}
where $\alpha\triangleq\alpha(\lambda)$ is as in~\eqref{eq:alpha-def} with $\lambda>0$. Up to the notation, the result is in agreement with, e.g.,~\cite[Theorem~3.1,~p.~185]{Donati-Martin+etal:RMI2001},~\cite[Proposition~3.2,~p.~261]{Linetsky:MF2006}. See also, e.g.,~\cite[Section~4.6,~pp.~276--282]{Linetsky:HandbookChapter2007}. Moreover, since $W_{1,b}(x)\sim x\, e^{-\tfrac{x}{2}}$ as $x\to+\infty$, then from~\eqref{eq:Greens-fcn-answer} for $r=y\to0+$ we obtain
\begin{align}\label{eq:Greens-fcn-answer-no-headstart}
G_{\lambda}(x,0)
&=
e^{-\tfrac{1}{\mu^2 x}}\,\dfrac{2}{\mu^2 x}\,\Gamma\left(\alpha-\frac{1}{2}\right)\mathcal{M}_{1,\alpha}\left(\dfrac{2}{\mu^2 x}\right),\;x\ge0,
\end{align}
which (up to the notation) is consistent with~\cite[Proposition~3.2,~pp.~261--262]{Linetsky:MF2006}.

Now that $G_{\lambda}(x,y)$ is available explicitly, we are all set to proceed to finding $p(x,t|r=y)$ as $\mathfrak{L}^{-1}\big\{G_{\lambda}(x,y);\lambda\to t\big\}$, for $\mathfrak{L}\big\{p(x,t|r=y);t\to\lambda\}=G_{\lambda}(x,y)$, as was mentioned earlier. Thus, the problem now is to find the Bromwich integral
\begin{align}\label{eq:Bromwich-int}
p(x,t|r=y)
&=
\mathfrak{L}^{-1}\big\{G_{\lambda}(x,y);\lambda\to t\big\}
\triangleq
\frac{1}{2\pi\imath}\lim_{T\to+\infty}\int_{\gamma-\imath T}^{\gamma+\imath T} e^{\lambda t}\,G_{\lambda}(x,y)\,d\lambda,
\end{align}
where $\gamma>0$.

To compute the Bromwich integral, let us devise the Residue Theorem, which is the standard approach to invert a Laplace transform. Specifically, recall that if $\mathcal{C}$ is a positively oriented simple closed path in the complex plane, then by the Residue Theorem we have
\begin{align}\label{eq:Bromwich-int-Cauchy-thm}
\oint_{\mathcal{C}} e^{\lambda t}\,G_{\lambda}(x,y)\,d\lambda
&=
2\pi\imath\sum_{k}e^{\lambda_k t}\,\Res_{\lambda=\lambda_k}G_{\lambda}(x,y),
\end{align}
where $\lambda_{k}\in\mathcal{C}$ are the singularity points of the Green's function $G_{\lambda}(x,y)$ viewed as a function of $\lambda$, and $\Res_{\lambda=\lambda_k}G_{\lambda}(x,y)\triangleq\lim_{\lambda\to\lambda_{k}}(\lambda-\lambda_{k})\,G_{\lambda}(x,y)$ are the respective residues; note that the poles $\lambda_{k}$ are to lie within the contour $\mathcal{C}$.

To design the contour $\mathcal{C}$, note that since the Whittaker functions $W_{a,b}(z)$ and $\mathcal{M}_{a,b}(z)$ are both analytic functions for all $a,b,z\in\mathbb{C}$, then from formula~\eqref{eq:Greens-fcn-answer} for the Green's function $G_{\lambda}(x,y)$ we see that ``trouble'' under the Browmich integral~\eqref{eq:Bromwich-int} can only come from two sources:\begin{inparaenum}[\itshape(a)]\item the Gamma function factor $\Gamma(\alpha(\lambda)-1/2)$, and \item the square root function lurking in $\alpha(\lambda)$ given by~\eqref{eq:alpha-def}\end{inparaenum}.

More specifically, with regard to the Gamma function factor $\Gamma(\alpha(\lambda)-1/2)$ in the formula~\eqref{eq:Greens-fcn-answer} for the Green's function $G_{\lambda}(x,y)$, recall that, for the Gamma function $\Gamma(z)$ a pole occurs when the argument $z$ is either zero or a negative integer. Therefore, since in our case the equation $\alpha(\lambda_n)-1/2=-n+1$, where $n\in\mathbb{N}$, is satisfied only for $n=1$ and the solution is $\lambda_1=0$, as can be seen from~\eqref{eq:alpha-def}, the Gamma function factor $\Gamma(\alpha(\lambda)-1/2)$ ends up having a (simple) pole at $\lambda=0$, and the Green's function $G_{\lambda}(x,y)$ has no other poles. However, since $\alpha(\lambda)$ given by~\eqref{eq:Bromwich-int} involves a square root, which is a multi-valued function when $\lambda\in\mathbb{C}$, in addition to the pole at $\lambda=0$, the Green's function $G_{\lambda}(x,y)$ also has a branch cut. That is the second ``source of trouble'' under the Browmich integral~\eqref{eq:Bromwich-int}. Specifically, the branch point opening the cut is $\lambda=-\mu^2/8$ which is the solution of the equation $\alpha(\lambda)=0$ and the cut extends left of the branch point to $-\infty$ along the negative real semiaxis. Formally, from~\eqref{eq:alpha-def} we see that the branch cut can be parameterized as follows:
\begin{align}\label{eq:beta-def}
\lambda_{\beta}
&\triangleq
-\frac{\mu^2}{8}(1+4\beta^2)
\;\;\text{with}\;\;\beta>0,
\;\;\text{so that}\;\;
\alpha
\triangleq
\alpha(\lambda_{\beta})
=
\pm\imath\beta.
\end{align}

\begin{figure}[h]
    \centering
    \includegraphics[width=0.9\textwidth]{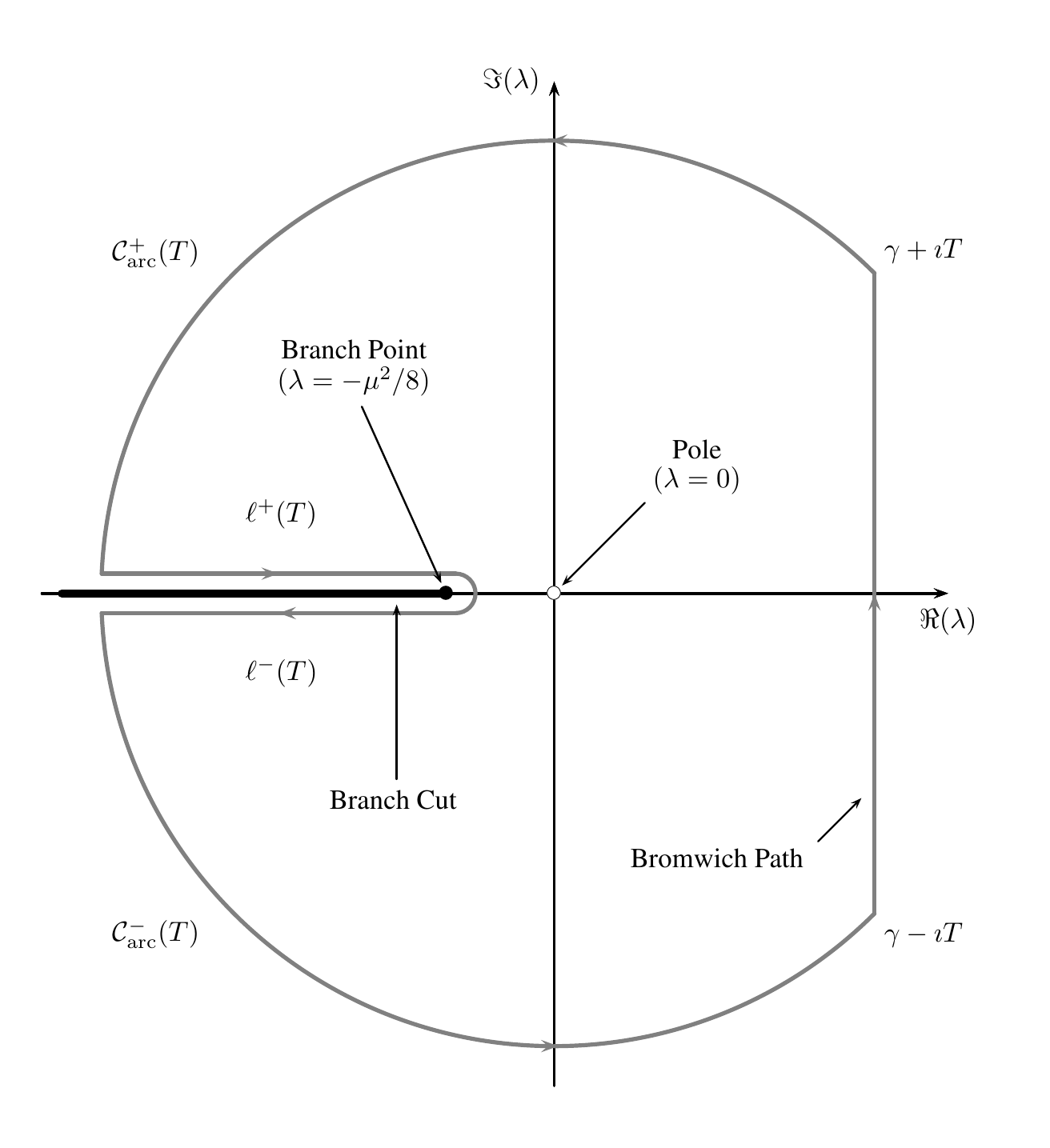}
    \caption{Bromwich contour $\mathcal{C}\triangleq\mathcal{C}(T)$, $T>0$, for Laplace transform inversion.}
    \label{fig:bromwich-contour}
\end{figure}

In view of the above, let us pick the contour $\mathcal{C}\triangleq\mathcal{C}(T)$, $T>0$, as shown in Figure~\ref{fig:bromwich-contour}. For this contour, we have
\begin{align}\label{eq:Bromwich-int-contour}
\begin{split}
\oint_{\mathcal{C}(T)} e^{\lambda t}\,G_{\lambda}(x,y)\,d\lambda
&=
\int_{\gamma-\imath T}^{\gamma+\imath T} e^{\lambda t}\,G_{\lambda}(x,y)\,d\lambda
+\\
&
\qquad\qquad\qquad
+\int_{\mathcal{C}_{\mathrm{arc}}^{+}(T)\cup\mathcal{C}_{\mathrm{arc}}^{-}(T)} e^{\lambda t}\,G_{\lambda}(x,y)\,d\lambda+\\
&\qquad+
\int_{\ell^{+}(T)}e^{\lambda t}\,G_{\lambda}(x,y)\,d\lambda
+
\int_{\ell^{-}(T)}e^{\lambda t}\,G_{\lambda}(x,y)\,d\lambda,
\end{split}
\end{align}
and, moreover, for the Bromwich integral~\eqref{eq:Bromwich-int} we can safely set $\gamma=0+$, because the Green's function $G_{\lambda}(x,y)$ is singularity-free to the right of the imaginary axis. As a matter of fact, the same conclusion can also be reached from the observation made above that the spectrum $\lambda$ of the operator $\mathcal{D}$ is concentrated on the negative real semiaxis (starting from zero).

Since the contour $\mathcal{C}(T)$ shown in Figure~\ref{fig:bromwich-contour} encloses only one pole, viz. at $\lambda=0$, whatever be $T>0$, the loop around the branch point
contributes nothing, and the integrals over the arc segments $\mathcal{C}_{\mathrm{arc}}^{+}(T)$ and $\mathcal{C}_{\mathrm{arc}}^{-}(T)$ vanish as $T\to+\infty$, from~\eqref{eq:Bromwich-int-Cauchy-thm} and~\eqref{eq:Bromwich-int-contour} we obtain
\begin{align}\label{eq:Bromwich-int-step1}
\begin{split}
\lim_{T\to+\infty}&\dfrac{1}{2\pi\imath}\int_{\gamma-\imath T}^{\gamma+\imath T} e^{\lambda t}\,G_{\lambda}(x,y)\,d\lambda
=
\Res_{\lambda=0}G_{\lambda}(x,y)-\\
&\quad-
\lim_{T\to+\infty}\dfrac{1}{2\pi\imath}\left\{
\int_{\ell^{+}(T)}e^{\lambda t}\,G_{\lambda}(x,y)\,d\lambda+\int_{\ell^{-}(T)}e^{\lambda t}\,G_{\lambda}(x,y)\,d\lambda
\right\}.
\end{split}
\end{align}

Let us first find $\Res_{\lambda=0}G_{\lambda}(x,y)$. To that end, observe that since $W_{a,b}(z)$ and $\mathcal{M}_{a,b}(z)$ are analytic for all $a,b,z\in\mathbb{C}$, from~\eqref{eq:Greens-fcn-answer} and the fact that $\lim_{\lambda\to0}\alpha(\lambda)=1/2$, we have
\begin{align}\label{eq:Res-at-zero1}
\begin{split}
\Res_{\lambda=0}G_{\lambda}(x,y)
&\triangleq
\lim_{\lambda\to0}\big\{\lambda\,G_{\lambda}(x,y)\big\}\\
&=
e^{-\tfrac{1}{\mu^2 x}}\,\dfrac{2}{\mu^2 x}\,e^{\tfrac{1}{\mu^2 y}}\,\dfrac{\mu^2 y}{2}\,\times\\
&\qquad\times\left\{\lim_{\lambda\to 0}\lambda\,\Gamma\left(\alpha(\lambda)-\frac{1}{2}\right)\right\}\times\\
&\qquad\qquad
\times W_{1,\tfrac{1}{2}}\left(\dfrac{2}{\mu^2\min\{x,y\}}\right)\mathcal{M}_{1,\tfrac{1}{2}}\left(\dfrac{2}{\mu^2\max\{x,y\}}\right)\\
&=
e^{-\tfrac{2}{\mu^2 x}}\,\dfrac{2}{\mu^2 x^2}\,\dfrac{2}{\mu^2}\,\lim_{\lambda\to0}\lambda\,\Gamma\left(\alpha(\lambda)-\frac{1}{2}\right),
\end{split}
\end{align}
for
\begin{align*}
M_{1,\tfrac{1}{2}}(z)
&=
W_{1,\tfrac{1}{2}}(z)=z\,e^{-\tfrac{z}{2}},
\;\;\text{so that}\;\;
\mathcal{M}_{1,\tfrac{1}{2}}(z)=\dfrac{M_{1,\tfrac{1}{2}}(z)}{\Gamma(2)}=z\,e^{-\tfrac{z}{2}},
\end{align*}
which can be concluded from
\begin{align*}
M_{a,a-\tfrac{1}{2}}(z)
&=
W_{a,a-\tfrac{1}{2}}(z)
=
z^{a}\,e^{-\tfrac{z}{2}},
\end{align*}
given, e.g., by~\cite[Identity~(28a),~p.~23]{Buchholz:Book1969}.

To find $\lim_{\lambda\to0}\big\{\lambda\Gamma(\alpha(\lambda)-1/2)\big\}$ and thus complete the evaluation of $\Res_{\lambda=0}G_{\lambda}(x,y)$ from~\eqref{eq:Res-at-zero1}, recall the so-called Euler limit representation of the Gamma function
\begin{align*}
\Gamma(z)
&=
\dfrac{1}{z}\prod_{n=1}^{\infty}\left[\left(1+\frac{1}{n}\right)^{z}\left/\,\left(1+\frac{z}{n}\right)\right.\right],
\end{align*}
given, e.g.,~\cite[Formula~8.322,~p.~894]{Gradshteyn+Ryzhik:Book2007}. This reforestation makes it apparent that the Gamma function's pole at $z=0$ is entirely due to the factor $1/z$ in front of the infinite product; the other poles at $z=-n$, $n\in\mathbb{N}$, come from the denominator under the infinite product, and this denominator is a well-behaved function around $z=0$. Hence, we obtain
\begin{align*}
\lim_{\lambda\to 0}\lambda\,\Gamma(\alpha(\lambda)-1/2)
&=
\lim_{\lambda\to 0}\Biggl\{\dfrac{\lambda}{\alpha(\lambda)-1/2}\times\\
&\quad\quad
\times\prod_{n=1}^{\infty}\left[\left(1+\frac{1}{n}\right)^{\alpha(\lambda)-\tfrac{1}{2}}\left/\,\left(1+\frac{\alpha(\lambda)-1/2}{n}\right)\right.\right]\Biggr\}\\
&=
\lim_{\lambda\to 0}\dfrac{\lambda}{\alpha(\lambda)-1/2}=\dfrac{\mu^2}{2},
\end{align*}
because from~\eqref{eq:alpha-def} it easy to see that
\begin{align*}
\alpha(\lambda)
&\triangleq
\dfrac{1}{2}\sqrt{1+\dfrac{8\lambda}{\mu^2}}
=
\dfrac{1}{2}\left\{1+\dfrac{4}{\mu^2}\lambda+O(\lambda^2)\right\}
\;\;\text{as}\;\;
\lambda\to0.
\end{align*}

We finally obtain
\begin{align}\label{eq:Bromwich-int-Res0}
\Res_{\lambda=0}G_{\lambda}(x,y)
&=
e^{-\tfrac{2}{\mu^2 x}}\,\dfrac{2}{\mu^2 x^2},
\;\;
x,y\ge0,
\end{align}
and we hasten to note the equality of $\Res_{\lambda=0}G_{\lambda}(x,y)$ to the stationary distribution $\rho(x)\triangleq\lim_{t\to+\infty}p(x,t|r)$ given by~\eqref{eq:rho-answer}, which we recall was obtained directly from the Kolmogorov forward equation~\eqref{eq:SR-Rt-pdf-Kolmogorov-fwd-PDE-2} by assuming that $p(x,t|r)$ is independent of time. This equality is no coincidence: the fact that $G_{\lambda}(x,y)=\mathfrak{L}\big\{p(x,t|r=y);t\to\lambda\big\}$ and the final value theorem for Laplace transforms, manifested in the statement
\begin{align*}
\lim_{t\to+\infty} f(x,t)
&=
\lim_{\lambda\to0} \lambda\,\mathfrak{L}\big\{f(x,t);t\to\lambda\big\},
\end{align*}
readily imply that $\rho(x)=\lim_{\lambda\to0}\big\{\lambda\,G_{\lambda}(x,y)\big\}\triangleq\Res_{\lambda=0}G_{\lambda}(x,y)$. This is an alternative way to find the stationary density $\rho(x)\triangleq\lim_{t\to+\infty}p(x,t|r)$.

Let us now evaluate the integrals above and below the branch cut, i.e., the integrals along the segments $\ell^{+}(T)$ and $\ell^{-}(T)$, respectively, as shown in Figure~\ref{fig:bromwich-contour}. To that end, under the parametrization $\lambda_{\beta}$ given by~\eqref{eq:beta-def} of the branch cut, the segment $\ell^{+}(T)$ is described by the equation $\lambda=|\lambda_{\beta}|\,e^{\imath\pi}$, $\beta\in[0,T]$, and the segment $\ell^{-}(T)$ is described by the equation $\lambda=|\lambda_{\beta}|\,e^{-\imath\pi}$, $\beta\in[0,T]$. Hence, along the segment $\ell^{+}(T)$ we have $\alpha(\lambda_{\beta})=\imath\beta$, while along the segment $\ell^{-}(T)$ we have $\alpha(\lambda_{\beta})=-\imath\beta$. This change of sign that the square root function buried in the definition~\eqref{eq:alpha-def} of $\alpha(\lambda)$ undergoes as we pass from the upper half-place to the lower half-plane was the reason to introduce the $\pm$ notation in $\ell^{\pm}(T)$. Since in either case from~\eqref{eq:beta-def} we have $d\lambda=d\lambda_{\beta}=-\mu^2\beta d\beta$, then for $T\to+\infty$ we obtain
\begin{align}\label{eq:Branch-cut-int-step0}
\begin{split}
\lim_{T\to+\infty}&\Biggl\{\int_{\ell^{+}(T)}e^{\lambda t}\,G_{\lambda}(x,y)\,d\lambda+\int_{\ell^{-}(T)}e^{\lambda t}\,G_{\lambda}(x,y)\,d\lambda\Biggr\}=\\
&=
\mu^2\,e^{-\tfrac{\mu^2 t}{8}}\Biggl\{\int_{0}^{\infty}e^{-\tfrac{\mu^2 t}{2}\beta^2}\Bigg[G_{\lambda}(x,y)\Bigg|_{\lambda=\tfrac{\mu^2}{8}(1+4\beta^2)e^{\imath\pi}}\Bigg]\beta\,d\beta-\\
&
\qquad\qquad\qquad\qquad
-\int_{0}^{\infty}e^{-\tfrac{\mu^2 t}{2}\beta^2}\Bigg[G_{\lambda}(x,y)\Bigg|_{\lambda=\tfrac{\mu^2}{8}(1+4\beta^2)e^{-\imath\pi}}\Bigg]\beta\,d\beta\Biggr\},
\end{split}
\end{align}
because the segment $\ell^{-}(T)$ is traversed ``backwards''.

Next, for the magnitude of the jump across the branch cut we obtain
\begin{align}\label{eq:Branch-cut-int-step1}
\begin{split}
G_{\lambda}&(x,y)\Bigg|_{\lambda=\tfrac{\mu^2}{8}(1+4\beta^2)e^{\imath\pi}}-G_{\lambda}(x,y)\Bigg|_{\lambda=\tfrac{\mu^2}{8}(1+4\beta^2)e^{-\imath\pi}}
=\\
&\qquad\qquad\qquad
=
e^{-\tfrac{1}{\mu^2 x}}\,\dfrac{2}{\mu^2 x}\,e^{\tfrac{1}{\mu^2 y}}\,\dfrac{\mu^2 y}{2}\,\times\\
&
\times\Biggl\{
\Gamma\left(\imath\beta-\frac{1}{2}\right)\,W_{1,\imath\beta}\left(\dfrac{2}{\mu^2\min\{x,y\}}\right)\mathcal{M}_{1,\imath\beta}\left(\dfrac{2}{\mu^2\max\{x,y\}}\right)-\\
&\quad
-
\Gamma\left(-\imath\beta-\frac{1}{2}\right)\,W_{1,-\imath\beta}\left(\dfrac{2}{\mu^2\min\{x,y\}}\right)\mathcal{M}_{1,-\imath\beta}\left(\dfrac{2}{\mu^2\max\{x,y\}}\right)
\Biggr\}\\
&\qquad\qquad\qquad
=
e^{-\tfrac{1}{\mu^2 x}}\,\dfrac{2}{\mu^2 x}\,e^{\tfrac{1}{\mu^2 y}}\,\dfrac{\mu^2 y}{2}\,W_{1,\imath\beta}\left(\dfrac{2}{\mu^2\min\{x,y\}}\right)\,\times\\
&
\times\Biggl\{
\Gamma\left(\imath\beta-\frac{1}{2}\right)\,\mathcal{M}_{1,\imath\beta}\left(\dfrac{2}{\mu^2\max\{x,y\}}\right)-\\
&\qquad\qquad\qquad\qquad
-
\Gamma\left(-\imath\beta-\frac{1}{2}\right)\,\mathcal{M}_{1,-\imath\beta}\left(\dfrac{2}{\mu^2\max\{x,y\}}\right)
\Biggr\},
\end{split}
\end{align}
where to pull out the Whittaker $W$ function we used the latter's symmetry property $W_{a,b}(z)=W_{a,-b}(z)$ valid for all $a,b,z\in\mathbb{C}$, and, as we explained above, this property is a consequence of the relation~\eqref{eq:Whittaker-fncs-identity} between the Whittaker $W$ and $M$ functions. As another implication of the relation~\eqref{eq:Whittaker-fncs-identity}, note that for $a=1$ and $b=\imath\beta$ it takes the form
\begin{align}\label{eq:Whittaker-fncs-identity-v2}
W_{1,\imath\beta}(z)
&=
\dfrac{\Gamma(-2\imath\beta)}{\Gamma(-\imath\beta-1/2)}\,M_{1,\imath\beta}(z)+\dfrac{\Gamma(2\imath\beta)}{\Gamma(\imath\beta-1/2)}\,M_{1,-\imath\beta}(z),
\end{align}
whence, because the two terms in the right-hand side are complex conjugates of each other, one may deduce that $W_{1,\imath\beta}(z)$ must necessarily be real-valued.

Moreover, with the help of~\eqref{eq:Whittaker-fncs-identity-v2}, we can simplify~\eqref{eq:Branch-cut-int-step1} as follows:
\begin{align}\label{eq:Branch-cut-int-step2}
\begin{split}
\Gamma(&\imath\beta-1/2)\,\mathcal{M}_{1,\imath\beta}(z)-\Gamma(-\imath\beta-1/2)\,\mathcal{M}_{1,-\imath\beta}(z)=\\
&\overset{\text{(a)}}{=}
\dfrac{\Gamma(\imath\beta-1/2)}{\Gamma(1+2\imath\beta)}\,{M}_{1,\imath\beta}(z)-\dfrac{\Gamma(-\imath\beta-1/2)}{\Gamma(1-2\imath\beta)}\,{M}_{1,-\imath\beta}(z)\\
&=
\dfrac{\Gamma(\imath\beta-1/2)\,\Gamma(-\imath\beta-1/2)}{\Gamma(2\imath\beta)\,\Gamma(-2\imath\beta)}\Biggl\{\dfrac{\Gamma(2\imath\beta)\,\Gamma(-2\imath\beta)}{\Gamma(-\imath\beta-1/2)\,\Gamma(1+2\imath\beta)}\,{M}_{1,\imath\beta}(z)-\\
&\qquad\qquad\qquad
-
\dfrac{\Gamma(2\imath\beta)\,\Gamma(-2\imath\beta)}{\Gamma(\imath\beta-1/2)\,\Gamma(1-2\imath\beta)}\,{M}_{1,-\imath\beta}(z)\Biggr\}\\
&\overset{\text{(b)}}{=}
\dfrac{\Gamma(\imath\beta-1/2)\,\Gamma(-\imath\beta-1/2)}{2\imath\beta\,\Gamma(2\imath\beta)\,\Gamma(-2\imath\beta)}\,\times\\
&\qquad\qquad
\times\Biggl\{\dfrac{\Gamma(-2\imath\beta)}{\Gamma(-\imath\beta-1/2)}\,{M}_{1,\imath\beta}(z)
+
\dfrac{\Gamma(2\imath\beta)}{\Gamma(\imath\beta-1/2)}\,{M}_{1,-\imath\beta}(z)\Biggr\}\\
&\overset{\text{(c)}}{=}
\dfrac{\Gamma(\imath\beta-1/2)\,\Gamma(-\imath\beta-1/2)}{2\imath\beta\,\Gamma(2\imath\beta)\,\Gamma(-2\imath\beta)}\,W_{1,\imath\beta}(z),
\end{split}
\end{align}
where $\text{(a)}$ is due to the fact that $\mathcal{M}_{a,b}(z)\triangleq M_{a,b}(z)/\Gamma(1+2b)$ which we mentioned previously, for $\text{(b)}$ we used the recurrence $\Gamma(1+z)=z\Gamma(z)$ rewritten in the form $\Gamma(1\pm 2\imath\beta)=\pm2\imath\beta\Gamma(\pm2\imath\beta)$, and $\text{(c)}$ is an implication of~\eqref{eq:Whittaker-fncs-identity-v2}.

Next, from the identity
\begin{align*}
\Gamma(\imath x)\,\Gamma(-\imath x)
&=
\dfrac{\pi}{x\sinh(\pi x)},\;\; x\in\mathbb{R},
\end{align*}
given, e.g., by~\cite[Formula~6.1.29,~p.~256]{Abramowitz+Stegun:Handbook1964} or by~\cite[Formula~8.332.1,~p.~896]{Gradshteyn+Ryzhik:Book2007}, we obtain
\begin{align}\label{eq:Gamma-fn-id1}
\Gamma(2\imath\beta)\,\Gamma(-2\imath\beta)
&=
\dfrac{\pi}{2\beta\sinh(2\pi\beta)},
\end{align}
which then goes over into~\eqref{eq:Branch-cut-int-step2} accompanied by the well-known fact that $\overline{\Gamma(z)}=\Gamma(\bar{z})$ given, e.g., by~\cite[Identity~6.1.23,~p.~256]{Abramowitz+Stegun:Handbook1964}, so that $|\Gamma(z)|^2\triangleq\Gamma(z)\,\overline{\Gamma(z)}=\Gamma(z)\,\Gamma(\bar{z})$, where $\bar{z}$ denotes the complex conjugate of $z\in\mathbb{C}$ and $|z|$ stands for the absolute value of $z\in\mathbb{C}$, to finally yield
\begin{align}\label{eq:Branch-cut-int-step3}
\begin{split}
G_{\lambda}&(x,y)\Bigg|_{\lambda=\tfrac{\mu^2}{8}(1+4\beta^2)e^{\imath\pi}}-G_{\lambda}(x,y)\Bigg|_{\lambda=\tfrac{\mu^2}{8}(1+4\beta^2)e^{-\imath\pi}}=\\
&\qquad
=
e^{-\tfrac{1}{\mu^2 x}}\,\dfrac{2}{\mu^2 x}\,e^{\tfrac{1}{\mu^2 y}}\,\dfrac{\mu^2 y}{2}\,\times\\
&\qquad\qquad\qquad
\times\left|\Gamma\left(\imath\beta-\frac{1}{2}\right)\right|^2\beta\sinh(2\pi\beta)\,\times\\
&\qquad\qquad\qquad\qquad\qquad
\times\,W_{1,\imath\beta}\left(\dfrac{2}{\mu^2\max\{x,y\}}\right)W_{1,\imath\beta}\left(\dfrac{2}{\mu^2\min\{x,y\}}\right)\\
&\qquad
=
e^{-\tfrac{1}{\mu^2 x}}\,\dfrac{2}{\mu^2 x}\,e^{\tfrac{1}{\mu^2 y}}\,\dfrac{\mu^2 y}{2}\,\times\\
&\qquad\qquad\qquad
\times\left|\Gamma\left(\imath\beta-\frac{1}{2}\right)\right|^2\beta\sinh(2\pi\beta)\,W_{1,\imath\beta}\left(\dfrac{2}{\mu^2 x}\right)W_{1,\imath\beta}\left(\dfrac{2}{\mu^2 y}\right),
\end{split}
\end{align}
where $x,y,\beta\ge0$.

We are now in a position to put all of the above together, namely, combine~\eqref{eq:Branch-cut-int-step3},~\eqref{eq:Branch-cut-int-step0},~\eqref{eq:Bromwich-int-Res0}, \eqref{eq:Bromwich-int-step1} and~\eqref{eq:Bromwich-int}, and write down the sought-after density, $p(x,t|r)$, in a closed form. Specifically, we obtain:
\begin{align}\label{eq:GSR-transition-pdf-formula}
\begin{split}
p(x,t|r=y)
&=
e^{-\tfrac{2}{\mu^2 x}}\,\dfrac{2}{\mu^2 x^2}\,\Biggl[1+e^{-\tfrac{\mu^2 t}{8}}\,e^{\tfrac{1}{\mu^2 x}}\,e^{\tfrac{1}{\mu^2 y}}\,\dfrac{\mu^2 x}{2}\,\frac{\mu^2 y}{2}\,\dfrac{1}{\pi^2}\times\\
&\quad
\times\int_0^\infty e^{-\tfrac{\mu^2 t}{2}\beta^2}\,\left|\Gamma\left(\imath\beta-\frac{1}{2}\right)\right|^2\beta\sinh(2\pi\beta)\,\times\\
&\qquad\qquad\qquad
\times\,W_{1,\imath\beta}\left(\dfrac{2}{\mu^2 x}\right)W_{1,\imath\beta}\left(\dfrac{2}{\mu^2 y}\right)d\beta\Biggr],\;x,y,t\ge0;
\end{split}
\end{align}
cf., e.g.,~\cite[Proposition~3.3,~p.~262]{Linetsky:MF2006},~\cite[Subsection~4.6.1,~pp.~276--279]{Linetsky:HandbookChapter2007}. We would like to reiterate that $W_{1,\imath\beta}(z)$ is purely real. We also observe that $p(x,t|r=y)/\mathfrak{m}(x)$, where $\mathfrak{m}(x)$ is the speed measure given by~\eqref{eq:speed+scale-answer}, is symmetric with respect to interchange of $x\ge0$ and $y\ge0$ for all $t\ge0$; that is, $p(x,t|r=y)/\mathfrak{m}(x)=p(y,t|r=x)/\mathfrak{m}(y)$, for all $x,y,t\ge0$. This is an expected property, and is a direct consequence of the aforementioned fact that $G_{\lambda}(x,y)/\mathfrak{m}(x)=G_{\lambda}(y,x)/\mathfrak{m}(y)$ for all $x,y,t\ge0$. Moreover, since $\rho(x)=\mathfrak{m}(x)$, where $\rho(x)\triangleq\lim_{t\to+\infty}p(x,t|r=y)$ is the stationary density given by~\eqref{eq:rho-answer}, one can conclude that $p(x,t|r=y)/\rho(x)$ is also symmetric with respect to interchange of $x\ge0$ and $y\ge0$, for all $t\ge0$.

Formula~\eqref{eq:GSR-transition-pdf-formula} is the main result of this paper. As complex as it may seem, the formula is {\em exact} and has a completely transparent structure: $p(x,t|r)$ is the sum of the stationary pdf $\rho(x)\triangleq\lim_{t\to+\infty}p(x,t|r)$ given by~\eqref{eq:rho-answer} and a term that dies off with time. Furthermore, despite the seemingly high complexity, formula~\eqref{eq:GSR-transition-pdf-formula} is amenable to numerical evaluation ``as is'' using {\em Mathematica}. The corresponding numerical study is offered in Section~\ref{sec:numerical-study} below.

As a basic ``sanity check'', let us now use formula~\eqref{eq:GSR-transition-pdf-formula} for $p(x,t|r)$ to show that the latter does, in fact, integrate to unity (the integration is with respect to $x$), i.e., satisfy the normalization constraint~\eqref{eq:normalization-constraint1}. To that end, from the identity
\begin{align*}
\int_0^\infty e^{-\tfrac{z}{2}}\, z^{c}\,W_{a,b}(z)\,\frac{dz}{z}
&=
\frac{\Gamma(c+1/2-b)\,\Gamma(c+1/2+b)}{\Gamma(c-a+1)},
\end{align*}
which holds for all complex $a,b$ and $c$ such that $\Re(c+1/2\pm b)>0$, as given, e.g., by~\cite[Identity~7.621.11,~p.~823]{Gradshteyn+Ryzhik:Book2007} or by~\cite[Identity~(3.6.14),~p.~53]{Slater:Book1960}, one may deduce that
\begin{align}\label{eq:W-identity3}
\int_{0}^{\infty} e^{-\tfrac{1}{\mu^2 x}}\,W_{1,\imath\beta}\left(\frac{2}{\mu^2 x}\right)\dfrac{dx}{x}
&=
\int_{0}^{\infty} e^{-\tfrac{u}{2}}\,W_{1,\imath\beta}(u)\,\dfrac{du}{u}
=
0
\end{align}
because $1/\Gamma(0)=0$. Therefore, by integrating both sides of~\eqref{eq:GSR-transition-pdf-formula} with respect to $x$ over $[0,\infty)$ and interchanging the order of integration, the contribution of the integral term in the right-hand side of~\eqref{eq:GSR-transition-pdf-formula} is easy to see to be zero. Hence, the normalization constraint~\eqref{eq:normalization-constraint1} does check out. Moreover, the same argument~\eqref{eq:W-identity3} can be also used to confirm that the stationary density $\rho(x)\triangleq\lim_{t\to+\infty}p(x,t|r)$ given by~\eqref{eq:rho-answer} solves the equation
\begin{align*}
\int_{0}^{\infty} p(x,t|r=y)\,\rho(y)\,dy
&=
\rho(x)
\;\;\text{for all}\;\; x\ge0,
\end{align*}
which can be used as a definition of the stationary distribution $\rho(x)$ alternative to the temporal limit definition $\rho(x)\triangleq\lim_{t\to+\infty}p(x,t|r)$.

To slightly lighten formula~\eqref{eq:GSR-transition-pdf-formula}, observe that since
\begin{align*}
\left|\Gamma\left(\imath x-\frac{1}{2}\right)\right|^2
&=
\frac{4\pi}{(1+4x^2)\cosh(\pi x)},
\end{align*}
as given, e.g., by~\cite[Identity~(51),~p.~769]{Becker:JMP2004}, and $\sinh(2x)=2\sinh(x)\cosh(x)$, we obtain that
\begin{align}\label{eq:Gamma-fn-id2}
\left|\Gamma\left(\imath\beta-\frac{1}{2}\right)\right|^2\,\beta\,\sinh(2\pi\beta)
&=
\frac{8\pi\beta\sinh(\pi\beta)}{1+4\beta^2},
\end{align}
and, therefore, formula~\eqref{eq:GSR-transition-pdf-formula} can be rewritten equivalently as:
\begin{align}\label{eq:GSR-transition-pdf-formula-v2}
\begin{split}
p(x,t|r=y)
&=
e^{-\tfrac{2}{\mu^2 x}}\,\frac{2}{\mu^2 x^2}\,\Biggl[1+e^{-\tfrac{\mu^2 t}{8}}\,e^{\tfrac{1}{\mu^2 y}}\,e^{\tfrac{1}{\mu^2 x}}\,\dfrac{\mu^2 y}{2}\,\dfrac{\mu^2 x}{2}\,\dfrac{8}{\pi}\,\times\\
&\quad\times\int_{0}^{\infty} e^{-\tfrac{\mu^2 t}{2}\beta^2}\,\dfrac{\beta\sinh(\pi\beta)}{1+4\beta^2}\times\\
&\qquad\qquad\qquad
\times\,W_{1,\imath\beta}\left(\dfrac{2}{\mu^2 x}\right)W_{1,\imath\beta}\left(\dfrac{2}{\mu^2 y}\right)d\beta\Biggr],\;\; x,y,t\ge0.
\end{split}
\end{align}

The formula is slightly easier to implement on a computer because the number of special functions involved in it is one fewer compared to that involved in the formula~\eqref{eq:GSR-transition-pdf-formula} that we obtained first. More importantly, as we shall demonstrate in the next section, representation~\eqref{eq:GSR-transition-pdf-formula-v2} of $p(x,t|r)$ is simpler to Laplace-transform (with respect to time) than representation~\eqref{eq:GSR-transition-pdf-formula}. Specifically, the aim of the next section is ``cross-check'' the results obtained in this section against similar results established previously elsewhere, and one of the obvious such ``cross-checks'' is to find $\mathfrak{L}\big\{p(x,t|r=y);t\to\lambda\big\}$ {\em directly} to show that the answer is indeed $G_{\lambda}(x,y)$ given by~\eqref{eq:Greens-fcn-answer}.

Let us now show that the obtained formula, namely formula~\eqref{eq:GSR-transition-pdf-formula}, for $p(x,t|r)$ can also be obtained from the expansion~\eqref{eq:pdf-spectral-expansion3} used by~\cite{Wong:SPMPE1964}. To that end, first note that on the two types of eigenfunctions $\psi_1(x,\lambda)$ and $\psi_2(x,\lambda)$ given by~\eqref{eq:eig-fn-answer-no-class}, only the first one fulfills~\eqref{eq:eig-fcn-int-to-zero}, which can be seen from~\eqref{eq:W-identity3}. More importantly, recall that, unlike the Green's function approach, formula~\eqref{eq:pdf-spectral-expansion3} requires the eigenfunctions to be normalized according to~\eqref{eq:eigenfun-orthonorm}. Therefore, $\psi_1(x,\lambda)$ has to be normalized according to~\eqref{eq:eigenfun-orthonorm}. To that end, we use the orthogonality property of the Whittaker $W$ function
\begin{align}\label{eq:whittaker-orthogonality}
\int_{0}^{\infty} W_{a,\imath b_1}(z)\,W_{a,\imath b_2}(z)\,\dfrac{dz}{z^2}
&=
\dfrac{2\pi\,\Gamma(2\imath b_1)\,\Gamma(-2\imath b_2)}{\Gamma(1/2-a+\imath b_1)\,\Gamma(1/2-a-\imath b_2)}\,\delta(b_1-b_2),
\end{align}
where $b_1,b_2>0$; cf.~\cite[Formula~III.23,~p.~125]{Grosche:AP1988} or~\cite[Formula~(1.2),~p.~740,~and~Formula~(3.18),~p.~743]{Szmytkowski+Bielski:ITSF2010}. Specifically, since
\begin{align*}
\int_{0}^{\infty}\rho(x)\,\psi(x,\lambda_{1})\,\psi(x,\lambda_{2})\,dx
&=
\int_{0}^{\infty}\dfrac{2}{\mu^2 x^2}\,e^{-\tfrac{2}{\mu^2 x}}\,\psi(x,\lambda_{1})\,\psi(x,\lambda_{2})\,dx\\
&=
\int_{0}^{\infty} e^{-u}\,\psi(u,\lambda_{1})\,\psi(u,\lambda_{2})\,du\;\;\text{with}\;\; u=\frac{2}{\mu^2 x},
\end{align*}
then using~\eqref{eq:whittaker-orthogonality} and~\eqref{eq:eig-fn-answer} with $\lambda_1=\lambda_{\beta_{1}}$ and $\lambda_2=\lambda_{\beta_{2}}$ selected from the branch cut parametrization~\eqref{eq:beta-def}, we readily obtain
\begin{align*}
\int_{0}^{\infty}\rho(x)\,\psi(x,\lambda_{\beta_1})\,\psi(x,\lambda_{\beta_2})\,dx
&=
\int_{0}^{\infty}W_{1,\imath\beta_1}(u)\,W_{1,\imath\beta_2}(u)\,\dfrac{du}{u^2}\\
&=
2\pi\dfrac{\Gamma(2\imath\beta_1)\,\Gamma(-2\imath\beta_2)}{\Gamma(-1/2+\imath\beta_1)\,\Gamma(-1/2-\imath\beta_2)}\,\delta(\beta_1-\beta_2),
\end{align*}
whence
\begin{align}\label{eq:eigfun-const-factor}
\|\psi(\cdot,\lambda=\lambda_{\beta})\|^2
&\triangleq
\dfrac{1}{2\pi}\,\left|\dfrac{\Gamma(\imath\beta-1/2)}{\Gamma(2\imath\beta)}\right|^2,
\end{align}
where we again used the well-known fact that $\overline{\Gamma(z)}=\Gamma(\bar{z})$ so that $|\Gamma(z)|^2\triangleq\Gamma(z)\,\overline{\Gamma(z)}=\Gamma(z)\,\Gamma(\bar{z})$. The obtained expression can be simplified further with the help of identity~\eqref{eq:Gamma-fn-id1}. Specifically, the latter plugged in to~\eqref{eq:eigfun-const-factor} gives
\begin{align*}
\|\psi(\cdot,\lambda=\lambda_{\beta})\|^2
&=
\frac{1}{\pi}\,\beta\sinh(2\pi\beta)\,\left|\Gamma\left(\imath\beta-\frac{1}{2}\right)\right|^2,
\end{align*}
and, as a result, we finally arrive at the formula for the eigenfunction
\begin{align}\label{eq:eigenfunction-formula}
\psi(x,\lambda_{\beta})
&=
e^{\tfrac{1}{\mu^2 x}}\,\frac{\mu^2 x}{2}\,\frac{1}{\pi}\,\left|\Gamma\left(\imath\beta-\frac{1}{2}\right)\right|\sqrt{\beta\sinh(2\pi\beta)}\,W_{1,\imath\beta}\left(\frac{2}{\mu^2 x}\right),
\;\; x\ge0,
\end{align}
where we reiterate that $\lambda_{\beta}$ and $\beta\ge 0$ are as in~\eqref{eq:beta-def} above. This is precisely the form obtained by~\cite{Wong:SPMPE1964} for the eigenfunctions. It is also worth reiterating that the symmetry $W_{a,b}(z)=W_{a,-b}(z)$ implies that $W_{1,\imath\beta}(x)$ is real-valued. Moreover, it also implies that $\varphi(x,\lambda_{\beta})$ is an even function of $\beta$.

Furthermore, if we use~\eqref{eq:Gamma-fn-id2} in~\eqref{eq:eigenfunction-formula}, then the following alternative representation of the eigenfunction $\psi(x,\lambda_{\beta})$ can be obtained:
\begin{align*}
\psi(x,\lambda_{\beta})
&=
e^{\tfrac{1}{\mu^2 x}}\,\frac{\mu^2 x}{2}\,\sqrt{\frac{8\beta\sinh(\pi\beta)}{\pi(1+4\beta^2)}}\,W_{1,\imath\beta}\left(\frac{2}{\mu^2 x}\right),
\;\; x\ge0,
\end{align*}
which, if plugged into the expansion~\eqref{eq:pdf-spectral-expansion3}, will yield formula~\eqref{eq:GSR-transition-pdf-formula-v2}.

To conclude this section, let us find $p(x,t|r=0)$ as the special case of formula~\eqref{eq:GSR-transition-pdf-formula} with $r=y\to0+$. The interest in $p(x,t|r=0)$ is obvious: it is the $\Pr_\infty$-transition pdf of the original SR statistics (with no headstart). Specifically, the asymptotic property $W_{1,b}(x)\sim x\, e^{-\tfrac{x}{2}}$ valid for $x\to+\infty$ implies that for $y\to0+$ formula~\eqref{eq:GSR-transition-pdf-formula} reduces to
\begin{align}\label{eq:GSR-transition-pdf-formula-zero-headstart}
\begin{split}
p(x,t|r=0)
&=
e^{-\tfrac{2}{\mu^2 x}}\frac{2}{\mu^2 x^2}\Biggl[1+
e^{-\tfrac{\mu^2 t}{8}}\,e^{\tfrac{1}{\mu^2 x}}\,\frac{\mu^2 x}{2}\,\frac{1}{\pi^2}\,\times\\
&\qquad\times\int_{0}^{\infty}e^{-\tfrac{\mu^2 t}{2}\beta^2}\left|\Gamma\left(\imath\beta-\frac{1}{2}\right)\right|^2\times\\
&\qquad\qquad\qquad\qquad
\times\beta\sinh(2\pi\beta)\,W_{1,\imath\beta}\left(\frac{2}{\mu^2 x}\right)d\beta\Biggr],\;\; x,t\ge0.
\end{split}
\end{align}
Up to the notation, this formula---though obtained in a completely different way and under an entirely different motivation---is a special case of~\cite[Formula~(5.5),~p.~643]{Monthus+Comtet:JPhIF1994} and of~\cite[Formulae~(72)\&(73),~p.~264]{Comtet+etal:JAP1998}.

By exactly the same argument as the one we used to rewrite formula~\eqref{eq:GSR-transition-pdf-formula} equivalently in the form given by~\eqref{eq:GSR-transition-pdf-formula-v2}, formula~\eqref{eq:GSR-transition-pdf-formula-zero-headstart} can too be equivalently rewritten as:
\begin{align}\label{eq:GSR-transition-pdf-formula-zero-headstart-v2}
\begin{split}
p(x,t|r=0)
&=
e^{-\tfrac{2}{\mu^2 x}}\,\dfrac{2}{\mu^2 x^2}\,\Biggl[1+e^{-\tfrac{\mu^2 t}{8}}\,e^{\tfrac{1}{\mu^2 x}}\,\dfrac{\mu^2 x}{2}\,\dfrac{8}{\pi}\times\\
&\qquad
\times\int_{0}^{\infty} e^{-\tfrac{\mu^2 t}{2}\beta^2}\dfrac{\beta\sinh(\pi\beta)}{1+4\beta^2}\, W_{1,\imath\beta}\left(\dfrac{2}{\mu^2 x}\right)d\beta\Biggr],\;\; x,t\ge0.
\end{split}
\end{align}
Alternatively, one can also pass~\eqref{eq:GSR-transition-pdf-formula-v2} to the limit as $r=y\to0+$ and use the fact that $W_{1,b}(x)\sim x\, e^{-\tfrac{x}{2}}$ as $x\to+\infty$ to make the transition from~\eqref{eq:GSR-transition-pdf-formula-zero-headstart} to~\eqref{eq:GSR-transition-pdf-formula-zero-headstart-v2}. Representation~\eqref{eq:GSR-transition-pdf-formula-zero-headstart-v2} will prove useful in the next section to demonstrate consistency with analogous results established earlier.


\section{Discussion of the Main Result}
\label{sec:discussion}

As we mentioned earlier, the main result of this work has been previously obtained in other disciplines. However, since not only the context and motivation for the result in these other disciplines were different, but also the methodology used to get it, the result was obtained in a form different from that we obtained. This brings about the obvious question of consistency of the result across its various versions available in the literature. It is to address this question that is the aim of this section.

Let us first show {\em directly} that the temporal Laplace transform of the density $p(x,t|r=y)$, which we found explicitly in Section~\ref{sec:main-result}, does, in fact, coincide with the Green's function $G_{\lambda}(x,y)$, which we also obtained in a closed form in Section~\ref{sec:main-result}. That is, let us pretend that we don't know that $\mathfrak{L}\big\{p(x,t|r=y);t\to\lambda\big\}$ is equal to $G_{\lambda}(x,y)$ given by~\eqref{eq:Greens-fcn-answer}, and attempt to arrive at this conclusion by finding $\mathfrak{L}\big\{p(x,t|r=y);t\to\lambda\big\}$ {\em directly} from our formula for $p(x,t|r=y)$. This sort of a basic ``sanity check''---if ``passed''---would reinforce the validity of formula~\eqref{eq:Greens-fcn-answer}, which was derived in a different way.

Specifically, from formula~\eqref{eq:GSR-transition-pdf-formula-v2} for $p(x,t|r=y)$, we have
\begin{align*}
\begin{split}
\mathfrak{L}\big\{p(x,t|r=y)&;t\to\lambda\big\}
=
e^{-\tfrac{2}{\mu^2 x}}\,\frac{2}{\mu^2 x^2}\,\Biggl[\mathfrak{L}\big\{\indicator{t\ge0};t\to\lambda\big\}+\\
&
+e^{\tfrac{1}{\mu^2 x}}\,e^{\tfrac{1}{\mu^2 y}}\,\dfrac{\mu^2 x}{2}\,\dfrac{\mu^2 y}{2}\,\dfrac{8}{\pi}\int_{0}^{\infty}\mathfrak{L}\left\{e^{-(1+4\beta^2)\tfrac{\mu^2 t}{8}}\indicator{t\ge0};t\to\lambda\right\}\times\\
&\qquad
\times\frac{\beta\sinh(\pi\beta)}{1+4\beta^2}\, W_{1,\imath\beta}\left(\dfrac{2}{\mu^2 x}\right)W_{1,\imath\beta}\left(\dfrac{2}{\mu^2 y}\right)d\beta\Biggr],
\end{split}
\end{align*}
and since $\mathfrak{L}\big\{\indicator{t\ge0};t\to\lambda\big\}=1/\lambda$, provided $\Re(\lambda)>0$, and
\begin{align*}
\mathfrak{L}\left\{e^{-(1+4\beta^2)\tfrac{\mu^2 t}{8}}\indicator{t\ge0};t\to\lambda\right\}
&=
\left[\lambda+\frac{\mu^2}{8}(1+4\beta^2)\right]^{-1},
\end{align*}
provided $\Re(\lambda)>-\mu^2(1+4\beta^2)/8$, we further obtain
\begin{align}\label{eq:direct-Laplace-transform}
\begin{split}
\mathfrak{L}\big\{p(x,t|r=y)&;t\to\lambda\big\}
=
e^{-\tfrac{2}{\mu^2 x}}\,\frac{2}{\mu^2 x^2}\,\Biggl[\frac{8}{\mu^2}\dfrac{1}{4\alpha^2-1}+\\
&
+e^{\tfrac{1}{\mu^2 x}}\,e^{\tfrac{1}{\mu^2 y}}\,\dfrac{\mu^2 x}{2}\,\dfrac{\mu^2 y}{2}\,\dfrac{8}{\pi}\,\dfrac{2}{\mu^2}\int_{0}^{\infty}\dfrac{\beta\,\sinh(\pi\beta)}{(1+4\beta^2)(\alpha^2+\beta^2)}\times\\
&\qquad\qquad\times\, W_{1,\imath\beta}\left(\dfrac{2}{\mu^2 x}\right)W_{1,\imath\beta}\left(\dfrac{2}{\mu^2 y}\right)d\beta\Biggr],
\end{split}
\end{align}
where we used the relation $\lambda=(4\alpha^2-1)\mu^2/8$ to switch from $\lambda$ to $\alpha(\lambda)$. To complete the derivation of $\mathfrak{L}\big\{p(x,t|r=y);t\to\lambda\big\}$, we turn to the work of~\cite{Becker:JMP2004}, who used complex plane contour integration and Cauchy's Residue Theorem to {\em directly} establish the identity
\begin{align}\label{eq:Becker-identity-v1}
\begin{split}
\int_0^\infty&\frac{\beta\sinh(\pi\beta)}{(1+4\beta^2)(s+\beta^2)}\,W_{1,\imath \beta}(x_1)\,W_{1,\imath \beta}(x_2)\,d\beta
=\\
&=
\frac{\pi}{8}\frac{\Gamma(-1/2+\sqrt{s})}{\Gamma(1+2\sqrt{s})}\,W_{1,\sqrt{s}}(\max\{x_1,x_2\})\,M_{1,\sqrt{s}}(\min\{x_1,x_2\})\\
&\quad\quad
-\frac{\pi}{2}e^{-\tfrac{1}{2}(x_1+x_2)}\frac{x_1x_2}{4s-1},\;\text{where}\; x_1,x_2>0\;\text{and}\;s\in\mathbb{C}\backslash\{1/4\}\cup\mathbb{R}^{-};
\end{split}
\end{align}
cf.~\cite[Formula~(52),~p.~769]{Becker:JMP2004}. The desired conclusion that $\mathcal{L}\big\{p(x,t|r=y);t\to\lambda\big\}$ and $G_{\lambda}(x,y)$ given by~\eqref{eq:Greens-fcn-answer} do, in fact, match can now be easily reached from~\eqref{eq:direct-Laplace-transform}, the fact that $\mathcal{M}_{a,b}(z)\triangleq M_{a,b}(z)/\Gamma(1+2b)$, and~\eqref{eq:Becker-identity-v1} with $s\triangleq\alpha^2$, $x_1\triangleq2/(\mu^2 x)\,(>0)$, and $x_2\triangleq2/(\mu^2 y)\,(>0)$.

Let us now use the formula~\eqref{eq:Greens-fcn-answer} for the Green's function $G_{\lambda}(x,y)$ to obtain yet another expression for the density $p(x,t|r=y)$, different from those given by formulae~\eqref{eq:GSR-transition-pdf-formula}--\eqref{eq:GSR-transition-pdf-formula-v2}. Specifically, the aim is to have $p(x,t|r=y)$ expressed in the form consistent with~\cite[Theorem~3.3,~p.~88]{DeSchepper+Goovaerts:IME1999}, which gives the transition pdf for the solution, $(S_{t})_{t\ge0}$, of the SDE $dS_t=(a+b S_t)\,dt+c\,S_t\,dB_t$. According to~\cite{DeSchepper+Goovaerts:IME1999}, this SDE is a continuous time GARCH-like econometric model for the interest rate term structure. To obtain the transition pdf for the process $S_t$,~\cite{DeSchepper+Goovaerts:IME1999} used Feynman--Kac functional integration. Since the SDE that the GSR statistic satisfies is a special case of the SDE for $S_t$, it makes sense to verify whether the pdf $p(x,t|r)$ of the GSR diffusion coincides with the special case of the pdf of the process $S_t$.

The key to establish consistency between our answer for $p(x,t|r)$ and the parallel result obtained (differently) in~\cite[Theorem~3.3,~p.~88]{DeSchepper+Goovaerts:IME1999} is to use the identity
\begin{align}\label{eq:Buchholz-identity}
\begin{split}
\Gamma(1/2+&b-a)\,W_{a,b}(c\,x_1)\,\mathcal{M}_{a,b}(c\,x_2)
=c\,\sqrt{x_1 x_2}\,\times\\
&\quad
\times\int_{0}^{\infty} e^{-\tfrac{c}{2}(x_1+x_2)\cosh(v)}I_{2b}(c \sinh(v) \sqrt{x_1 x_2})\coth^{2a}\left(\dfrac{v}{2}\right)dv,\\
&\qquad\text{provided $\Re(b)>0$, $x_1>x_2$, and $\Re(1/2+b-a)>0$};
\end{split}
\end{align}
cf.~\cite[Identity~(5c),~p.~86]{Buchholz:Book1969}\footnote{This identity is also cited by~\cite{Gradshteyn+Ryzhik:Book2007} under the number 6.669.4 on p.~716. However,~\cite[Identity~6.669.4,~p.~716]{Gradshteyn+Ryzhik:Book2007} has a misprint. Specifically, Gradshteyn and Ryzhik's~\citeyearpar{Gradshteyn+Ryzhik:Book2007} version of~\cite[Identity~(5c),~p.~86]{Buchholz:Book1969} uses the fact that $\mathcal{M}_{a,b}(z)\triangleq M_{a,b}(z)/\Gamma(1+2b)$ and reads
\begin{align*}
\begin{split}
\frac{\Gamma(1/2+b-a)}{\Gamma(1+2b)}&\,W_{a,b}(c\,x_1)\,M_{a,b}(c\,x_2)
=\\
&=
c\sqrt{x_1x_2}\int_{0}^{\infty} e^{-\tfrac{1}{2}(x_1x_2)\cosh(v)}I_{2b}(c\sinh(v)\sqrt{x_1x_2})\coth^{2a}\left(\dfrac{v}{2}\right)dv,\\
&\qquad\text{provided $\Re(b)>0$, $x_1>x_2$, and $\Re(1/2+b-a)>0$},
\end{split}
\end{align*}
i.e., the exponential factor under the integral contains the product of $x_1$ and $x_2$ while it is supposed to be the sum, as is correctly given by the original~\cite[Identity~(5c),~p.~86]{Buchholz:Book1969}.}. Here and onward $I_{a}(z)$ denotes the modified Bessel function of the first kind. For a brief introduction to modified Bessel functions, see, e.g.,~\cite[Section~9.6]{Abramowitz+Stegun:Handbook1964}. As stated, this identity is not applicable to our case, because we have $a=1$ and $b=\imath\beta$, $\beta>0$, and, therefore, $\Re(b)=0$ and $\Re(1/2+b-a)=-1/2<0$. However, as pointed out by~\cite[Part~I,~Chapter~3,~p.~34]{Hostler:PhD-Thesis1963} and then also by~\cite{Hostler:JMPh1964}, the condition $\Re(b)>0$ is a misprint: the actual condition should read $\Re(c)>0$; for a detailed proof, see~\cite[Appendix~I,~pp.~246--264]{Hostler:PhD-Thesis1963}. Since in our case we have $c=2/\mu^2$, the condition $\Re(c)>0$ is clearly fulfilled.

Moreover, the condition $\Re(1/2+b-a)>0$, as also pointed out by~\cite[Part~I,~Chapter~3,~p.~34]{Hostler:PhD-Thesis1963} and then by~\cite{Hostler:JMPh1964}, can be lifted. Specifically, the reason this condition is in place is to make sure the integral in the right-hand side of~\eqref{eq:Buchholz-identity} converges at the lower limit (i.e., at zero). However, we can cast the integral as a contour one in the complex plane, and choose the path of integration to go not through the origin but around it. By the standard analytic continuation argument, this would take care of the possible singularity of the function under the integral at zero. As a result, the restriction $\Re(1/2+b-a)>0$ can be dropped altogether. For a complete proof of this generalization of~\cite[Identity~(5c),~p.~86]{Buchholz:Book1969}, see~\cite[Appendix~I,~pp.~246--264]{Hostler:PhD-Thesis1963}.

Formally, Hostler's~\citeyearpar{Hostler:PhD-Thesis1963,Hostler:JMPh1964} more general version of Buchholz's~\citeyearpar{Buchholz:Book1969} identity~\eqref{eq:Buchholz-identity} is as follows:
\begin{align}\label{eq:Hostler-identity}
\begin{split}
\Gamma(1/2&+b-a)\,W_{a,b}(c\,x_1)\,\mathcal{M}_{a,b}(c\,x_2)
=\\
&=
c\,\sqrt{x_1 x_2}\int_{0}^{\infty} e^{-\tfrac{c}{2}(x_1+x_2)\cosh(v)}I_{2b}(c \sinh(v) \sqrt{x_1 x_2})\coth^{2a}\left(\dfrac{v}{2}\right)dv,\\
&\quad\text{provided $\Re(b)>0$, $\Re(c)>0$, and $x_1>x_2>0$};
\end{split}
\end{align}
cf.~\cite[Appendix~I,~pp.~246--264]{Hostler:PhD-Thesis1963}. This identity, applied to our case, yields
\begin{align*}
\Gamma\Big(\alpha&-\dfrac{1}{2}\Big)\,W_{1,\alpha}\left(\dfrac{2}{\mu^2\min\{x,y\}}\right)\mathcal{M}_{1,\alpha}\left(\dfrac{2}{\mu^2\max\{x,y\}}\right)=\\
&=
\dfrac{2}{\mu^2\sqrt{xy}}\int_{0}^{\infty} e^{-\tfrac{1}{\mu^2}\left(\tfrac{1}{x}+\tfrac{1}{y}\right)\cosh(v)}I_{2\alpha}\left(\sinh(v)\dfrac{2}{\mu^2\sqrt{x y}}\right)\coth^{2}\left(\dfrac{v}{2}\right)dv,
\end{align*}
and, therefore, the Green's function $G_{\lambda}(x,y)$ given by~\eqref{eq:Greens-fcn-answer} can be expressed alternatively as
\begin{align*}
\begin{split}
G_{\lambda}(x,y)
&=
e^{-\tfrac{1}{\mu^2}\left(\tfrac{1}{x}-\tfrac{1}{y}\right)}\,\dfrac{2}{\mu^2 x}\,\sqrt{\dfrac{y}{x}}\,\times\\
&
\times\int_{0}^{\infty}e^{-\tfrac{1}{\mu^2}\left(\tfrac{1}{x}+\tfrac{1}{y}\right)\cosh(v)}I_{2\alpha}\left(\sinh(v)\dfrac{2}{\mu^2\sqrt{x y}}\right)\coth^{2}\left(\dfrac{v}{2}\right)dv,\;x,y\ge0,
\end{split}
\end{align*}
where $\alpha=\alpha(\lambda)$ is as in~\eqref{eq:alpha-def}.

From this representation of $G_{\lambda}(x,y)$ it is clear that to compute $\mathfrak{L}^{-1}\big\{G_{\lambda}(x,y);\lambda\to t\big\}$ to recover $p(x,t|r=y)$ we are to find $\mathfrak{L}^{-1}\left\{I_{2\alpha(\lambda)}(z);\lambda\to t\right\}$; recall again that $\alpha\triangleq\alpha(\lambda)$ is given by~\eqref{eq:alpha-def}. To that end, the inverse Laplace transform of $I_{2\alpha(\lambda)}(z)$ with respect to $\lambda$ can be found from the results of~\cite{DeSchepper+etal:IME1994}, who employed complex-plane contour integration together with the Residue Theorem and showed that
\begin{align}\label{eq:DeSchepper-inverse-Laplace-transform}
\begin{split}
\mathfrak{L}^{-1}\left\{I_{2\sqrt{2s}}(z);s\to t\right\}
&=
\dfrac{z}{2\pi\sqrt{2\pi t}}\,e^{\tfrac{2}{t}\pi^2}\,\times\\
&\qquad\times\int_{0}^{\infty} e^{-\tfrac{2}{t}v^2-z\cosh(v)}\sinh(v)\sin\left(\dfrac{4\pi v}{t}\right)dv;
\end{split}
\end{align}
cf.~\cite[Section~2,~pp.~34--35]{DeSchepper+etal:IME1994}, although for an alternative proof of this inversion formula, see also, e.g.,~\cite[pp.~86--87]{Yor:ZW1980}.

To apply~\eqref{eq:DeSchepper-inverse-Laplace-transform} to find $\mathfrak{L}^{-1}\left\{I_{2\alpha(\lambda)}(z);\lambda\to t\right\}$ we reparameterize $\lambda$ to $s$ through $\lambda(s)=\mu^2(s-1/8)$ so as to bring $I_{2\alpha(\lambda)}(z)$ to the form $I_{2\sqrt{2s}}(z)$.  As a result, we obtain
\begin{align*}
\begin{split}
\mathfrak{L}^{-1}\left\{I_{2\alpha(\lambda)}(z);\lambda\to t\right\}
&=
\dfrac{\mu z}{2\pi\sqrt{2\pi t}}\,e^{-\tfrac{\mu^2 t}{8}}\,e^{\tfrac{2\pi^2}{\mu^2 t}}\,\times\\
&\qquad\qquad\times\int_{0}^{\infty} e^{-\tfrac{2v^2}{\mu^2 t}-z\cosh(v)}\sinh(v)\sin\left(\dfrac{4\pi v}{\mu^2t}\right)dv,
\end{split}
\end{align*}
whence
\begin{align*}
\begin{split}
p(x,&t|r=y)
=
\dfrac{\sqrt{2}}{\pi\sqrt{\pi}}\,\dfrac{1}{\mu^3 x^2 \sqrt{t}}\,e^{-\tfrac{1}{\mu^2}\left(\tfrac{1}{x}-\tfrac{1}{y}\right)}\,e^{-\tfrac{\mu^2 t}{8}}\,e^{\tfrac{2\pi^2}{\mu^2 t}}\times\\
&
\qquad\qquad
\times\int_{0}^{\infty}e^{-\tfrac{1}{\mu^2}\left(\tfrac{1}{x}+\tfrac{1}{y}\right)\cosh(v)}\times\\
&
\times\Biggl\{\int_{0}^{\infty}e^{-\tfrac{2u^2}{\mu^2 t}-\tfrac{2}{\mu^2\sqrt{xy}}\sinh(v)\cosh(u)}\sinh(u)\sin\left(\dfrac{4\pi u}{\mu^2t}\right)du\Biggr\}\coth^{2}\left(\dfrac{v}{2}\right)dv.
\end{split}
\end{align*}

If we now reverse the order of integration and introduce $\omega\triangleq\omega(v)$ such that
\begin{align}\label{eq:Grosche-u-sub1}
\sinh(2\omega/\mu^2)
=
\dfrac{1}{\sinh(v)}
\;\;
\text{so that}
\;\;
\cosh(v)=\coth(2\omega/\mu^2),
\end{align}
then
\begin{align}\label{eq:Grosche-u-sub2}
dv
&=
-\dfrac{2d\omega}{\mu^2\sinh(2\omega/\mu^2)}
\;\;\text{and}\;\;
\coth^{2}\left(\dfrac{v}{2}\right)dv
=
-\dfrac{2e^{2\omega}d\omega}{\mu^2\sinh(2\omega/\mu^2)},
\end{align}
and, consequently, the above double-integral formula for $p(x,t|r=y)$ takes the form
\begin{align}\label{eq:DeSchepper-pdf-formula}
\begin{split}
p(x,t|&r=y)=
\dfrac{2\sqrt{2}}{\pi\sqrt{\pi}}\,\dfrac{1}{\mu^5 x^2 \sqrt{t}}\,e^{-\tfrac{1}{\mu^2}\left(\tfrac{1}{x}-\tfrac{1}{y}\right)}\,e^{-\tfrac{\mu^2 t}{8}}\,e^{\tfrac{2\pi^2}{\mu^2 t}}\times\\
&
\quad
\times\int_{0}^{\infty}\int_{0}^{\infty}\exp\left\{-\frac{1}{\mu^2}\left[\left(\frac{1}{x}+\frac{1}{y}\right)\coth(2\omega/\mu^2)-4\omega\right]\right\}\times\\
&
\quad
\times
\exp\left\{-\frac{2}{\mu^2}\left(\frac{u^2}{t}+\frac{\cosh(u)}{\sinh(2\omega/\mu^2)\sqrt{xy}}\right)\right\}\sinh(u)\sin\left(\dfrac{4\pi u}{\mu^2 t}\right)du\,d\omega,
\end{split}
\end{align}
where $x,y,t\ge0$. Up to the notation, this is in accord with~\cite[Theorem~3.3,~p.~88]{DeSchepper+Goovaerts:IME1999}, although the latter has a misprint: in our notation,~\cite[Theorem~3.3,~p.~88]{DeSchepper+Goovaerts:IME1999} states that the overall denominator of the second exponent under the double integral is a multiple of $\mu$ while it should be a multiple of $\mu^2$.

Let us now take a closer look at the work of~\cite{Peskir:Shiryaev2006}, which, as was mentioned in Section~\ref{sec:literature-overview}, is of high relevance to the present work. Specifically, recall that Peskir's~\citeyearpar{Peskir:Shiryaev2006} goal was to solve a problem even more general than that considered in this work. However, in spite of the original and rather elegant solution strategy, Peskir~\citeyearpar{Peskir:Shiryaev2006} ended up solving the ``more general problem'' only partially, and obtained the explicit solution only in a few special cases. Since one of these special cases is also a special case of our problem, it makes sense to find out whether the answer we obtained in this work (for that special case) coincides with the corresponding result of Peskir~\citeyearpar{Peskir:Shiryaev2006}. We shall now demonstrate that the two answers are, in fact, equivalent. We shall stick to our notation, and, for simplicity, recite Peskir's~\citeyearpar{Peskir:Shiryaev2006} result using our notation.

Suppose $\mu=1$ and $r=0$. This is the special case that Peskir~\citeyearpar{Peskir:Shiryaev2006} gave a closed-form answer for. However, contrary to this work's,  Peskir's~\citeyearpar{Peskir:Shiryaev2006} answer was not for the pdf $p(x,t|r)$ but for the respective cumulative distribution function (cdf), i.e., for the quantity
\begin{align*}
P(x,t|r=y)
&\triangleq
\int_{0}^{x} p(u,t|r=y)\,du,
\;\;
x,y,t\ge0,
\end{align*}
assuming $\mu=1$ and $r=0$. We would like to remark parenthetically that Peskir's~\citeyearpar{Peskir:Shiryaev2006} idea to deal with the cdf $P(x,t|r)$ instead of the pdf $p(x,t|r)$ was a clever one, for, unlike the former, the latter is a generalized function due to the initial condition $\lim_{t\to0+}p(x,t|r)=\delta(x-r)$.

Specifically,~\cite[Formula~(3.29),~p.~542]{Peskir:Shiryaev2006} together with~\cite[Formula~(3.36),~p.~543]{Peskir:Shiryaev2006} yield
\begin{align}\label{eq:cdf-Laplace-image-Peskir}
\mathfrak{L}\Big\{P(x,t|r=0);t\to\lambda\Big\}
&=
\frac{1}{\lambda}\left[1-\sqrt{\dfrac{2\pi}{x}}\, e^{-\tfrac{1}{x}}I_{\alpha(\lambda)}\left(\dfrac{1}{x}\right)\right],\; x\ge0,
\end{align}
where we reiterate that $I_{a}(z)$ denotes the modified Bessel function of the first kind.

Let us first apply the final value theorem for Laplace transforms to Peskir's~\citeyearpar{Peskir:Shiryaev2006} formula~\eqref{eq:cdf-Laplace-image-Peskir} to see whether this would yield the cdf of the stationary density $\rho(x)\triangleq\lim_{t\to+\infty}p(x,t|r=y)$ given explicitly by~\eqref{eq:rho-answer}. To that end, since from~\eqref{eq:alpha-def} we have $\lim_{\lambda\to0}\alpha(\lambda)=1/2$, it follows at once that
\begin{align*}
\lim_{\lambda\to0}\lambda\,\mathfrak{L}\Big\{P(x,t|r=0);t\to\lambda\Big\}
&=
1+\sqrt{\dfrac{2\pi}{x}}\, e^{-\tfrac{1}{x}}I_{\tfrac{1}{2}}\left(\dfrac{1}{x}\right),
\end{align*}
and in light of the identity
\begin{align*}
I_{\tfrac{1}{2}}(z)
&=
\sqrt{\dfrac{2}{\pi z}}\sinh(z),
\end{align*}
we further obtain
\begin{align*}
\lim_{\lambda\to0}\lambda\,\mathfrak{L}\Big\{P(x,t|r=0);t\to\lambda\Big\}
&=
1-2e^{-\tfrac{1}{x}}\sinh\left(\dfrac{1}{x}\right)
=
e^{-\tfrac{2}{x}},
\end{align*}
which is exactly the cdf of the stationary pdf $\rho(x)$ given by~\eqref{eq:rho-answer} with $\mu=1$. This necessarily validates Peskir's~\citeyearpar{Peskir:Shiryaev2006} formula~\eqref{eq:cdf-Laplace-image-Peskir}.

Next, let us show that $\mathfrak{L}\big\{P(x,t|r=0);t\to\lambda\big\}$ given by~\eqref{eq:cdf-Laplace-image-Peskir}, which is an expression obtained by~\cite{Peskir:Shiryaev2006}, can alternatively be found from formula~\eqref{eq:Greens-fcn-answer} for $G_{\lambda}(x,0)=\mathcal{L}\{p(x,t|r=0);t\to\lambda\}$ obtained by us in Section~\ref{sec:main-result} above. To show this, let first observe that for $\mu=1$ our formula~\eqref{eq:Greens-fcn-answer} reduces to
\begin{align*}
\mathcal{L}\{p(x,t|r=0);t\to\lambda\}
&=
G_{\lambda}(x,0)
=
2\,\dfrac{e^{-\tfrac{1}{x}}}{x}\,\dfrac{\Gamma(\alpha-1/2)}{\Gamma(1+2\alpha)}\,M_{1,\alpha}\left(\dfrac{2}{x}\right),\;x,t\ge0,
\end{align*}
where we used the fact that $\mathcal{M}_{a,b}(z)\triangleq M_{a,b}(z)/\Gamma(1+2b)$. As a result, we obtain
\begin{align}\label{eq:cdf-Laplace-transform-step1}
\begin{split}
\mathfrak{L}\Big\{P(x,t|r=0);t\to\lambda\Big\}
&=
\int_0^{x}\mathcal{L}\{p(v,t|r=0);t\to\lambda\}\,dv\\
&=
\int_{0}^{x}G_{\lambda}(v,0)\,dv\\
&=
2\,\dfrac{\Gamma(\alpha-1/2)}{\Gamma(1+2\alpha)}\int_{2/x}^{+\infty}e^{-\tfrac{u}{2}}M_{1,\alpha}(u)\,\dfrac{du}{u}.
\end{split}
\end{align}

By the indefinite integral formula
\begin{align*}
\int x^{a-2}\,e^{-\tfrac{x}{2}}\,M_{a,b}(x)\,dx
&=
\frac{1}{a+b-1/2}\,x^{a-1}\,e^{-\tfrac{x}{2}}\,M_{a-1,b}(x),
\end{align*}
given, e.g., by~\cite[Identity~1.13.1.6,~p.~39]{Prudnikov+etal:Book1990}, we have
\begin{align*}
\int e^{-\tfrac{u}{2}}\,M_{1,\alpha}(u)\,\dfrac{du}{u}
&=
\frac{1}{\alpha+1/2}\,e^{-\tfrac{u}{2}}\,M_{0,\alpha}(u),
\end{align*}
whence, in view of the asymptotic property $M_{0,b}(x)\sim e^{x/2}\,\Gamma(1+2b)\,/\,\Gamma(b-1/2)$ valid as $x\to+\infty$, from~\eqref{eq:cdf-Laplace-transform-step1} we therefore obtain
\begin{align}\label{eq:cdf-Laplace-transform-step2}
\begin{split}
\mathfrak{L}\Big\{P(x,t|r=0);t\to\lambda\Big\}
&=
\dfrac{2\,\Gamma(\alpha-1/2)}{(\alpha+1/2)\Gamma(\alpha+1/2)}-\\
&\qquad\qquad
-\dfrac{2\,\Gamma(\alpha-1/2)}{(\alpha+1/2)\Gamma(1+2\alpha)}\,e^{-\tfrac{1}{x}}\,{M}_{0,\alpha}\left(\dfrac{2}{x}\right),
\end{split}
\end{align}
and let us next simplify the right-hand side one term at a time.

Let us begin with the first term. To that end, from the identity $\Gamma(\alpha+1/2)=(\alpha-1/2)\,\Gamma(\alpha-1/2)$, which is a consequence of the recurrence $\Gamma(z+1)=z\,\Gamma(z)$, for the first term in the right-hand side of~\eqref{eq:cdf-Laplace-transform-step2} we obtain
\begin{align}\label{eq:cdf-Laplace-transform-step3}
\dfrac{2\,\Gamma(\alpha-1/2)}{(\alpha+1/2)\,\Gamma(\alpha+1/2)}
&=
\dfrac{2}{\alpha^2-1/4}
=
\dfrac{1}{\lambda},
\end{align}
since $\alpha^2-1/4=2\lambda$, as is easy to see from~\eqref{eq:alpha-def}.

The second term in the right-hand side of~\eqref{eq:cdf-Laplace-transform-step2} can be simplified by virtue of, e.g.,~\cite[Identity~9.235.1,~p.~1027]{Gradshteyn+Ryzhik:Book2007}, as per which ${M}_{0,b}(2z)=2^{2b}\Gamma(1+b)\,\sqrt{2z}\,I_{b}(z)$, and therefore
\begin{align}\label{eq:cdf-Laplace-transform-step4}
\dfrac{2\,\Gamma(\alpha-1/2)}{(\alpha+1/2)\,\Gamma(1+2\alpha)}\,{M}_{0,\alpha}\left(\dfrac{2}{x}\right)
&=
\dfrac{2^{2\alpha-2}\,\Gamma(\alpha-1/2)\,\Gamma(\alpha)}{\lambda\,\Gamma(2\alpha-1)}\sqrt{\dfrac{2}{x}}\,I_{\alpha}\left(\dfrac{1}{x}\right),
\end{align}
because $\Gamma(\alpha+1)=\alpha\,\Gamma(\alpha)$ and $\Gamma(2\alpha+1)=(2\alpha)\,\Gamma(2\alpha)=(2\alpha)\,(2\alpha-1)\,\Gamma(2\alpha-1)$---as can be deduced from the recurrence $\Gamma(z+1)=z\,\Gamma(z)$ applied twice---and also $\alpha^2-1/4=2\lambda$.

At this point note that by the duplication formula for the Gamma function given, e.g., by~\cite[Formula~6.1.18,~p.~256]{Abramowitz+Stegun:Handbook1964}, which states that $\sqrt{2\,\pi}\,\Gamma(2z)=2^{2z-1/2}\,\Gamma(z)\,\Gamma(z+1/2)$, we have $\sqrt{2\,\pi}\,\Gamma(2\alpha-1)=2^{2\alpha-3/2}\Gamma(\alpha-1/2)\,\Gamma(\alpha)$, and, therefore, from~\eqref{eq:cdf-Laplace-transform-step2},~\eqref{eq:cdf-Laplace-transform-step3} and~\eqref{eq:cdf-Laplace-transform-step4} combined, we can finally conclude that the formula~\eqref{eq:Greens-fcn-answer-no-headstart} that we obtained in Section~\ref{sec:main-result} above for $G_{\lambda}(x,0)=\mathfrak{L}\big\{p(x,t|r=0);t\to\lambda\big\}$ {\em does} lead to precisely the expression~\eqref{eq:cdf-Laplace-image-Peskir} for $\mathfrak{L}\big\{P(x,t|r=0);t\to\lambda\big\}$ obtained by~\cite{Peskir:Shiryaev2006} through a different approach.

We now show that direct integration with respect to $x$ of $p(x,t|r=0)$ given by formula~\eqref{eq:GSR-transition-pdf-formula-zero-headstart-v2} but tailored to the case when $\mu=1$ does yield the cdf $P(x,t|r=0)$ of the form consistent with that obtained by~\cite{Peskir:Shiryaev2006}, i.e.,
\begin{align}\label{eq:Peskir-cdf-formula}
\begin{split}
P(x,t|r=0)
&=
1-
\dfrac{1}{\pi x^{3/2}}\,e^{-\tfrac{1}{x}}\int_{0}^{t} e^{-\tfrac{s}{4}+\tfrac{\pi^2}{2s}}\times\\
&
\quad
\times
\Biggl\{\int_{0}^{\infty} e^{-\tfrac{v^2}{2s}-\frac{1}{x}\cosh(v)}\,\sinh(v)\sin\left(\dfrac{\pi v}{s}\right)dv\Biggr\}\,ds,\;\; x,t\ge0;
\end{split}
\end{align}
cf.~\cite[Formula~(3.41),~p.~544]{Peskir:Shiryaev2006}. We would like to reiterate that in addition to the assumption $r=0$ this formula also requires $\mu=1$. The formula suggests that $P(x,t|r=0)$ is the superposition of a term that captures the initial condition $\lim_{t\to0+}P(x,t|r=0)=1$, valid for all $x\ge0$, and a term that depends on $t$. We also recall that $\lim_{t\to+\infty}P(x,t|r=0)=e^{-2/x}$ for all $x\ge0$, as we established above from the result~\eqref{eq:cdf-Laplace-image-Peskir} obtained by~\cite{Peskir:Shiryaev2006}.

Since for $\mu=1$ our formula~\eqref{eq:GSR-transition-pdf-formula-zero-headstart-v2} gives
\begin{align*}
\begin{split}
p(x,t|r=0)
&=
e^{-\tfrac{2}{x}}\,\dfrac{2}{x^2}\,\Biggl[1+e^{-\tfrac{t}{8}}\,e^{\tfrac{1}{x}}\,\dfrac{x}{2}\,\dfrac{8}{\pi}\times\\
&\qquad
\times\int_{0}^{\infty} e^{-\tfrac{t}{2}\beta^2}\dfrac{\beta\sinh(\pi\beta)}{1+4\beta^2}\, W_{1,\imath\beta}\left(\dfrac{2}{x}\right)d\beta\Biggr],
\;\; x,t\ge0,
\end{split}
\end{align*}
we have
\begin{align}\label{eq:cdf-step1}
\begin{split}
P(x,t|r=0)
&\triangleq
\int_{0}^{x}
p(v,t|r=0)\,dv\\
&=
e^{-\tfrac{2}{x}}
+
\dfrac{1}{\pi}\int_{t}^{\infty}\int_{0}^{\infty}e^{-(1+4\beta^2)\tfrac{s}{8}}\beta\sinh(\pi\beta)\times\\
&\qquad\qquad\qquad
\times\Biggl\{\int_{2/x}^{\infty}e^{-\tfrac{u}{2}}\,W_{1,\imath\beta}(u)\,\dfrac{du}{u}\Biggr\}\,d\beta\,ds,
\;\;
x,t\ge0,
\end{split}
\end{align}
where in addition to swapping the order of integration and one obvious change of variables we also made use the identity
\begin{align*}
\dfrac{8}{1+4\beta^2}e^{-(1+4\beta^2)\tfrac{t}{8}}
&=
\int_{t}^{\infty}e^{-(1+4\beta^2)\tfrac{s}{8}}\,ds.
\end{align*}

The inner $du$-integral in the right-hand side of~\eqref{eq:cdf-step1} can be computed using~\cite[Identity~1.13.2.6,~p.~40]{Prudnikov+etal:Book1990}, according to which
\begin{align*}
\int x^{a-2}\,e^{-\tfrac{x}{2}}\,W_{a,b}(x)\,dx
&=
-x^{a-1}\,e^{-\tfrac{x}{2}}\,W_{a-1,b}(x),
\end{align*}
so that
\begin{align}\label{eq:int1}
\int_{2/x}^{\infty}e^{-\tfrac{u}{2}}\,W_{1,\imath\beta}(u)\,\dfrac{du}{u}
&=
e^{-\tfrac{1}{x}}\,W_{0,\imath\beta}\left(\dfrac{2}{x}\right)
=
e^{-\tfrac{1}{x}}\dfrac{\sqrt{2}}{x^{3/2}\sqrt{\pi}}\,K_{\imath\beta}\left(\dfrac{1}{x}\right),
\end{align}
because $W_{0,b}(x)\sim e^{-\tfrac{x}{2}}$ as $x\to+\infty$, and
\begin{align*}
W_{0,b}(2z)
&=
\sqrt{\frac{2z}{\pi}}K_{b}(z),
\end{align*}
which is given, e.g., by~\cite[Identity~9.235.2,~p.~1027]{Gradshteyn+Ryzhik:Book2007}. Here and throughout the sequel, $K_{b}(z)$ denotes the modified Bessel function of the second kind, also known as the MacDonald function; see, e.g.,~\cite[Section~9.6]{Abramowitz+Stegun:Handbook1964}. By inserting~\eqref{eq:int1} back into the right-hand side of~\eqref{eq:cdf-step1} we obtain
\begin{align*}
\begin{split}
P(x,t|r=0)
&=
e^{-\tfrac{2}{x}}
+
e^{-\tfrac{1}{x}}\dfrac{\sqrt{2}}{(\pi x)^{3/2}}\,\times\\
&\qquad
\times\int_{t}^{\infty}\Biggl\{\int_{0}^{\infty}e^{-(1+4\beta^2)\tfrac{s}{8}}\beta\sinh(\pi\beta)\,K_{\imath\beta}\left(\dfrac{1}{x}\right)d\beta\Biggr\}\,ds,
\;\;
x,t\ge0.
\end{split}
\end{align*}

The inner $d\beta$-integral can be found from the identity
\begin{align}\label{eq:K-identity-DeSchepper}
\begin{split}
\int_{0}^{\infty} e^{-\tfrac{\mu^2 t}{2}\beta^2}&\beta\sinh(2\pi\beta)\,K_{2\imath\beta}(z)\,d\beta
=\\
&=
e^{\tfrac{2\pi^2}{\mu^2 t}}\,\frac{z}{4}\sqrt{\frac{2\pi}{\mu^2 t}}\int_{0}^{\infty} e^{-\tfrac{2v^2}{\mu^2 t}-z\cosh(v)}\sinh(v)\sin\left(\dfrac{4\pi v}{\mu^2 t}\right)dv,
\end{split}
\end{align}
which was effectively established, e.g., by~\cite[p.~35]{DeSchepper+etal:IME1994}, using integration by parts along with properties of the MacDonald function.

We finally arrive at the formula
\begin{align*}
\begin{split}
P(x,t|r=0)
&=
e^{-\tfrac{2}{x}}+
\dfrac{1}{\pi x^{3/2}}\,e^{-\tfrac{1}{x}}\int_{t}^{\infty} e^{-\tfrac{s}{4}+\tfrac{\pi^2}{2s}}\times\\
&
\qquad\qquad
\times
\Biggl\{\int_{0}^{\infty} e^{-\tfrac{v^2}{2s}-\tfrac{1}{x}\cosh(v)}\,\sinh(v)\sin\left(\dfrac{\pi v}{s}\right)dv\Biggr\}\,ds,
\;\; x,t\ge0,
\end{split}
\end{align*}
which indicates that the cdf $P(x,t|r=0)$ is a combination of the stationary cdf $e^{-2/x}$ and a term that vanishes as $t\to+\infty$. This is dual to the structure of Peskir's~\citeyearpar{Peskir:Shiryaev2006} formula~\eqref{eq:Peskir-cdf-formula}. To see that the two formulae coincide, observe that either one leads to exactly the same temporal differential $\partial_{t} P(x,t|r=0)$. Therefore, to arrive at Peskir's~\citeyearpar{Peskir:Shiryaev2006} formula~\eqref{eq:Peskir-cdf-formula}, it is enough to integrate $\partial_{s} P(x,s|r=0)$ with respect to $s$ from $0$ to $t$ and use the initial condition $\lim_{t\to0+}P(x,t|r=0)=1$ valid for all $x\ge0$. However, if one integrates $\partial_{s} P(x,s|r=0)$ with respect to $s$ from $t$ to $+\infty$ instead, and uses the fact that $\lim_{t\to+\infty}P(x,t|r=0)=e^{-2/x}$, $x\ge0$, which we already established as well, then one will recover our formula.

To wrap up this section, we would like to mention the formula~\eqref{eq:DeSchepper-pdf-formula} for the pdf $p(x,t|r)$ quoted in our notation from~\cite[Theorem~3.3,~p.~88]{DeSchepper+etal:IME1994} can also be obtained {\em directly} from our formula~\eqref{eq:GSR-transition-pdf-formula}. To that end, the key is the identity
\begin{align}\label{eq:Buchholz-identity2}
\begin{split}
\Gamma(&1/2+b-a)\,\Gamma(1/2-b-a)\,W_{a,b}(c\,x_1)\,W_{a,b}(c\,x_2)
=\\
&=
2c\sqrt{x_1x_2}\int_{0}^{\infty}e^{-\tfrac{c}{2}(x_1+x_2)\cosh(v)}K_{2b}(c\sinh(v)\sqrt{x_1x_2})\coth^{2a}\left(\frac{v}{2}\right)dv,\\
&\qquad\text{provided $\Re(c[\sqrt{x_1}+\sqrt{x_2}\,]^2)>0$ and $\Re(1/2\pm b-a)>0$};
\end{split}
\end{align}
cf.~\cite[Identity~(4a),~p.~85]{Buchholz:Book1969}\footnote{This identity is also cited by~\cite{Gradshteyn+Ryzhik:Book2007} under the number 6.669.3 on p.~716. However,~\cite[Identity~6.669.3,~p.~716]{Gradshteyn+Ryzhik:Book2007} suffers from the same misprint as~\cite[Identity~6.669.4,~p.~716]{Gradshteyn+Ryzhik:Book2007} does. Specifically, Gradshteyn and Ryzhik's~\citeyearpar{Gradshteyn+Ryzhik:Book2007} version of~\cite[Identity~(4a),~p.~85]{Buchholz:Book1969} reads
\begin{align*}
\begin{split}
\Gamma(&1/2+b-a)\,\Gamma(1/2-b-a)\,W_{a,b}(c\,x_1)\,W_{a,b}(c\,x_2)
=\\
&=
c\sqrt{x_1x_2}\int_{0}^{\infty} e^{-\tfrac{1}{2}(x_1x_2)\cosh(v)}K_{2b}(c\sinh(v)\sqrt{x_1x_2})\coth^{2a}\left(\dfrac{v}{2}\right)dv,\\
&\qquad\text{provided $\Re(c[\sqrt{x_1}+\sqrt{x_2}\,]^2)>0$ and $\Re(1/2\pm b-a)>0$},
\end{split}
\end{align*}
i.e., the exponential factor under the integral contains the product of $x_1$ and $x_2$ while it is supposed to be the sum, as is correctly given by the original~\cite[Identity~(4a),~p.~85]{Buchholz:Book1969}.}. As was the case with~\cite[Identity~(4c),~p.~86]{Buchholz:Book1969} mentioned above as identity~\eqref{eq:Buchholz-identity}, the problem with this identity is also that the condition $\Re(1/2\pm b-a)>0$ required for it to hold is violated in our case, for we have $a=1$ and $b=\imath\beta$, $\beta>0$, so that $\Re(1/2\pm b-a)=-1/2<0$. However, since identity~\eqref{eq:Buchholz-identity2} derives from identity~\eqref{eq:Buchholz-identity} and the property of the modified Bessel functions
\begin{align*}
K_{a}(z)
&=
\frac{I_{-a}(z)-I_{a}(z)}{2\pi\sin(\pi z)},
\end{align*}
given, e.g., by~\cite[Identity~9.6.2,~p.~375]{Abramowitz+Stegun:Handbook1964}, the argument given by~\cite{Hostler:PhD-Thesis1963} to generalize the identity~\eqref{eq:Buchholz-identity} to the identity~\eqref{eq:Hostler-identity} can be used again to eliminate the constraint $\Re(1/2\pm b-a)>0$ in identity~\eqref{eq:Buchholz-identity2}. As a result, we obtain
\begin{align*}
\left|\Gamma\left(\imath\beta-\frac{1}{2}\right)\right|^2&\,W_{1,\imath\beta}\left(\dfrac{2}{\mu^2 x}\right)W_{1,\imath\beta}\left(\dfrac{2}{\mu^2 y}\right)=\\
&\qquad
=\dfrac{4}{\mu^2\sqrt{x\,y}}\int_{0}^{\infty} e^{-\tfrac{1}{\mu^2}\left(\tfrac{1}{x}+\tfrac{1}{y}\right)\cosh(v)}\times\\
&\qquad\qquad\qquad
\times K_{2\imath\beta}\left(\sinh(v)\dfrac{2}{\mu^2\sqrt{x y}}\right)\coth^{2}\left(\dfrac{v}{2}\right)dv,
\end{align*}
so that
\begin{align*}
\begin{split}
\int_{0}^{\infty} &e^{-\tfrac{\mu^2 t}{2}\beta^2}\,\left|\Gamma\left(\imath\beta-\frac{1}{2}\right)\right|^2\beta\sinh(2\pi\beta)\,W_{1,\imath\beta}\left(\dfrac{2}{\mu^2 x}\right)W_{1,\imath\beta}\left(\dfrac{2}{\mu^2 y}\right)d\beta=\\
&
\qquad\qquad
=\frac{4}{\mu^2\sqrt{x\,y}}\int_{0}^{\infty} e^{-\tfrac{1}{\mu^2}\left(\tfrac{1}{x}+\tfrac{1}{y}\right)\cosh(v)}\times\\
&
\times
\Biggl\{\int_{0}^{\infty} e^{-\tfrac{\mu^2 t}{2}\beta^2}\beta\,\sinh(2\pi\beta)\,K_{2\imath\beta}\left(\sinh(v)\dfrac{2}{\mu^2\sqrt{x y}}\right)d\beta\Biggr\}\times\\
&
\qquad\qquad\qquad\qquad\qquad\qquad
\times\coth^{2}\left(\dfrac{v}{2}\right)dv,
\end{split}
\end{align*}
which upon insertion into the formula~\eqref{eq:GSR-transition-pdf-formula} for the pdf $p(x,t|r)$ will give a new expression for the latter. To bring the new expression to the form given by formula~\eqref{eq:DeSchepper-pdf-formula}, it suffices to evaluate the inner $d\beta$ integral using~\eqref{eq:K-identity-DeSchepper} and then perform the change of variables~\eqref{eq:Grosche-u-sub1}--\eqref{eq:Grosche-u-sub2}.

\section{A Numerical Study}
\label{sec:numerical-study}

We now exploit the obtained expression for $p(x,t|r)$ numerically to get an idea as to the dynamics of the GSR statistic in the pre-change regime. To that end, as can be seen from formula~\eqref{eq:GSR-transition-pdf-formula}, the behavior of the GSR statistic depends on:\begin{inparaenum}[\itshape(a)]\item magnitude of the drift $\mu\neq0$, \item time $t\ge0$, and \item the headstart $R_0^r=r\ge0$\end{inparaenum}. We have implemented the formula for $p(x,t|r)$ in a {\em Mathematica} script as a function of all these parameters to study the effect of each one of the parameters on the distribution of the GSR statistic. We also note the symmetry with respect to $\mu$: the formula for $p(x,t|r)$ is indifferent to whether $\mu<0$ or $\mu>0$. Hence, without loss of generality, we shall consider only positive $\mu$'s. We also note that numerical evaluation of $p(x,t|r)$ is problematic for small values of $t$ because $\lim_{t\to0+}p(x,t|r)=\delta(x-r)$. However, as we shall see shortly, the Delta-function-shape of $p(x,t|r)$ at $t=0$ quickly ``dissolves'' across the $(x,r)$-plane as $t$ increases, and {\em Mathematica} successfully handles the integral in the formula for $p(x,t|r)$ when $t$ is as small as $0.1$.

We picked two values of $\mu$: $\mu=1$ and $\mu=1.5$. Our study is organized as follows. For each of the two $\mu$'s we vary $t$ over the set $\{0.1,0.5,1,2,5,10\}$ and plot $p(x,t|r)$ as a function of $x$ and $r$, each confined to the interval $[0,3]$. As we shall see below, restricting $x$ and $r$ to lie inside the square $[0,3]\times[0,3]$ is harmless in the sense that no important features of $p(x,t|r)$ will be left out.

We would like to structure our study chronologically, i.e., start with the case when $t=0.1$, and then gradually move forward in time, with specific time moments taken from the set $\{0.1,0.5,1,2,5,10\}$. For each value of $t$, the respective 3d-plots of $p(x,t|r)$ as a function of $x\in[0,3]$ and $r\in[0,3]$ for $\mu=1$ and for $\mu=1.5$ will be presented next to each other to conveniently demonstrate the effect of $\mu$.

\begin{figure}[!htb]
    \centering
    \subfloat[$\mu=1.0$.]{\label{fig:p_inf__t_0_1_m_1_0}
        \includegraphics[width=0.47\textwidth]{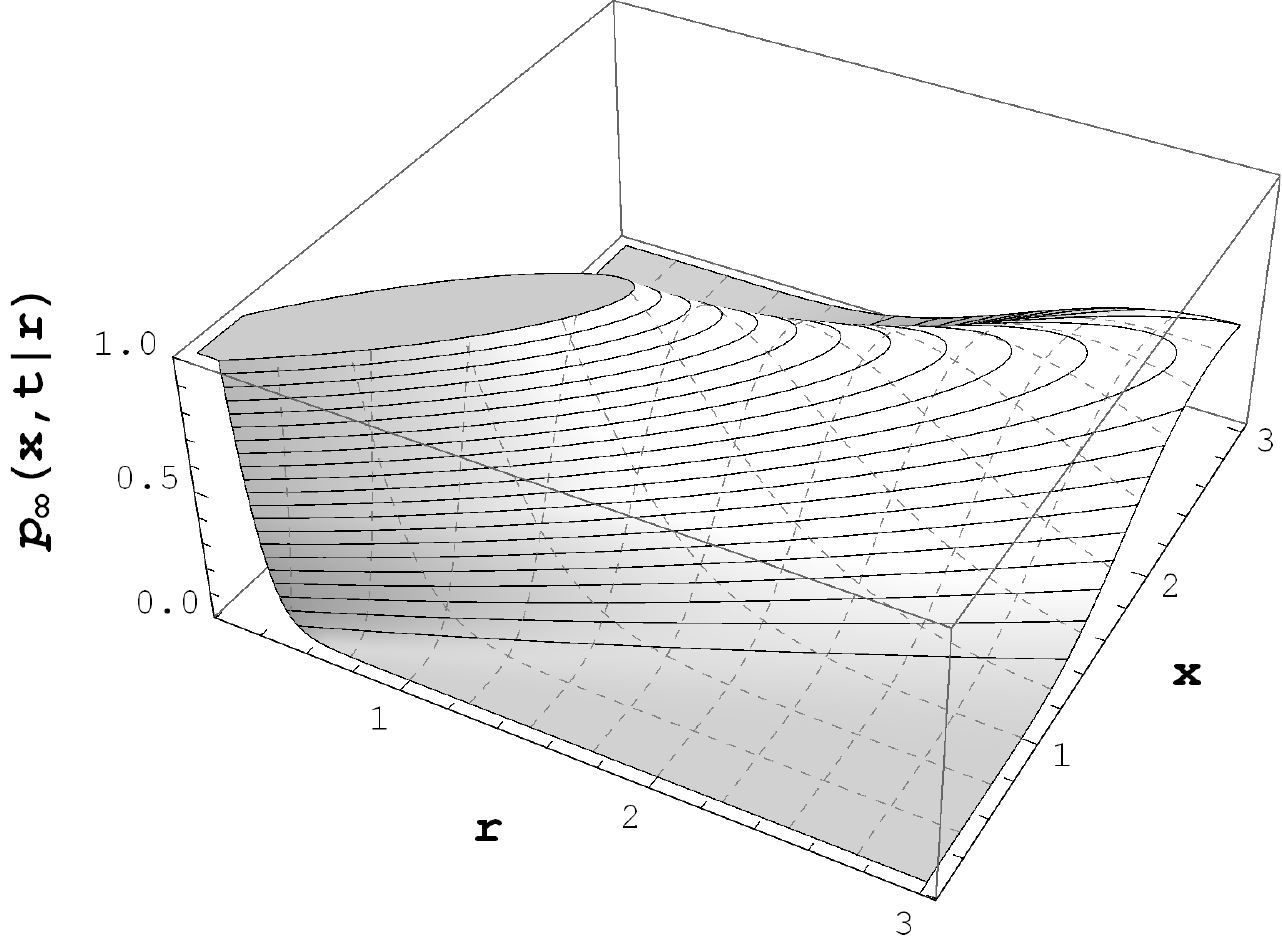}
    }
    \subfloat[$\mu=1.5$.]{\label{fig:p_inf__t_0_1_m_1_5}
        \includegraphics[width=0.47\textwidth]{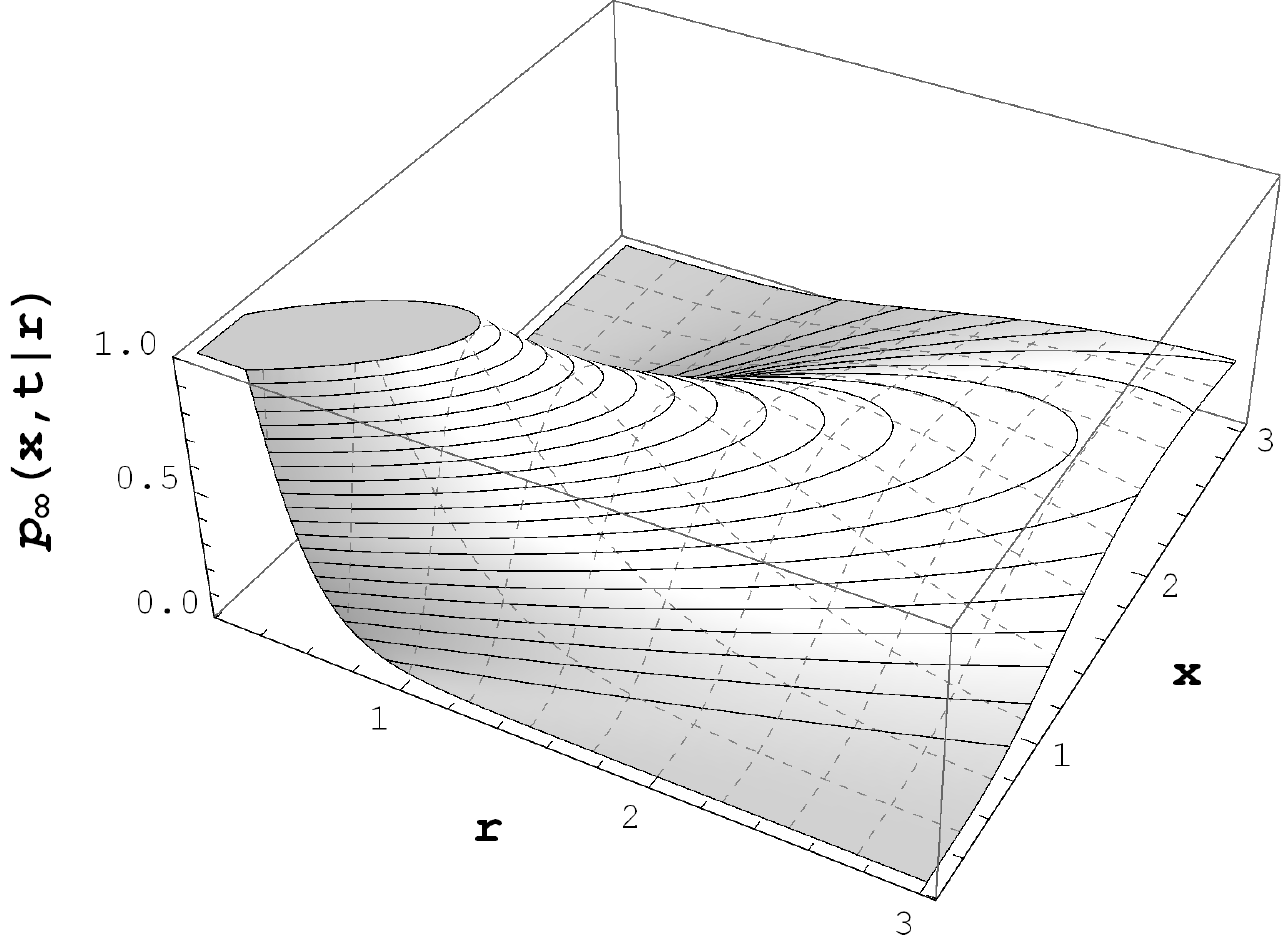}
    }
    \caption{$p_\infty(x,t|r)$ as a function of $x\in[0,3]$ and $r\in[0,3]$ for $\mu=1$ and $\mu=1.5$ when $t=0.1$.}
    \label{fig:p_inf__t_0_1}
\end{figure}

To get started with the study, suppose first that $t=0.1$. For this case, $p(x,t|r)$ as a function of $x\in[0,3]$ and $r\in[0,3]$ is shown in Figure~\ref{fig:p_inf__t_0_1}. Specifically, Figure~\ref{fig:p_inf__t_0_1_m_1_0} is for $\mu=1$ and Figure~\ref{fig:p_inf__t_0_1_m_1_5} is for $\mu=1.5$, and we note that, for convenience, both figures have identical scales along the respective axes. One immediate observation to be made from these figures is that in either one (i.e., no matter whether $\mu=1$ or $\mu=1.5$) the shape of $p(x,t|r)$ resembles a ridge which is formed along the identity line $x=r$. This is expected and is a direct consequence of the initial condition $\lim_{t\to0+}p(x,t|r)=\delta(x-r)$. Upon a closer look, one may notice that the ridge is uniformly ``taller'' and ``wider'' for $\mu=1$ than for $\mu=1.5$. To explain this phenomenon we recall that the GSR statistic, $(R_t^r)_{t\ge0}$, by definition
\begin{align*}
R_t^r
&\triangleq
r\LR_t+\int_0^{t}\frac{\LR_t}{\LR_s}\,dt,\;\text{where}\;\LR_t=\exp\left\{\mu X_t-\frac{\mu^2}{2} t\right\},
\end{align*}
is effectively the temporal average of the likelihood ratio $(\LR_t)_{t\ge0}$, with each time moment treated as equally likely to be the change-point. This makes $R_t^r$ a {\em linear} functional of $\LR_t$. As a result, it is reasonable to expect the dynamics of $R_t^r$ to be similar to that of $\LR_t$. This observation is crucial to understanding the behavior of $R_t^r$. When the pre- and post-change distributions of the observed process are close to one another, i.e., when the change is faint ($\abs{\mu}$ is close to zero), the distribution of $\LR_t$ under any probability measure is almost entirely concentrated around 1. Moreover, the fainter the change, the smaller the variance of $\LR_t$ (with respect to any probability measure). Hence, for small changes, it is reasonable to expect the variance (under any probability measure) of $R_t^r$ to be small as well, and, in fact, the less contrast the change, the smaller the variance of $R_t^r$. As a result, one can conclude that the fainter the change, the narrower and the higher the density $p(x,t|r)$. This is the reason why the ridge is uniformly taller and wider for $\mu=1$ than for $\mu=1.5$: the latter corresponds to a more contrast change.

An alterative (and perhaps easier) explanation for the above can be obtained by appealing directly to the SDE: $dR_t^r=dt+\mu R_t^r\,dB_t$, $t\ge0$, where $R_0^r=r\ge0$. It can be seen from this SDE that the only source of randomness there is the second term in the right-hand side, i.e., the diffusion term. It is this term that is ``responsible'' for the very fact that $(R_t^r)_{t\ge0}$ is random (as opposed to just being a deterministic function). When $\mu$ is small, so is the contribution of the diffusion term, and the behavior of the GSR statistic is ``more deterministic'', which accounts for all of the observations we made above from Figures~\ref{fig:p_inf__t_0_1_m_1_0} and~\ref{fig:p_inf__t_0_1_m_1_5}. One more curious observation that can be made from these figures is that the height of the ridge is greater around the origin, i.e., when $x$ is small. This can be explained by again appealing to the SDE that the GSR diffusion satisfies. It can be seen from this equation when $R_t^r$ is small, so is the contribution of the diffusion term, which translates to $R_t^r$ being ``less random'' in the sense that its variance is smaller. Hence, the greater height and the narrower width of $p(x,t|r)$ around the origin.

\begin{figure}[!htb]
    \centering
    \subfloat[$\mu=1.0$.]{\label{fig:p_inf__t_0_5_m_1_0}
        \includegraphics[width=0.47\textwidth]{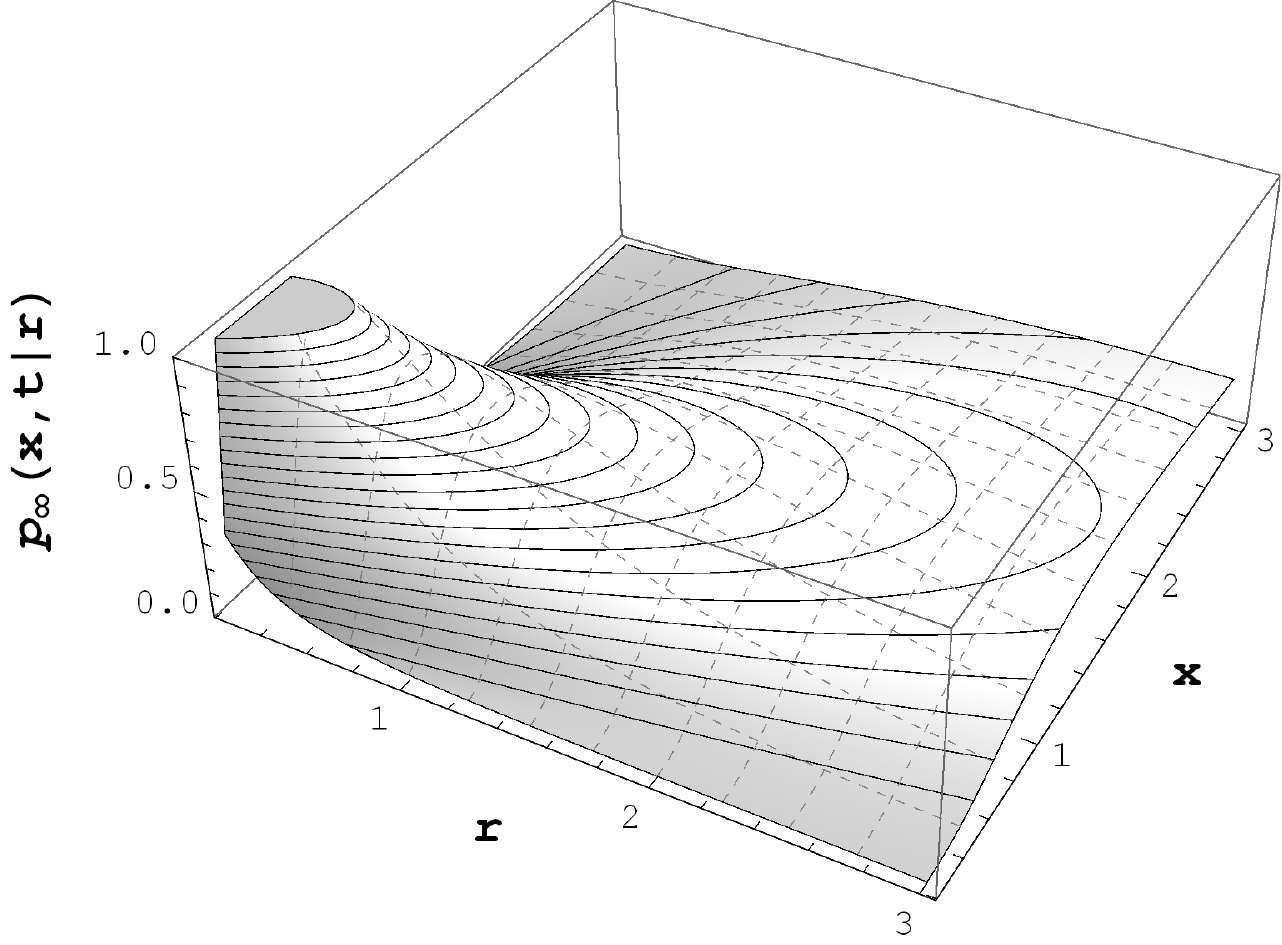}
    }
    \subfloat[$\mu=1.5$.]{\label{fig:p_inf__t_0_5_m_1_5}
        \includegraphics[width=0.47\textwidth]{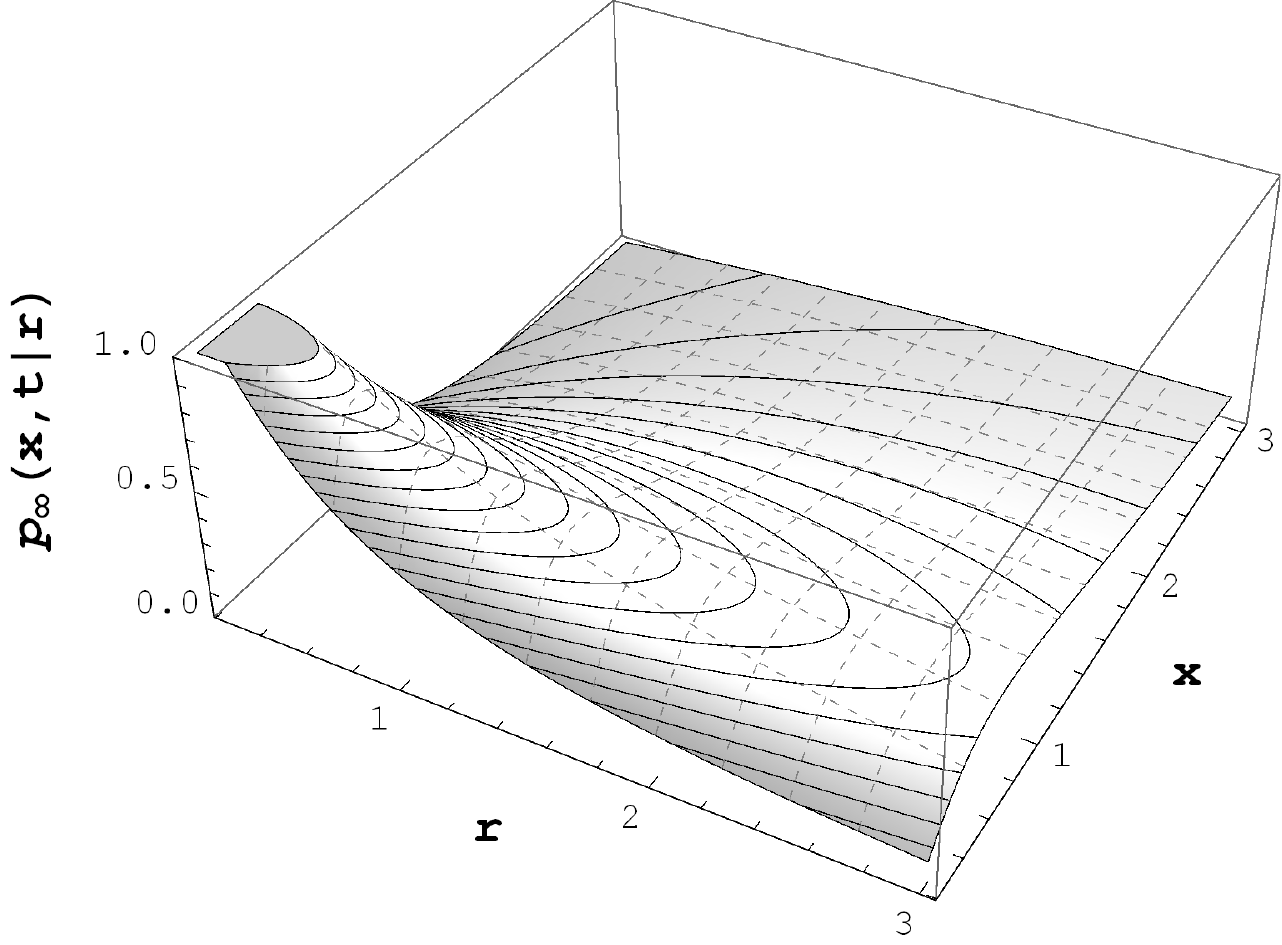}
    }
    \caption{$p_\infty(x,t|r)$ as a function of $x\in[0,3]$ and $r\in[0,3]$ for $\mu=1$ and $\mu=1.5$ when $t=0.5$.}
    \label{fig:p_inf__t_0_5}
\end{figure}

Moving forward in time, shown in Figure~\ref{fig:p_inf__t_0_5} is the density $p(x,t|r)$ as a function of $x\in[0,3]$ and $r\in[0,3]$ when $t=0.5$. Specifically, Figure~\ref{fig:p_inf__t_0_5_m_1_0} corresponds to $\mu=1$ and Figure~\ref{fig:p_inf__t_0_5_m_1_5} corresponds to $\mu=1.5$. The main difference between these figures and their counterparts for $t=0.1$ (see Figure~\ref{fig:p_inf__t_0_1}) is that the ridge-shape exhibited by $p(x,t|r)$ for $t=0.1$ is no longer there. In other words, the ridge has essentially dissolved across the $(x,r)$-plane. This is an indication that the effect of the Delta-initial condition has diminished significantly. However, the influence of the initial condition on the shape of $p(x,t|r)$ remains strong around the origin, where $p(x,t|r)$ is still Delta-shaped, as it was for $t=0.1$, although to a much lesser extent than for $t=0.1$. Moreover, judging by the level-contours superimposed on the 3d-surface representing $p(x,t|r)$, not only did the ridge lose much of its hight, it also curled toward the $r$-axis. This is an indication that the headstart began to matter less. That is, $p(x,t|r)$ started to get closer to the limiting stationary distribution $\rho(x)$.

\begin{figure}[!htb]
    \centering
    \subfloat[$\mu=1.0$.]{\label{fig:p_inf__t_1_0_m_1_0}
        \includegraphics[width=0.47\textwidth]{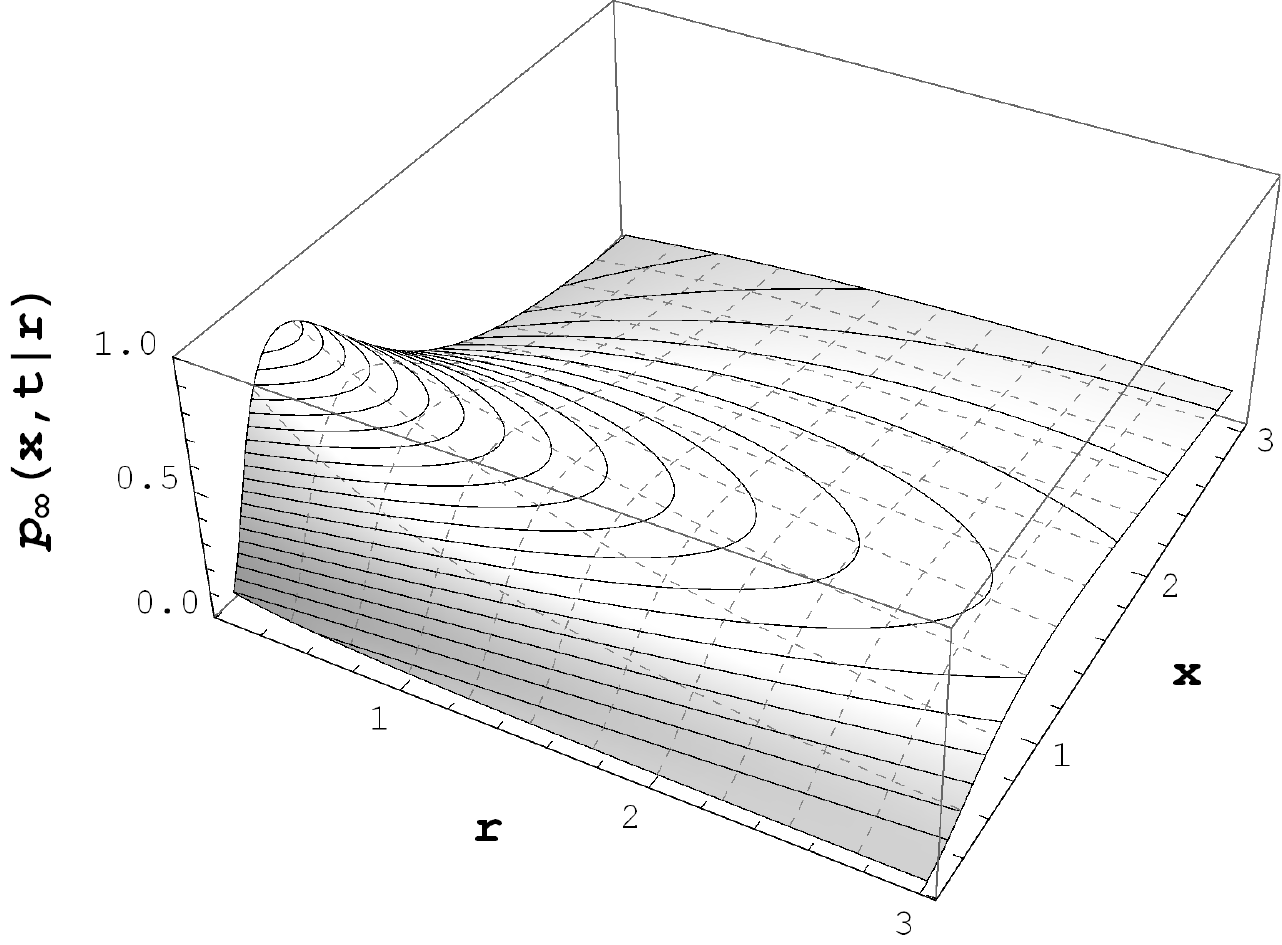}
    }
    \subfloat[$\mu=1.5$.]{\label{fig:p_inf__t_1_0_m_1_5}
        \includegraphics[width=0.47\textwidth]{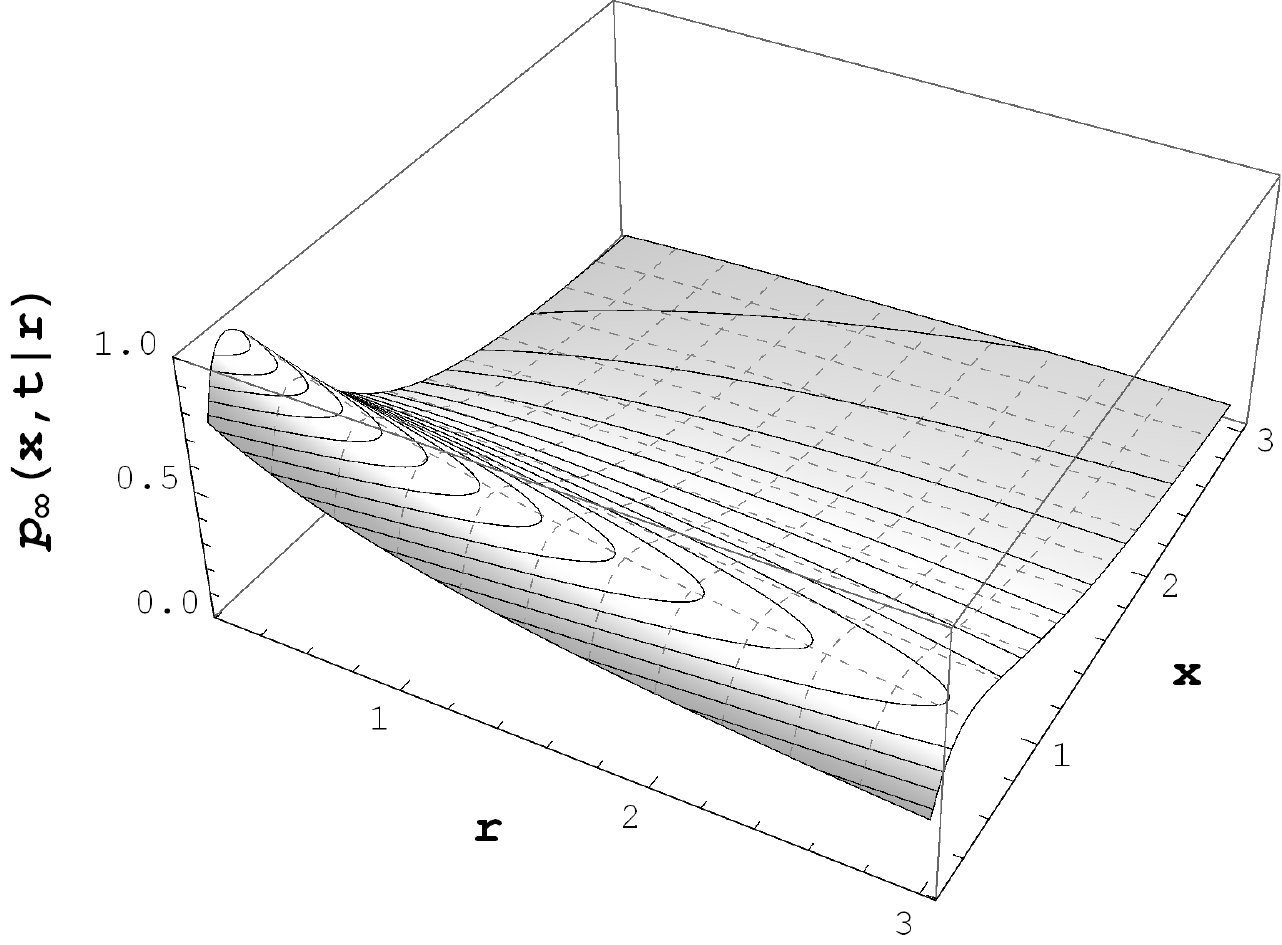}
    }
    \caption{$p_\infty(x,t|r)$ as a function of $x\in[0,3]$ and $r\in[0,3]$ for $\mu=1$ and $\mu=1.5$ when $t=1$.}
    \label{fig:p_inf__t_1_0}
\end{figure}

Figure~\ref{fig:p_inf__t_1_0} shows $p(x,t|r)$ as a function of $x\in[0,3]$ and $r\in[0,3]$ for $\mu=1$ (Figure~\ref{fig:p_inf__t_1_0_m_1_0}) and for $\mu=1.5$ (Figure~\ref{fig:p_inf__t_1_0_m_1_5}) when $t=1$. The difference, as compared to Figure~\ref{fig:p_inf__t_0_5} showing the respective results obtained for $t=0.5$, is that now the spike around the origin is less contrast, and moreover, it shows signs of prolonging across $r$ even more. This results in the shape of $p(x,t|r)$ resembling a ``dam'' parallel to the $r$-axis. This effect is an indication of the diminished role played by the headstart $r$ as compared to the case of $t=0.5$. The diminished dependence on $r$, in turn, is a sign of convergence of $p(x,t|r)$ to the stationary distribution $\rho(x)\triangleq\lim_{t\to\infty}p(x,t|r)$. Also, note that for $\mu=1$, the ``dam'' is slightly further away from the $r$-axis than for $\mu=1.5$. The reason is because the stationary distribution $\rho(x)$ has a maximum at $x_*=1/\mu^2$ which is smaller for higher values of $\mu$.

\begin{figure}[!htb]
    \centering
    \subfloat[$\mu=1.0$.]{
        \includegraphics[width=0.47\textwidth]{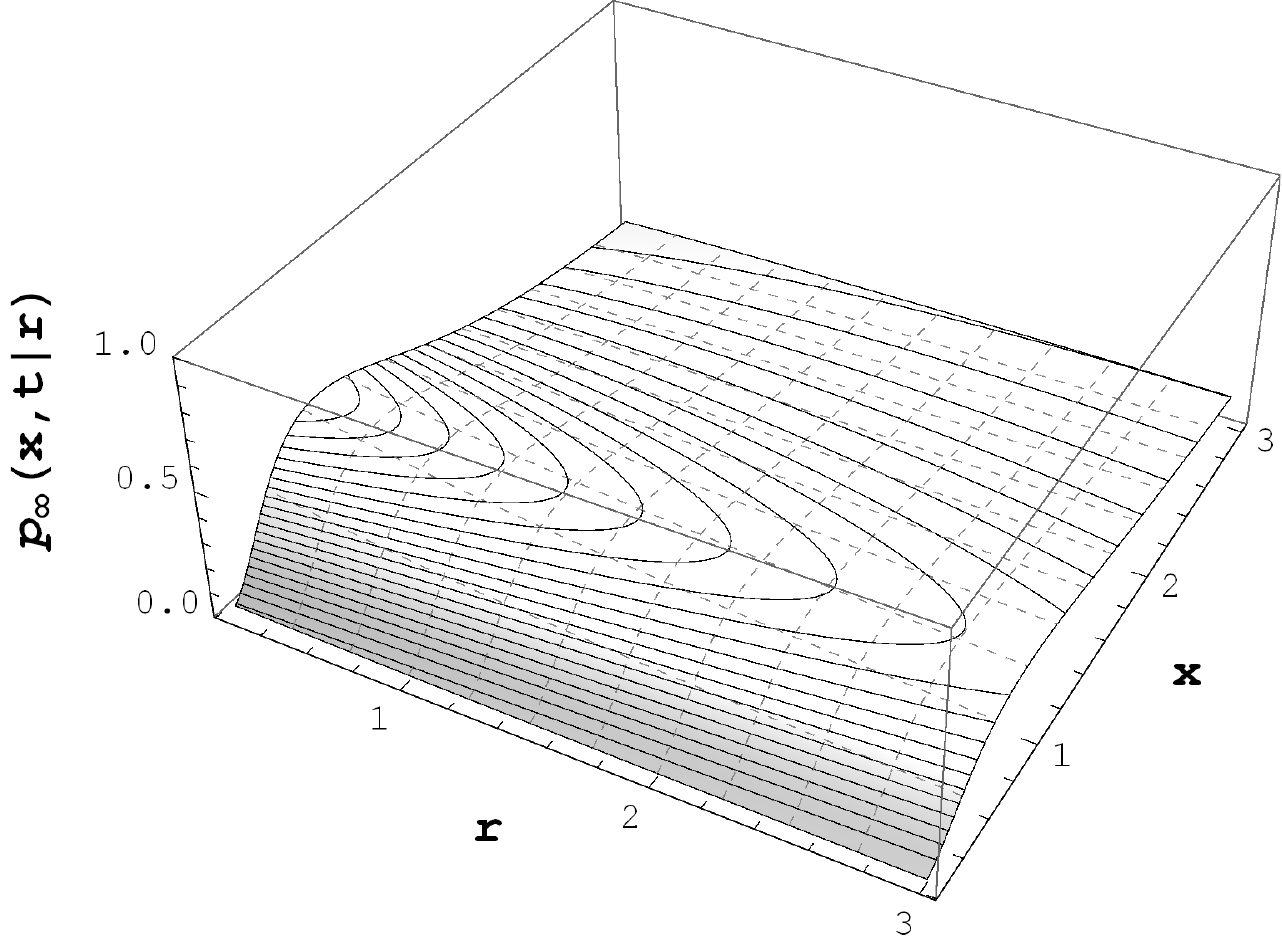}
    }
    \subfloat[$\mu=1.5$.]{
        \includegraphics[width=0.47\textwidth]{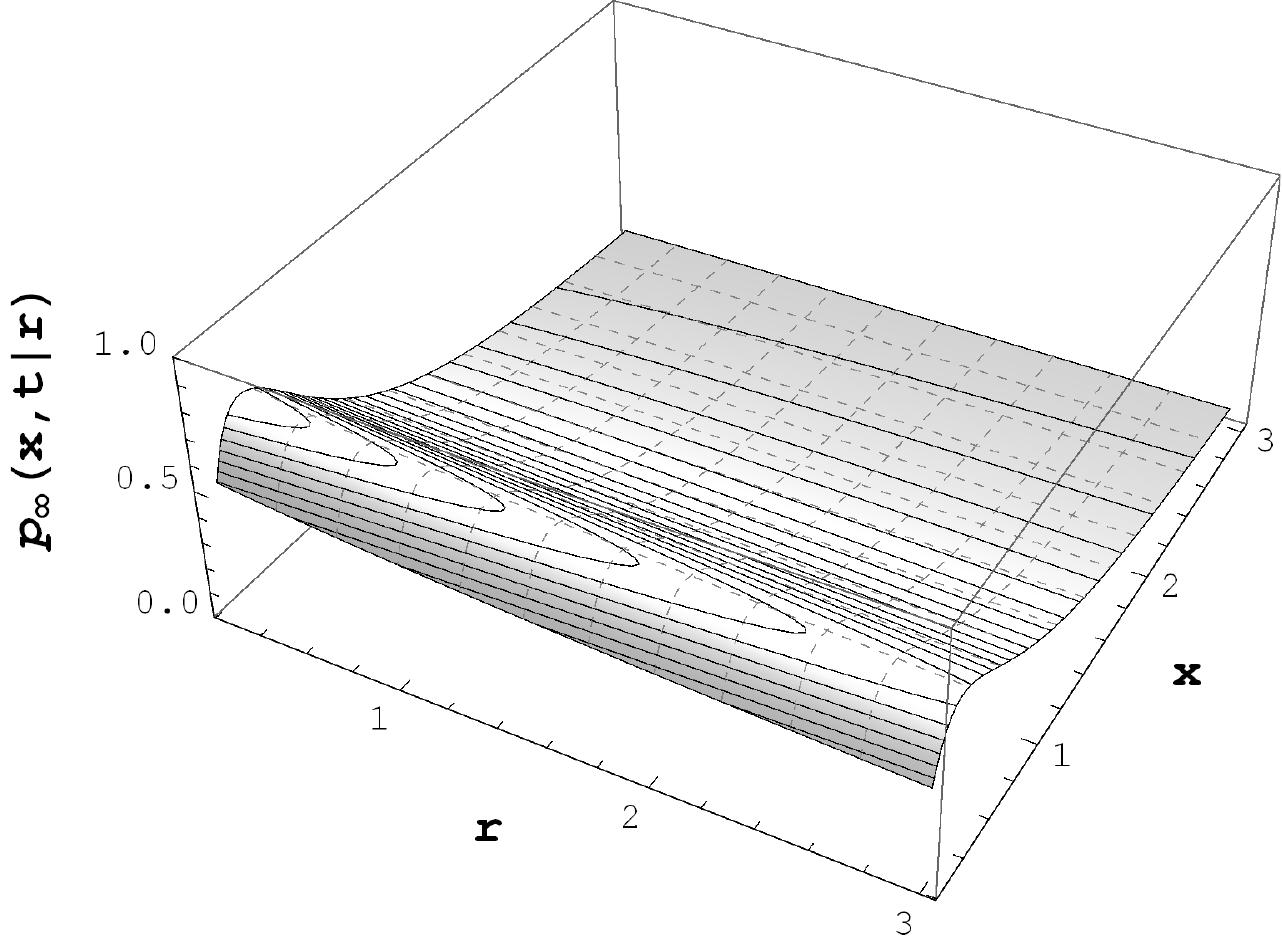}
    }
    \caption{$p_\infty(x,t|r)$ as a function of $x\in[0,3]$ and $r\in[0,3]$ for $\mu=1$ and $\mu=1.5$ when $t=2$.}
    \label{fig:p_inf__t_2_0}
\end{figure}

When $t=2$ (Figure~\ref{fig:p_inf__t_2_0}), the ``dam''-shape is much more pronounced suggesting even greater proximity of $p(x,t|r)$ to the stationary distribution $\rho(x)$. Note that for $\mu=1.5$ the ``dam''-shape is more apparent than for $\mu=1$. This is because for higher values of $\mu$ the convergence to the stationary distribution is quicker.

\begin{figure}[!htb]
    \centering
    \subfloat[$\mu=1.0$.]{\label{fig:p_inf__t_5_0_m_1_0}
        \includegraphics[width=0.47\textwidth]{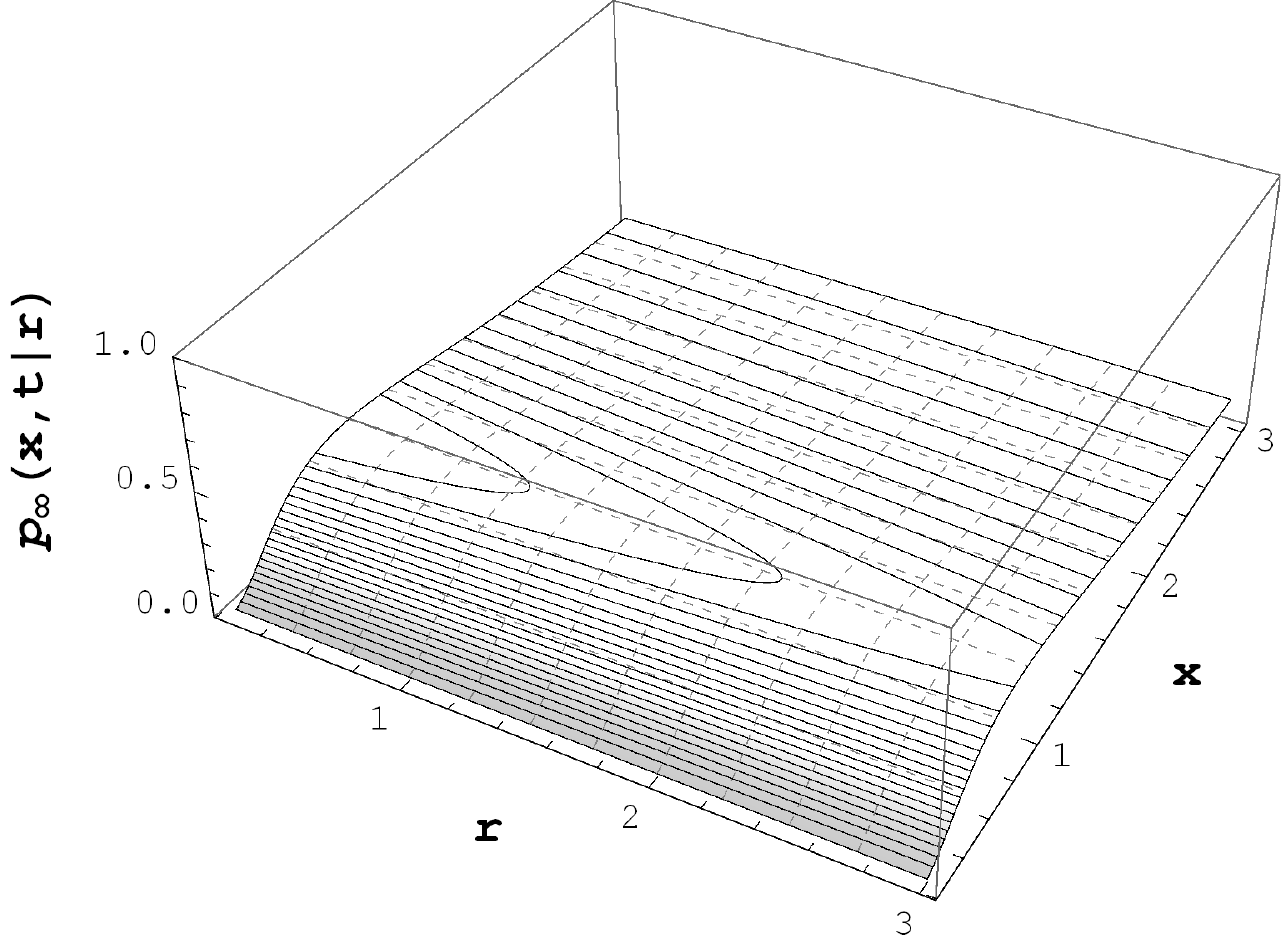}
    }
    \subfloat[$\mu=1.5$.]{\label{fig:p_inf__t_5_0_m_1_5}
        \includegraphics[width=0.47\textwidth]{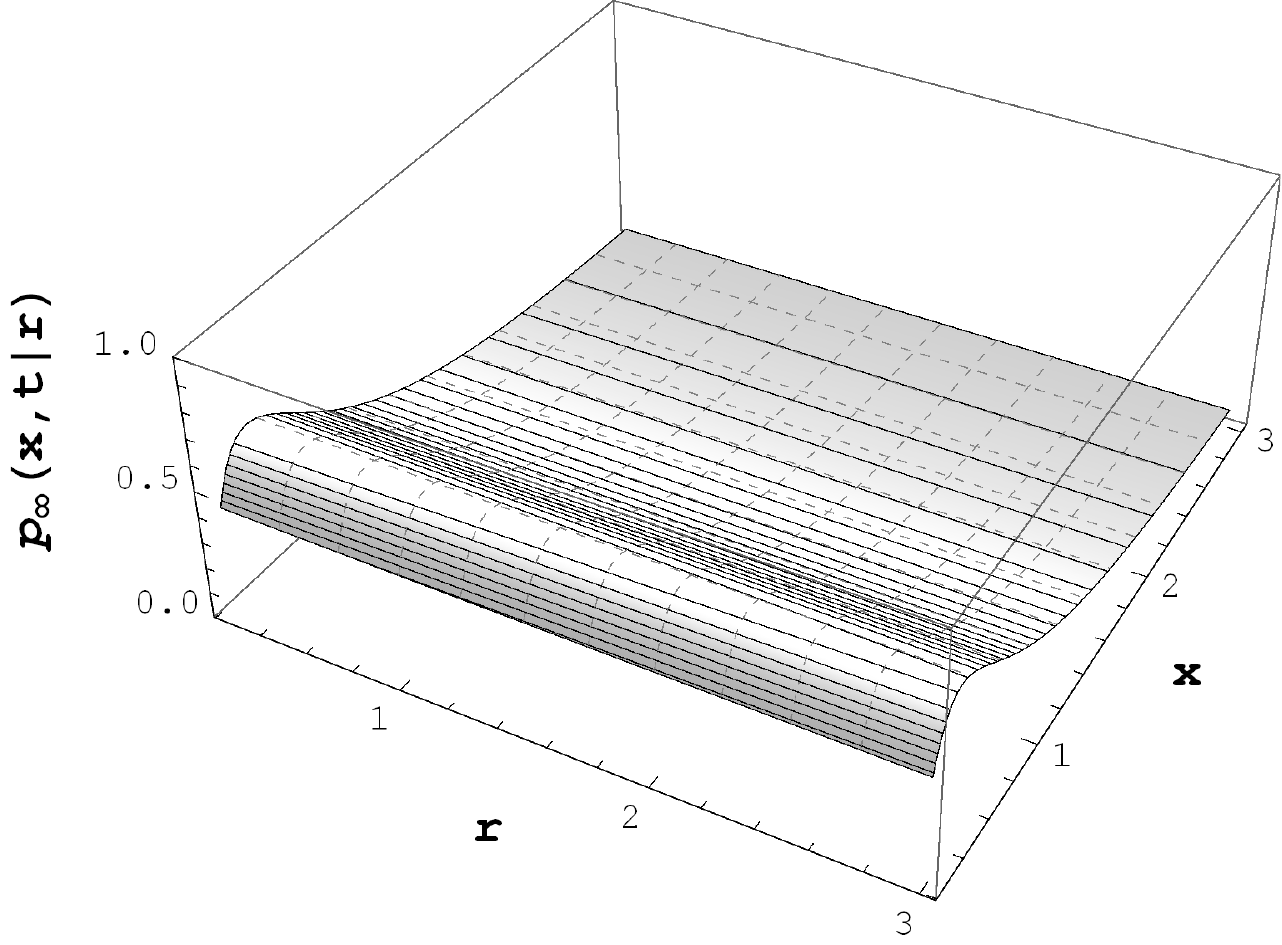}
    }
    \caption{$p_\infty(x,t|r)$ as a function of $x\in[0,3]$ and $r\in[0,3]$ for $\mu=1$ and $\mu=1.5$ when $t=5$.}
    \label{fig:p_inf__t_5_0}
\end{figure}

Figure~\ref{fig:p_inf__t_5_0} shows the results for $t=5$. We observe that for $\mu=1.5$ (Figure~\ref{fig:p_inf__t_5_0_m_1_5}) the transition pdf $p(x,t|r)$ essentially ceased to depend on $r$, which is to say that the stationary distribution has taken effect. However, for $\mu=1$ (Figure~\ref{fig:p_inf__t_5_0_m_1_0}) the stationary distribution hasn't quite taken effect yet, because some residual dependence on $r$ is still visible, although it is not as strong as it was for $t=2$.

\begin{figure}[!htb]
    \centering
    \subfloat[$\mu=1.0$.]{\label{fig:p_inf__t_10_0_m_1_0}
        \includegraphics[width=0.47\textwidth]{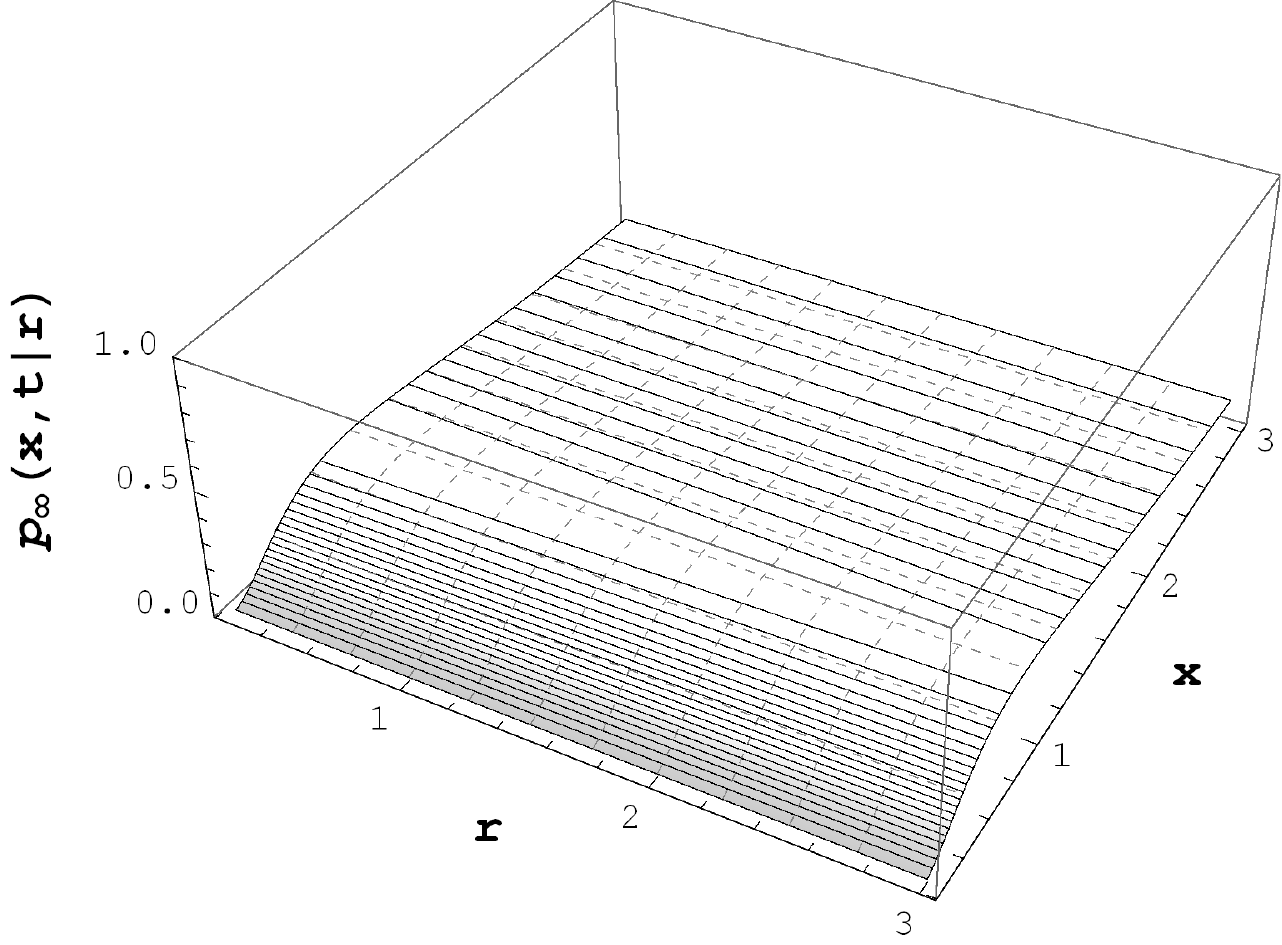}
    }
    \subfloat[$\mu=1.5$.]{\label{fig:p_inf__t_10_0_m_1_5}
        \includegraphics[width=0.47\textwidth]{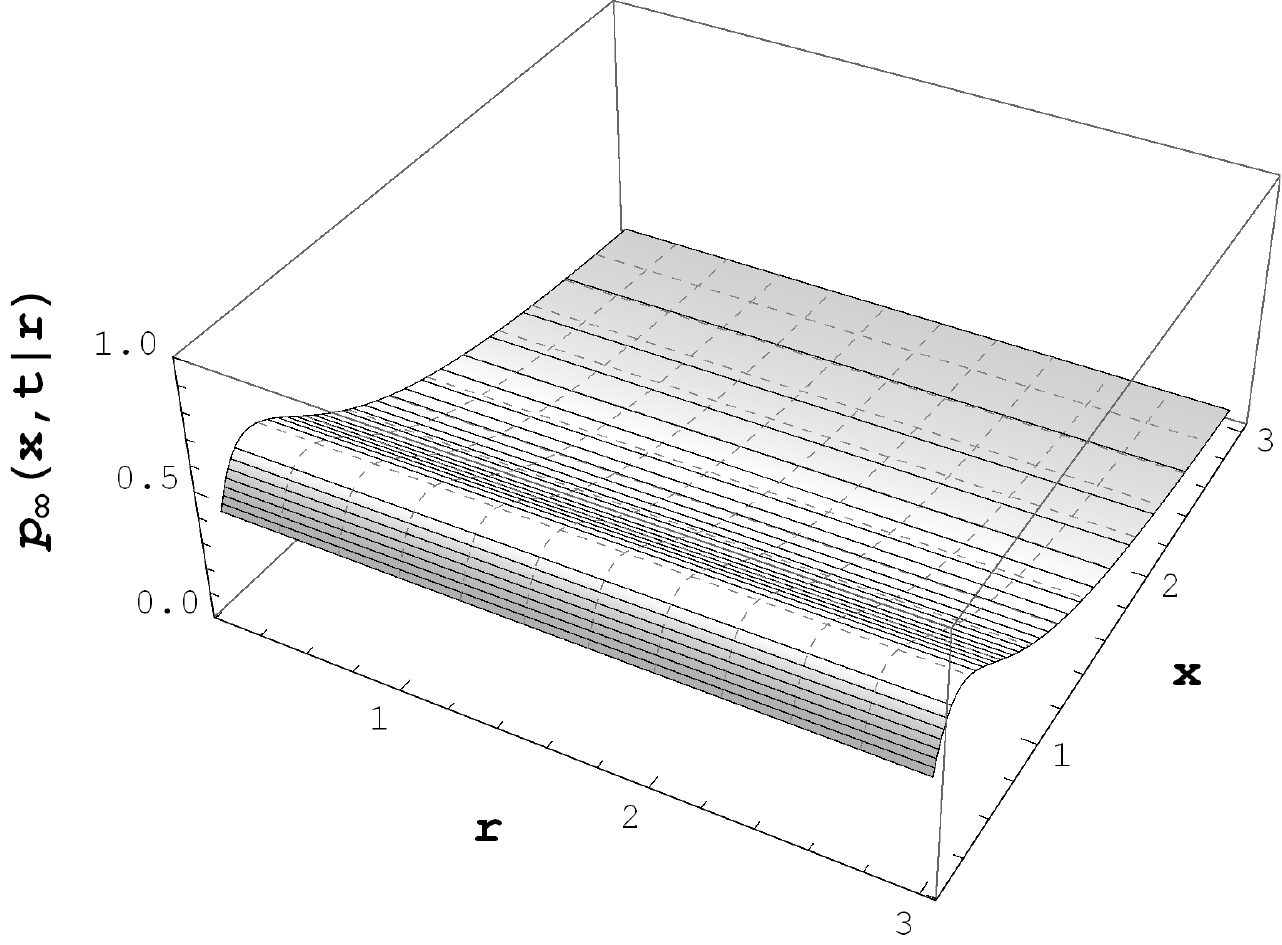}
    }
    \caption{$p_\infty(x,t|r)$ as a function of $x\in[0,3]$ and $r\in[0,3]$ for $\mu=1$ and $\mu=1.5$ when $t=10$.}
    \label{fig:p_inf__t_10_0}
\end{figure}

Finally, the density when $t=10$ is shown in Figure~\ref{fig:p_inf__t_10_0}. For $\mu=1.5$ (Figure~\ref{fig:p_inf__t_10_0_m_1_5}) the plot is essentially no different from its counterpart for $t=5$ (Figure~\ref{fig:p_inf__t_5_0_m_1_5}). This confirms that $p(x,t|r)$ has converged to its stationary distribution. For $\mu=1$ (Figure~\ref{fig:p_inf__t_10_0_m_1_0}) we see that the headstart is no longer a factor. That is, $p(x,t|r)$ has gotten very close to the stationary distribution as well, and increasing $t$ further will not change the picture.

\section{Conclusion}
\label{sec:conclusion}
The aim of this work was to derive a closed-form formula for the transition pdf of the diffusion process generated by the Generalized Shiryaev--Roberts (GSR) change-point detection statistic set up to react to a spontaneous onset of a drift in ``live''-observed standard Brownian motion. Specifically, the transition pdf of interest was under the assumption that the observed Brownian motion is still ``driftless''. We set up the respective Kolmogorov forward equation for the sought transition pdf and solved it analytically with the aid of the Fourier spectral method, which allowed to separate the spacial and the temporal variables. The obtained pdf formula was then implemented in a {\em Mathematica} script to study (numerically) the behavior of the GSR statistic when it is ``running idle'', i.e., when the Brownian motion under surveillance is still drift-free.

Methodologically, the Fourier spectral method can be used to get other characteristics of the GSR statistic as well. For example, the authors are currently in pursuit of obtaining the distribution of the GSR diffusion when the latter is confined to the strip $[0,A]$ with $A>0$, also under the ``no drift'' hypothesis. To the best of our knowledge, this distribution has not yet been obtained, and is needed to evaluate the performance of the actual stopping time associated with the GSR procedure. However, although the solution strategy is no different from the one used in this paper, the condition $R_t^r\in[0,A]$ makes the steps far more involved, and therefore this result will appear elsewhere.

\begin{acknowledgements}
The authors are grateful to Dr.~Olympia Hadjiliadis (Department of Mathematics at Brooklyn College and Department of Mathematics and Department of Computer Science, Graduate Center, City University of New York) for the interest in this work. The authors would also like to thank Dr.~Evgeny Burnaev (PreMoLab.Ru, Kharkevich Institute for Information Transmission Problems, Moscow, Russia) for bringing the work of~\cite{Peskir:Shiryaev2006} to their attention. One more equally big thank you is to go out to the anonymous referee for the detailed and valuable feedback provided on an earlier version of the manuscript. The feedback helped substantially improve the paper's quality.

The effort of A.S.~Polunchenko was supported, in part, by the Simons Foundation (see on the Web at~\url{www.simonsfoundation.org}) via a Collaboration Grant in Mathematics (Award \#\,304574) and by the Research Foundation for the State University of New York at Binghamton via an Interdisciplinary Collaboration Grant (Award \#\,66761).

Last but not least, A.S.~Polunchenko is also personally indebted to the Office of the Dean of the Harpur College of Arts and Sciences at the State University of New York (SUNY) at Binghamton for the support provided through the Dean's Research Semester Award for Junior Faculty granted for the Fall semester of 2014. The Award allowed to focus on this research more fully.
\end{acknowledgements}

\bibliographystyle{spbasic}      
\bibliography{main,physics,special-functions,finance,stochastic-processes,pearson-distributions,differential-equations}

\end{document}